\documentclass[aps,pra,superscriptaddress,twocolumn,floatfix]{revtex4-1}
\usepackage{amsmath}
\usepackage{dcolumn}
\usepackage{latexsym}
\usepackage{graphicx}
\usepackage{amsfonts}
\usepackage{braket}
\usepackage{natbib}
\usepackage{amssymb}
\usepackage{nicefrac}
\usepackage{fancyhdr}
\usepackage[utf8]{inputenc}
\usepackage{dsfont}
\usepackage{colortbl}
\usepackage{xcolor}
\usepackage{subfigure}
\usepackage{psfrag}
\usepackage{tikz}
\usepackage{paralist}
\usepackage{nicefrac}
\usepackage{ulem}
\usepackage{placeins}
\usepackage{mathtools}
\usepackage{bbm}
\usepackage{stfloats}
\usepackage[utf8]{inputenc}
\usepackage{array}
\usepackage{wrapfig}
\usepackage{multirow}
\usepackage{tabularx}
\usepackage{hyperref}

\makeindex
\begin{document}
\title{Roadmap for quantum simulation of the fractional quantum Hall effect}
\author{Michael P. Kaicher}
\affiliation{Theoretical Physics, Saarland University, 66123 Saarbr{\"u}cken, Germany}
\author{Simon B. J\"ager}
\altaffiliation{present address: JILA, NIST, and Department of Physics, University of Colorado, Boulder, Colorado 80309-0440, USA.}
\affiliation{Theoretical Physics, Saarland University, 66123 Saarbr{\"u}cken, Germany}
\author{Pierre-Luc Dallaire-Demers}
\affiliation{Zapata Computing, Inc., 1 Yonge Street, Toronto, Canada}
\author{Frank K. Wilhelm}
\affiliation{Theoretical Physics, Saarland University, 66123 Saarbr{\"u}cken, Germany}
\newcommand*{\simon}{\textcolor{blue}} 
\newcommand*{\pierre}{\textcolor{cyan}} 
\newcommand*{\ryan}{\textcolor{red}} 
\newcommand*{\frank}{\textcolor{green}} 
\newcommand*{\michael}{\textcolor{orange}} 
\begin{abstract}
A major motivation for building a quantum computer is that it provides a tool to efficiently simulate strongly correlated quantum systems. In this work, we present a detailed roadmap on how to simulate a two-dimensional electron gas---cooled to absolute zero and pierced by a strong transversal magnetic field---on a quantum computer. This system describes the setting of the Fractional Quantum Hall Effect (FQHE), one of the pillars of modern condensed matter theory. We give analytical expressions for the two-body integrals that allow for mixing between $N$ Landau levels at a cutoff $M$ in angular momentum and give gate count estimates for the efficient simulation of the energy spectrum of the Hamiltonian on an error-corrected quantum computer. We then focus on studying efficiently preparable initial states and their overlap with the exact ground state for noisy as well as error-corrected quantum computers. By performing an imaginary time evolution of the covariance matrix we find the generalized Hartree-Fock solution to the many-body problem and study how a multi-reference state expansion affects the state overlap. We perform small-system  numerical simulations to study the quality of the two initial state Ans\"{a}tze in the Lowest Landau Level (LLL) approximation.
\end{abstract}
\maketitle

\section{Introduction and overview}
Feynman's conjecture that quantum computers could provide a means for efficiently simulating other quantum systems was proven by Lloyd in 1996 \cite{lloyd1996universal}, where a simulation is considered to be \textit{efficient}, if the computational cost scales at most polynomially with the system size. The following year, Abrams and Lloyd~\cite{abrams1997simulation} showed how a fermionic quantum system could be simulated on such a device in either first or second quantization. 25 years after the proposal of quantum computing \cite{benioff1980computer,feynman1982simulating}, Aspuru-Guzik et al. ~\cite{aspuru2005simulated} demonstrated that the calculation time for the energy of atoms and molecules scales polynomially using quantum algorithms given an initial state with sufficient support on the desired eigenstate. This provided the initial spark to ignite a plethora of studies on molecular electronic systems using quantum computers (see e.g.~\cite{cao2018quantum} for a recent summary). Until then, quantum computing was more famously known for being able to break RSA-encryption \cite{shor1994algorithms} but with the proposed simulation of quantum mechanical systems, quantum computing gained a lot of interest across various fields.

While the study of strongly correlated fermionic systems has been  advocated as a strong suit for quantum computers, one of its most prominent phenomena, the FQHE, has so far been rather sparsely covered ~\footnote{With the exception of Ref.~\cite{johri2017entanglement}, where a quantum algorithm to compute the entanglement spectrum of a quantum state such as the Laughlin state on a quantum computer is presented, but a detailed state creation analysis is not included.}. This effect occurs when electrons are confined to two dimensions \footnote{Only the movement of the electrons is restricted to be (approximately) two-dimensional, we are not referring to the electrons living in a universe with two spatial dimensions, where the form of the Coulomb potential would be quite different from the three dimensional version that we are studying.}, cooled to near absolute zero and are subject to a strong transversal magnetic field. The FQHE manifests itself by a quantization of the Hall conductance over a finite range of the applied magnetic field for certain electron densities and led to various theories and proposed new quasi-particles, such as composite fermions, aimed at describing the observed patterns \cite{jain2007composite}. The plateaus appear at integer or fractional values of $e^2/h$ (where $e$ is the electron charge and $h$ is Planck's constant) and while the integer value plateaus can be well explained by Landau quantization and the effect of disorder (without having to take into account interactions), the Coulomb interaction between electrons plays a key role for the understanding of the observation of plateaus at fractional values of $e^2/h$. Deriving a microscopic theory to explain the fractional plateaus is an active field of research in condensed matter physics. It is believed that quasi-hole and -particle excitations of the ground state of Fractional Quantum Hall (FQH) systems display anyonic statistics, which form the building blocks of a topological quantum computer \cite{freedman2003topological}.

It is not known whether a quantum computer will help us find underlying universal principles that enable us to explain the phenomena of the simulated correlated quantum system. However, a quantum computer does provide a tool to test such theories against exact and approximate solutions for system sizes far beyond what any classical computer will be able to simulate. Our aim is to give an \textit{ab-initio} roadmap that paves the way towards a digital quantum simulation of FQH systems. 

We will consider two different types of quantum computers, on the one hand those which are error-corrected and potentially able to perform millions of gate operations and on the other hand those available today, i.e. error-prone quantum processors, which are limited to execute quantum operations well within their coherence times. 

Within the context of error-corrected quantum computers, we study the scaling of current state-of-the-art quantum algorithms based on the Linear Combination of Unitaries (LCU) method, which is designed to compute the energy spectrum of a given Hamiltonian $H$ to desired precision $\Delta E$ \cite{childs2012hamiltonian}. These quantum algorithms realize a unitary alternative to the usual time evolution operator  \cite{Abrams1999} of the quantum phase estimation algorithm \cite{Kitaev1995} and allow one to efficiently extract information about the Hamiltonian's spectrum. 

While the quantum phase estimation algorithm has a theoretically proven exponential speedup in sampling a Hamiltonian or eigenvalue sampling of a  unitary matrix generated by the exponential of a sparse matrix, current and near-term quantum computers are not fault tolerant and applying the quantum phase estimation algorithm is impossible due to the tremendous amount of gate operations that need to be applied coherently. On the other hand, algorithms which are applicable to Noisy Intermediate-Scale Quantum (NISQ) \cite{preskill2018quantum} devices, i.e. non-error-corrected quantum computers---such as the Variational Quantum Eigensolver (VQE) \cite{peruzzo2014variational,mcclean2016theory}---are restricted to coherence time limited circuit depths and are of heuristic nature. Such heuristic algorithms are intuitively compelling and capable of systematic refinement, but lack rigorous bounds on their performance \footnote{It is a topic of current discussion which type of shallow circuit Ansatz might provide an advantage over classical algorithms \cite{napp2019efficient} and the study of VQE-type algorithms revealed other challenges, such as exponentially vanishing gradients \cite{mcclean2018barren}.}. 

\squeezetable
\begin{table}[h!]
	\caption{List containing all abbreviations used in main text.}
	\renewcommand{\arraystretch}{1.2}
	\begin{tabular}{|c|l|}
		\hline
		Abbreviation &\\ \hline 
		FQH(E) & Fractional Quantum Hall (Effect)\\
		(L)LL & (Lowest) Landau Level\\
		NISQ & Near Intermediate-Scale Quantum\\ 
		VQE & Variational Quantum Eigensolver\\
		LCU & Linear Combination of Unitaries\\
		FGS & Fermionic Gaussian State\\
		CM & Covariance Matrix\\
		ASCI & Adaptive Sampling Configuration Interaction\\
		FCI & Full Configuration Interaction\\
		\hline
	\end{tabular}
	\label{tab:abbreviations}
\end{table}

A large part of our work will focus on finding an initial state $\ket{\Psi_{\text{init}}}$ (sometimes also called a trial-, or reference state) which approximates the ground state $\ket{\Psi_0}$ of $H$. We restrict ourselves to initial states which are \textit{efficiently computable on a classical}- and \textit{efficiently preparable on a quantum computer} and need to possess a non-vanishing overlap with the desired eigenstate of the Hamiltonian. We engage in the task of finding an initial state which would serve as the starting point of a given quantum algorithm to approximate the ground state of the Hamiltonian describing the FQH system and how one could then extract physically meaningful properties from it, e.g. by means of computing the one- and two-particle correlation functions. The problem of finding an initial state $\ket{\Psi_{\text{init}}}$ with above mentioned prerequisites has largely been ignored in literature and has only recently been studied thoroughly for a variety of electronic systems \cite{tubman2018postponing}, with the exception of FQH systems. Such initial states are not only of interest for NISQ algorithms, but also for quantum-error-corrected algorithms such as in Refs.~\cite{berry2018improved,ge2019faster}.

This work is structured as follows. In Section~\ref{system_hamiltonian} we present the Hamiltonian of interacting electrons in a disk geometry pierced by a strong magnetic field. We provide efficiently computable analytical expressions for the two-body coefficients of the Hamiltonian in second quantization and describe how this Hamiltonian can be mapped from the fermionic to the spin basis using the Jordan-Wigner transformation. In Section~\ref{lcu}, we present an efficient strategy for simulating the FQHE on an error-corrected quantum computer using a quantum algorithm proposed in Ref.~\cite{Berry2019} based on the LCU method \cite{childs2012hamiltonian}. In Section~\ref{init_state} we discuss the classically efficient computation of initial states from the family of Fermionic Gaussian States (FGS), which can be implemented on NISQ devices. We extend our discussion by including a multi-reference state approach suited for error-corrected quantum computers, which is based on linear combination of Slater determinants using a state-of-the-art quantum chemistry algorithm \cite{tubman2018postponing}. The results of the numerical simulations are presented in Section~\ref{numerics}, where we compare fidelities of the respective initial state and the actual ground state $\ket{\Psi_0}$ for small system sizes (which corresponds to the typical size of current cloud-based quantum computing hardware). In Section~\ref{discussion} we discuss possible avenues one could explore in order to improve the FQH Hamiltonian model. We sum up our findings in Section~\ref{conclusion}. The Appendix provides further details mainly on the derivations of the Coulomb matrix elements, an alternative Hamiltonian simulation strategy based on the self-inverse matrix decomposition strategy of ~\cite{Berry2013}, the equations of motion for the imaginary time evolution of the Covariance Matrix (CM), and helpful relations for the implementation of the multi-reference state approach. We provide a list of the main abbreviations in Table~\ref{tab:abbreviations} and of symbols used throughout the main text in Table~\ref{tab:list_of_symbols}.

\squeezetable
\begin{table*}[t]
	\caption{This table lists and explains the most important symbols appearing in the main text and refers to their respective definitions or appearances in the last column. We include a listing of all abbreviations used in the main text at the end of the table.   \label{tab:list_of_symbols}}
	\renewcommand{\arraystretch}{1.2}
	\begin{tabular}{|c||l||c|}
		\hline
		Symbol & Explanation & Equation\\ \hline\hline
		$O(f(x))\ (\tilde O(f(x)))$ & Limiting behaviour of a function $f(x)$ for large values of $x$  (suppressing polylogarithmic factors)& \\
		$H$ & System Hamiltonian (both in first and second quantized representation)  $H=H_1+H_2$, where $H_1$ & \eqref{sh},\eqref{sec1},\eqref{measurement1},\\& contains all single particle terms and $H_2$ contains all interaction terms between particles &\eqref{ham_1},\eqref{single2}\\
		$\ket{\Psi_0}$&Exact ground state energy of the system Hamiltonian $H$&\\
		$\ket{\Psi_{\text{init}}}$&Initial state / reference state that approximates $\ket{\Psi_0}$&\\
		$N$ & Denotes largest considered Landau level (LL). Individual LLs are indexed by $P_1=0,1,\dots, N$ & \\
		$M$ & Denotes cutoff in angular momentum, individual angular momenta are indexed by $P_2=0,1,\dots, M$ & \\
		$N_{\text{so}}\approx NM$, $N_{\text{el}}$ & Number of spin-orbitals and electrons---in numerical simulations one chooses $N\ll M$ & \\
		$\mathbf P,\mathbf Q,\mathbf R,\mathbf S$ & Quantum number tuples, $\mathbf P = (P_1,P_2)$, ..., $\mathbf S=(S_1,S_2)$. Convention: $P_\Sigma=P_1+P_2$ &\\
		$\psi_{\mathbf P}(\mathbf r)$ & Single particle wave function, depending on $\mathbf P$ and particle coordinate $\mathbf r$, eigenfunctions of $H_1$ & \eqref{bs2}\\
		$f_{\mathbf P\mathbf Q}$ & One-body Hamiltonian coefficients of $H_1$ in its second quantized representation & \eqref{sec2},\eqref{EP1}\\
		$h_{\mathbf P\mathbf Q\mathbf R\mathbf S}$ & Coulomb-matrix elements of $H_2$ in its second quantized representation & \eqref{sec3}\\
		$h_{\mathbf P\mathbf Q\mathbf R\mathbf S}^{(i)}$ & Coulomb-matrix elements for case (i) where $P_2-S_2\geq 0$ (case (ii) $P_2-S_2<0$ follows from symmetry) & \eqref{lauri2}\\
		$c_{\mathbf P},c_{\mathbf P}^\dag,c_p^\dag,c_p$ & Fermionic annihilation and creation operators satisfying the anticommutation relations. & \eqref{sec1},\eqref{single2}\\
		$F_A^{(4)}[\dots]$ & Lauricella function (here, a finite hypergeometric series) & \eqref{lauri}, \eqref{llauri2}\\
		$(\lambda)_n$ & Pochhammer symbol (also known as rising factorial), result of division of two Gamma functions & \eqref{poch}\\
		$R_d$ & Radius of simulated 2D disk. Describes the disk boundary due to the cutoff in angular momenta at $M$ & \eqref{Rd}\\
		$\nu$ & Filling factor, defined as the number of electrons per flux quantum penetrating the disk  & \eqref{nu1},\eqref{nu2}\\
		$M_{mn}^l$ & Coulomb matrix elements in the LLL approx., identical to $h_{\mathbf P\mathbf Q\mathbf R\mathbf S}$ for $P_1=Q_1=R_1=S_1=0$. & \eqref{coul2}\\
		$l,m,n$ & Coefficients of $M_{mn}^l$, corresponding to $l=P_2-S_2$, $m=S_2$ and $n=Q_2$ & \\
		$\Sigma$& Number of non-zero terms of the system Hamiltonian $H$ in Jordan-Wigner representation, $\Sigma\propto L$ &\eqref{measurement1}\\
		$L$ & Two meanings: Either number of terms in LCU expansion, or number of determinants in ASCI state &\eqref{ham_1},\eqref{asci1} \\
		$U_\ell$ & Unitary matrices from the linear combination of unitaries (LCU) method & \eqref{ham_1}\\
		$\omega_\ell$ & Coefficients of the LCU method, related to Hamiltonian coefficients, $H=\sum_\ell \omega_\ell U_\ell$ & \eqref{ham_1}\\
		$\lambda$ & Sum of absolute values of all $\omega_\ell$ values. Important for determining complexity of LCU method & \eqref{ham_1}\\
		$\textsc{select}$ & Oracle of LCU method efficiently implementing the $U_\ell$ in superposition & \eqref{select} \\
		$\textsc{prepare}$ & Oracle of LCU method generating a linear combination of states indexed by $\ell$ and weighted by $\omega_\ell/\lambda$ & \eqref{prepare} \\
		$C_S,C_P$ & Gate complexity of $\textsc{select}$ and $\textsc{prepare}$ oracles & \eqref{comp1} \\ 
		$\Delta E$ & Target precision of the energy in phase estimation &\eqref{scaling_1}\\
		$\Gamma$ & Covariance matrix (CM) characterizing the FGS & \eqref{single3}\\
		$f_{pq},h_{pqrs}$ & One- and two-body coefficients of $H$ (the latter being chosen to fulfill Eq.~\eqref{sym2}) in LLL &\eqref{single2}\\
		$h_m(\Gamma),E_m$ & Mean-field matrix of system Hamiltonian $H$ and corresponding mean-field energy & \eqref{single9},\eqref{single11}\\ 
		$\textit{cdtes}$, $\textit{tdtes}$ & Number of core- and target-space determinants of ASCI algorithm & \\
		$C_i$ & Expansion coefficients of the ASCI algorithm & \eqref{asci1}\\ 
		$\ket{D_i}$ & Expansion determinants of the ASCI algorithm & \eqref{asci1}\\ 
		$A_i$ & Perturbed wave functions amplitudes over all single and double excitations in ASCI & \eqref{asci2}\\
		\hline 
	\end{tabular}
\end{table*}

\section{System Hamiltonian\label{system_hamiltonian}}
This Section presents the considered Hamiltonian of electrons under Coulomb repulsion in a strong magnetic field. We present analytical solutions in symmetric gauge disk geometry for the one- and two-body matrix elements of the second quantized Hamiltonian which allows for Landau Level (LL) mixing. A similar result has been reported for a spherical geometry in \cite{wooten2014configuration}.

We introduce Landau levels and the single-particle basis states in Section~\ref{sec_sh}. Since Section~\ref{secd} presents very technical results which are only required when one wants to study large systems with various LLs and are not needed for the understanding of the rest of this work, the non-interested reader can skip to section~\ref{LLL} where we describe the system Hamiltonian in the LLL which is the setting of our numerical simulations. We conclude this section by showing how to map the fermionic Hamiltonian through the Jordan-Wigner transformation to a spin Hamiltonian in Section~\ref{jw}.
\subsection{The Hamiltonian in first quantization\label{sec_sh}}
We analyze a two-dimensional electron gas in the $x$-$y$ plane with no disorder and in a strong magnetic field ${\bf B}=(0, 0 ,B)^T$ allowing for discarding the spin degrees of freedom. It is described by the Hamiltonian \cite{jain2007composite}
\begin{align}
H=&H_1+H_2,\label{sh}
\end{align}
that is the sum of the single-particle terms
\begin{align}
H_1=&\sum_j\frac{1}{2m_b}\left(-i\hbar\mathbf \nabla_j+\frac{e}{c}\mathbf A(\mathbf r_j)\right)^2\label{H1},
\end{align}
and the two-particle interactions described by
\begin{align}
H_2=&\frac{e^2}{\epsilon}\sum_{j<k}\frac{1}{|\mathbf r_j-\mathbf r_k|}\label{H2}.
\end{align}
Eq.~\eqref{H1} describes the energy of the electrons with effective band mass $m_b$ in absence of interactions and in a constant magnetic field $\mathbf B = \boldsymbol{\nabla}\times\mathbf A$. We use the vector potential in symmetric gauge \cite{jain2007composite}
\begin{align}
\mathbf A=\frac{{\bf B}\times {\bf r}}{2},
\end{align}
which breaks translational symmetry in $x$- and $y$-direction, but preserves the rotational symmetry about the origin, which makes the angular momentum a good quantum number. Here,
${\bf r}_j=(x_j,y_j,0)$ is  the position of the electron $j$ in the $x$-$y$ plane and $e$, $c$ the electron charge and speed of light, respectively. The Hamiltonian in Eq.~\eqref{H2} describes the Coulomb interactions between the atoms where $\epsilon=4\pi\epsilon_0$ and $\epsilon_0$ is the dielectric constant. 

\subsubsection{Eigenfunctions of single-particle Hamiltonian\label{disk}}
The eigenfunctions and energies of $H_1$ (Eq.~\eqref{H1}) are known analytically and will be used later to describe the full Hamiltonian $H$ (Eq.~\eqref{sh}) in second quantization. The corresponding single-particle states are the basis of choice for the second quantized Hamiltonian and are described by a set of two quantum numbers $\mathbf P =(P_1,P_2)$, where  $P_1$ denotes the LL and the second quantum number $P_2$ denotes the angular momentum. 

For a given LL $P_1=0,1,\dots$, the angular momentum can take the values $P_2 = -P_1, -P_1+1,... $. The single-particle wave function are given by
\begin{align}
\psi_{\mathbf P}(\mathbf r)=&\frac{(-1)^{P_1}}{\sqrt{2\pi}}\sqrt{\frac{P_1!}{2^{P_2}(P_1+P_2)!}}L_{P_1}^{(P_2)}\left(\tfrac{r^2}{2}\right)z^{P_2}e^{-\frac{1}{4}r^2},\label{bs2}
\end{align}
with  $z_j=x_j-iy_j=r_je^{-i\theta_j}$ and $r_j=\sqrt{z_jz_j^*}$ being the complex particle coordinates and where we defined the associated Laguerre polynomials of degree $n$ and order $\alpha$
\begin{align}
L_n^{(\alpha)}(x)=\sum_{i=0}^n(-1)^i\binom{n+\alpha}{n-i}\frac{x^i}{i!}\label{bs3}.
\end{align}
The functions $\psi_{\mathbf P}(\mathbf r)$ fulfill 
\begin{align}
H_1\psi_{\mathbf P}(\mathbf r)=E_{P_1}\psi_{\mathbf P}(\mathbf r),    
\end{align}
with eigenenergy
\begin{align}
E_{P_1}=\hbar \omega_c\left(P_1+\frac{1}{2}\right), \label{EP1}
\end{align}
where $\omega_c=eB/(\hbar m_b c)$ is called the cyclotron frequency. The discrete energy levels of the kinetic terms---the LLs---are the workhorse of the quantum Hall problem. The formation of Landau levels provide the key insight for the understanding of the integer quantum Hall effect and the fractional quantum Hall effect  can be explained by a splitting of a Landau level into ``Landau-like'' energy levels in presence of interactions \cite{jain2007composite}. We note that other basis choices might provide a more compact representation of the system Hamiltonian (even though it is unclear how simple restrictions to single LLs would be possible in such representations), however the Landau level basis is a reasonable representation of the FQH problem. 
\subsection{The Hamiltonian in second quantization\label{secd}}
For the purpose of simulating the quantum mechanical system on a quantum computer we derive the Hamiltonian in second quantization. The second quantized form of $H$  (as in Eq.~\eqref{sh}) in the single-particle basis of Eq.~\eqref{bs2} is given by \cite{helgaker2014molecular} 
\begin{align}
H=\sum_{{\bf P},{\bf Q}}f_{{\bf P}{\bf Q}}c_{{\bf P}}^\dag c_{{\bf Q}} +\frac{1}{2}\sum_{{\bf P},{\bf Q},{\bf R},{\bf S}}h_{{\bf P}{\bf Q}{\bf R}{\bf S}}c_{{\bf P}}^\dag c_{{\bf Q}}^\dag c_{{\bf R}} c_{{\bf S}},\label{sec1}
\end{align}
where the one- and two-body coefficients are given by
\begin{align}
f_{{\bf P}{\bf Q}}=&\int d\mathbf r\psi_{\mathbf P}(\mathbf r)^*H_1\psi_{\mathbf Q}(\mathbf r),\label{sec2}\\
h_{{\bf P}{\bf Q}{\bf R}{\bf S}}=&\iint d\mathbf r_1d\mathbf r_2\psi_{\mathbf P}(\mathbf r_1)^*\psi_{\mathbf Q}(\mathbf r_2)^*H_2\psi_{\mathbf S}(\mathbf r_1)\psi_{\mathbf R}(\mathbf r_2),\label{sec3},
\end{align}
using Eq.~\eqref{H1} and Eq.~\eqref{H2}, respectively and $\int d\mathbf r_j =\int_{-\infty}^{\infty}dx_j\int_{-\infty}^{\infty} dy_j$ for $j=1,2$. The operators $c_{{\bf P}}$ and $c_{{\bf P}}^{\dag}$ are the fermionic annihilation and creation operators fulfilling the anticommutator relations $\{c_{{\bf P}},c_{{\bf Q}}\}=0$ and $\{c_{{\bf P}},c_{{\bf Q}}^{\dag}\}=\delta_{{\bf P}{\bf Q}}$ with the Kronecker delta $\delta_{{\bf P}{\bf Q}}$. The total number of terms in Eq.~\eqref{sec1} scales as $O(N_{\text{so}}^4)$, where the number of spin orbitals is approximately given by $N_{\text{so}}\approx NM$, with $N$ and $M$ denoting the cutoff in the number of LLs and angular momentum and $N\ll M$.
\subsubsection{Kinetic term\label{kinetic}}
Since we used an eigenbasis of $H_1$ for the representation of the Hamiltonian in second quantization the one-particle coefficients are diagonal. They are given by
\begin{align}
f_{{\bf P}{\bf Q}}=E_{P_1}\delta_{{\bf P}{\bf Q}},\label{kin1}
\end{align}
where the eigenenergies are given by Eq.~\eqref{EP1}.
\subsubsection{Coulomb term\label{coulomb}}
In order to derive the second quantized representation of the Coulomb operator, we will give an analytical solution of Eq.~\eqref{sec3} which is valid for all possible values of $\mathbf P,\mathbf Q,\mathbf R,\mathbf S$. For the evaluation of Eq.~\eqref{sec3}, we use the Fourier representation \cite{tsiper2002analytic},
\begin{align}
\frac{1}{|\mathbf r_{1}-\mathbf r_2|}=\frac{1}{2\pi}\int d\mathbf q \frac{1}{q}e^{i\bf q(\bf r_1-\bf r_2)},\label{four}
\end{align}
where $\int d\mathbf q = \int_{-\infty}^{\infty} dq_x\int_{-\infty}^{\infty} dq_y$. 
We insert Eq.~\eqref{four} into Eq.~\eqref{sec3}, using polar coordinates we obtain
\begin{align}
h_{\mathbf{PQRS}}
=&\frac{e^2\mathcal C}{\epsilon}\int_0^\infty dq K_{\mathbf P, \mathbf S}(q) K_{\mathbf R, \mathbf Q}(q)^* \delta_{P_2-S_2,R_2-Q_2},\label{hpqrs}
\end{align}
where the delta-function on the right-hand side reflects the conservation of total angular momentum and we defined the coefficient
\begin{align}
\mathcal{C}=& \frac{(-1)^{P_1+Q_1+S_1+R_1}}{\pi^2 2^{(P_2+Q_2+S_2+R_2+4)/2}}\sqrt{\frac{P_1!Q_1!S_1!R_1!}{P_\Sigma!Q_\Sigma!S_\Sigma!R_\Sigma!}}\label{derivation4}
\end{align}
and the integral
\begin{align}
K_{\mathbf P, \mathbf S}(q)=&\frac{2\pi}{i^{S_2-P_2}}\int_0^\infty dr_1r_1^{P_2+S_2+1}e^{-\tfrac{1}{2}r_1^2}\nonumber\\ &\times L_{P_1}^{(P_2)}\left(\tfrac{r_1^2}{2}\right)L_{S_1}^{(S_2)}\left(\tfrac{r_1^2}{2}\right)J_{P_2-S_2}(qr_1).\label{derivation9}
\end{align}
In the above derivation we made use of the integral representation of the Bessel function \cite{de2017integral}
\begin{align}
J_n(x)=&\frac{i^n}{2\pi}\int_0^{2\pi}d\theta e^{i(n\theta-x\cos(\theta))}.\label{derivation8}
\end{align}
Integrating the right-hand side of Eq.~\eqref{derivation9} leads to
\begin{align}
K_{\mathbf P, \mathbf S}(q)
=&2^{S_2+1}\pi i^{P_2-S_2}\tfrac{S_\Sigma!}{S_1!}(-1)^{P_1+S_1}q^{P_2-S_2}e^{-\tfrac{1}{2}q^2}\nonumber\\&\times L_{P_1}^{(S_1-P_1)}\left(\tfrac{q^2}{2}\right)L_{S_\Sigma}^{(P_\Sigma-S_\Sigma)}\left(\tfrac{q^2}{2}\right)\label{derivation12}.
\end{align}
By substituting $x_j=q_j^2/2$ for $j=1,2$ and using  Eq.~\eqref{derivation12}, the solution for Eq.~\eqref{hpqrs} is given by
\begin{widetext}
	\begin{align}
	h_{\mathbf{PQRS}}^{(i)}=&\frac{e^2\mathcal C^{(i)}}{\epsilon}
	\frac{\Gamma(p)}{2^p}\left(\prod_{k=1}^4\binom{n_k+\alpha_k}{n_k}\right) F_A^{(4)}\left[\begin{matrix}
	p,-n_1,-n_2,-n_3,-n_4;\\\alpha_1+1,\alpha_2+1,\alpha_3+1,\alpha_4+1;
	\end{matrix}\frac{1}{2},\frac{1}{2},\frac{1}{2},\frac{1}{2}\right]
	\delta_{P_2-S_2,R_2-Q_2}\label{lauri2},
	\end{align}
\end{widetext}
	where $\Gamma(x)$ denotes the Gamma function, $\mathcal C^{(i)}$, $p$, and $n_j$ and $\alpha_j$ with $j=1,2,3,4$, are given in Table~\eqref{tab:integ1} for all possible values of the quantum numbers $P_1,P_2,Q_1,\dots,S_2$. The superscript $(i)$ indicates that we consider the case where $P_2-S_2\geq 0$, the remaining case $(ii)$, where $P_2-S_2< 0$ can be obtained from symmetry, as indicated in the last row of Table~\ref{tab:integ1}. The function
	\begin{align}
	&F_A^{(4)}\left[\begin{matrix}
	p,-n_1,-n_2,-n_3,-n_4;\\\alpha_1+1,\alpha_2+1,\alpha_3+1,\alpha_4+1;
	\end{matrix}\frac{1}{2},\frac{1}{2},\frac{1}{2},\frac{1}{2}\right]\nonumber\\
	=&\sum_{k_1,k_2,k_3,k_4=0}^{\infty}\frac{(p)_{k_1+k_2+k_3+k_4}(-n_1)_{k_1}(-n_2)_{k_2}(-n_3)_{k_3}}{(\alpha_1+1)_{k_1}(\alpha_2+1)_{k_2}(\alpha_3+1)_{k_3}}\nonumber\\&\times \frac{(-n_4)_{k_4}}{(\alpha_4+1)_{k_4}2^{k_1+k_2+k_3+k_4}k_1!k_2!k_3!k_4!},\label{lauri}
	\end{align}
is known as the Lauricella function, where 
\begin{align}
(\lambda)_n= \frac{\Gamma(\lambda+n)}{\Gamma(\lambda)}\label{poch}
\end{align}
is the rising factorial (Pochhammer symbol). Since $-n_1,-n_2,-n_3,-n_4$ in Eq.~\eqref{lauri} are non-positive integers, the series terminates after a finite number of terms. One can represent the Lauricella function as an integral of a product of lower-order hypergeometric functions \cite{padmanabham2000summation}, which results in
\begin{align}
F_A^{(4)}\left[\begin{matrix}
p,-n_1,-n_2,-n_3,-n_4;\\\alpha_1+1.\alpha_2+1,\alpha_3+1,\alpha_4+1;
\end{matrix}\frac{1}{2},\frac{1}{2},\frac{1}{2},\frac{1}{2}\right]
= \boldsymbol \xi \cdot \boldsymbol{(p)},\label{llauri2}
\end{align}
where we defined the two column vectors
\begin{align}
\boldsymbol{(p)}=&((p)_0,(p)_1,\dots,(p)_{n_1+n_2+n_3+n_4})^T\label{vec_1}\\
\boldsymbol \xi=&(\xi_0,\xi_1,\dots,\xi_{n_1+n_2+n_3+n_4})^T,\label{vec_2}
\end{align}
with convolution coefficients
\begin{align}
\xi_k=\sum_{p=0}^k&\sum_{q=0}^p\sum_{r=0}^q \frac{(-n_1)_r(-n_2)_{q-r}}{(\alpha_1+1)_{r}(\alpha_2+1)_{q-r}(\alpha_3+1)_{p-q}}\nonumber\\
&\times\frac{(-n_3)_{p-q}(-n_4)_{k-p}}{(\alpha_4+1)_{k-p}r!(q-r)!(p-q)!(k-p)!2^k}.\label{vec_el}
\end{align}
A detailed derivation of the results of this section can be found in Appendix~\ref{app:analytical_matrix_elements}. While recent work provided analytic expressions for the two-body matrix elements in finite spherical quantum Hall systems \cite{wooten2014configuration}, we are not aware of prior analytic expressions for the two-body matrix elements that include general LL mixing for a two-dimensional disk geometry setting. 

\squeezetable
\begin{table*}
	\caption{\raggedleft This table defines the coefficients $\mathcal C^{(i)}$, $p$, $n_j$ and $\alpha_j$ for $j=1,2,3,4$ in Eq.~\eqref{lauri2} and defines the explicit integral form of the Coulomb matrix elements of Eq.~\eqref{derivation20}. The various sub-cases (i.i)-(i.ix) are defined in Table~\ref{tab:regime} in the appendix. Note that the values for $\alpha_j$ follow from the definition of $L_{[n_1,n_2,n_3,n_4]}$ in Eq.~\eqref{derivation2} and we defined the compact notation $P_\Sigma=P_1+P_2$. The expressions for case (ii) do not need to be calculated, as they follow from $h_{\mathbf{PQRS}}^{(ii)}=h_{\mathbf{SRQP}}^{(i)}$ as indicated by the last row of this table.\raggedright}
	\renewcommand{\arraystretch}{1.2}
	\begin{tabular}{|c||c|c|c|c|c|c|c|}
		\hline
		\multicolumn{8}{|c|}{$h_{\mathbf{PQRS}}^{(i)}=\mathcal C^{(i)}\int_0^\infty dx x^{p-1}e^{-2x}L_{[n_1,n_2,n_3,n_4]}$} \\ \cline{1-8}
		\hline\hline
		case & $\mathcal C^{(i)}$& $p$& $[n_1,n_2,n_3,n_4]$ & $\alpha_1$&$\alpha_2$&$\alpha_3$&$\alpha_4$\\ \hline
		(i.i)&$(-1)^{S_1-P_1+Q_1-R_1}\sqrt{\tfrac{P_1!P_\Sigma!R_\Sigma!R_1!}{2S_1!S_\Sigma!Q_\Sigma!Q_1!}}$&$S_1-P_1+Q_\Sigma-R_\Sigma+1/2$&$[P_1,P_\Sigma,R_1,R_\Sigma]$& $S_1-P_1$&$S_\Sigma-P_\Sigma$&$Q_1-R_1$&$Q_\Sigma-R_\Sigma$ \\ \hline
		(i.ii)&$(-1)^{S_\Sigma-P_\Sigma}\sqrt{\tfrac{P_\Sigma!P_1!Q_\Sigma!R_1!}{2Q_1!S_1!S_\Sigma!R_\Sigma!}}$&$S_1-P_1+1/2$&$[P_1,P_\Sigma,R_1,Q_\Sigma]$& $S_1-P_1$&$S_\Sigma-P_\Sigma$&$Q_1-R_1$&$R_\Sigma-Q_\Sigma$\\ \hline
		(i.iii)&$(-1)^{S_\Sigma-P_\Sigma+R_1-Q_1}\sqrt{\tfrac{P_1!P_\Sigma!Q_1!Q_\Sigma!}{2S_1!S_\Sigma!R_1!R_\Sigma!}}$&$S_1-P_1+R_1-Q_1+1/2$&$[P_1,P_\Sigma,Q_1,Q_\Sigma]$& $S_1-P_1$&$S_\Sigma-P_\Sigma$&$R_1-Q_1$&$R_\Sigma-Q_\Sigma$\\ \hline
		(i.iv)&$(-1)^{Q_\Sigma-R_\Sigma}\sqrt{\tfrac{P_1!S_\Sigma!R_1!(R_\Sigma!}{2P_\Sigma!Q_1!Q_\Sigma!S_1!}}$&$Q_1-R_1+1/2$&$[P_1,S_\Sigma,R_1,R_\Sigma]$& $S_1-P_1$&$P_\Sigma-S_\Sigma$&$Q_1-R_1$&$Q_\Sigma-R_\Sigma$\\ \hline
		(i.v)&$\mathcal B= \sqrt{\frac{P_1!Q_\Sigma!S_\Sigma!R_1!}{2P_\Sigma!Q_1!S_1!R_\Sigma!}}$&$P_2-S_2+1/2$&$[P_1,S_\Sigma,R_1,Q_\Sigma]$& $S_1-P_1$&$P_\Sigma-S_\Sigma$&$Q_1-R_1$&$R_\Sigma-Q_\Sigma$\\ \hline
		(i.vi)&$(-1)^{R_1-Q_1}\sqrt{\tfrac{P_1!Q_\Sigma!Q_1!S_\Sigma!}{2P_\Sigma!S_1!R_\Sigma!R_1!}}$&$R_1-Q_1+P_2-S_2+1/2$&$[P_1,S_\Sigma,Q_1,Q_\Sigma]$& $S_1-P_1$&$P_\Sigma-S_\Sigma$&$R_1-Q_1$&$R_\Sigma-Q_\Sigma$\\ \hline
		(i.vii)&$(-1)^{P_1-S_1+Q_\Sigma-R_\Sigma}\sqrt{\tfrac{S_1!S_\Sigma!R_1!R_\Sigma!}{2P_\Sigma!P_1!Q_\Sigma!Q_1!}}$&$P_1-S_1+Q_1-R_1+1/2$&$[S_1,S_\Sigma,R_1,R_\Sigma]$& $P_1-S_1$&$P_\Sigma-S_\Sigma$&$Q_1-R_1$&$Q_\Sigma-R_\Sigma$\\ \hline
		(i.viii)&$(-1)^{P_1-S_1}\sqrt{\tfrac{Q_\Sigma!S_1!S_\Sigma!R_1!}{2P_\Sigma!P_1!Q_1!R_\Sigma!}}$&$P_\Sigma-S_\Sigma+1/2$&$[S_1,S_\Sigma,R_1,Q_\Sigma]$& $P_1-S_1$&$P_\Sigma-S_\Sigma$&$Q_1-R_1$&$R_\Sigma-Q_\Sigma$\\ \hline
		(i.ix)&$(-1)^{P_1-S_1+R_1-Q_1}\sqrt{\tfrac{Q_1!Q_\Sigma!S_1!S_\Sigma!}{2P_\Sigma!P_1!R_1!R_\Sigma!}}$&$(P_\Sigma-S_\Sigma+R_1-Q_1+1/2)$&$[S_1,S_\Sigma,Q_1,Q_\Sigma]$& $P_1-S_1$&$P_\Sigma-S_\Sigma$&$R_1-Q_1$&$R_\Sigma-Q_\Sigma$\\ \hline\hline
		\multicolumn{8}{|c|}{$h_{\mathbf{PQRS}}^{(ii)}=h_{\mathbf{SRQP}}^{(i)}$} \\ \cline{1-8}
		\hline
	\end{tabular}
	\label{tab:integ1}
\end{table*}

Due to the conservation of angular momentum in Eq.~\eqref{lauri2}, the number of terms in the Hamiltonian scales at most as $O(N^4M^3)$, reducing the order of the polynomial by one in $M$ (which is the most costly parameter, since $N\ll M$).
\subsection{System Hamiltonian in the LLL\label{LLL}}
As indicated by the absence of any $P_2$ dependence in Eq.~\eqref{EP1}, each LL is degenerate in the absence of interactions. For the LLL, the single-particle wave functions in symmetric gauge in Eq.~\eqref{bs2} simplify to
\begin{align}
\psi_{(0,P_2)}(\mathbf r)=&\frac{1}{\sqrt{2\pi2^{P_2}P_2!}}z^{P_2}e^{-\frac{1}{4}r^2}.
\end{align}
The above wave functions are peaked on concentric rings, whose distance from the origin is proportional to the square root of the the angular coordinate, 
\begin{align}
R_d=l_B\sqrt{2(M+1)},    \label{Rd}
\end{align}
where $l_B=\sqrt{\hbar c/(eB)}$. Not only for computational purposes is it of interest to introduce a cut-off for the angular momentum $M$ that fulfills
\begin{align}
P_2\leq M.
\end{align}
Physically, this can be interpreted as a confinement of the electrons on a disk with radius $R_d$. Coulomb repulsion will typically force the electrons to fly away from each other while a confinement counteracts this repulsion by forcing them to stay within a confined region of space. Note that by fixing the Radius $R_d$, the cut-off $M$---and therefore the degeneracy of the LLs---can be tuned by changing the magnetic field $B$. 

The cyclotron energy $\hbar \omega_c$ is proportional to the transversal magnetic field $B$ and sets the spacing between the LLs. From the form of the wave functions one can deduce that the degeneracy within each LL is approximately given by $N_{\text{deg}}=AB/\phi_0$, where $A$ is the area spanned by the confinement and $B$ is the transversal magnetic field \cite{jain2007composite}.  We will restrict our simulations to systems where the number of particles $N_{\text{el}}$ is smaller than the degeneracy $N_{\text{deg}}$ within each LL. For such a configuration, in the limit of sufficiently large magnetic field, only states in the LLL will be occupied and we can neglect coupling to states to higher LLs. 

The filling factor $\nu$ is the number of electrons per flux quantum penetrating the sample and defined as \cite{jain2007composite} 
\begin{align}
\nu = \rho\frac{\phi_0}{B},\label{nu1}
\end{align}
where $\rho$ is the 2D electron density and the flux quantum $\phi_0= h c/e$. Assuming a homogeneous density of the $N_{\mathrm{el}}$ electrons on a disk with radius $R_d$ (Eq.~\eqref{Rd}) and restricting to the LLL, we get the density  $\rho=N_{\mathrm{el}}/(\pi R_d^2)$. With this we can derive 
\begin{align}
\nu = \frac{N_{\mathrm{el}}}{M+1}.\label{nu2}
\end{align}
Fixing the filling factor $\nu$ at constant magnetic field thus also results in a constant electron density $\rho$. The factor $e^2/\epsilon$ which appears in front of the Coulomb term in Eq.~\eqref{hpqrs} merely sets the overall energy scale when working in the LLL, which is why we set it equal to one in our numerical simulations and the integral expressions. For simulations that incorporate the effect of LL mixing, one would have to include the factor $e^2/\epsilon$ in front of the Coulomb terms again, as well as the cyclotron energy $\hbar\omega_c$, since it sets the energy spacing between LLs and depends on the strength of the transversal magnetic field. 

For the disk geometry in the LLL approximation, we use the compact result of Ref.~\cite{tsiper2002analytic}, where the matrix elements $h_{\mathbf P\mathbf Q\mathbf R\mathbf S}=M_{mn}^l$ are expressed as finite sums of fractions of factorials
\begin{align}
M_{mn}^l=&C_{lmn}\left(A_{mn}^lB_{nm}^l+B_{mn}^lA_{nm}^l\right)\label{coul2}\\
C_{lmn}=&\sqrt{\frac{(m+l)!(n+l)!}{m!n!}}\frac{\Gamma(l+m+n+3/2)}{\pi 2^{l+m+n+2}}\label{coul3}\\
A_{mn}^l=&\sum_{i=0}^m\binom{m}{i}\frac{\Gamma(i+1/2)\Gamma(l+i+1/2)}{(l+i)!\Gamma(l+n+i+3/2)}\label{coul4}\\
B_{mn}^l=&\sum_{i=0}^m\binom{m}{i}\frac{\Gamma(i+1/2)\Gamma(l+i+1/2)}{(l+i)!\Gamma(l+n+i+3/2)}\nonumber\\&\times(l+2i+1/2),\label{coul5}  
\end{align}
where all indices $P_1,Q_1,R_1,S_1$ are equal to zero, $l=P_2-S_2$, $m=S_2$, $n=Q_2$. The fact that instead of the four angular momentum quantum numbers $P_2,Q_2,R_2,S_2$, only three such numbers, $l,m,n$, appear in Eq.~\eqref{coul2} is a manifestation of the  conservation of angular momentum. Due to the appearance of fractions of large integers in the coefficients in Eqs.~\eqref{coul3}-\eqref{coul5}, numerical implementation of Eq.~\eqref{coul2} for large system sizes has to be performed with great caution. A rash implementation will lead to numerical instabilities already below one hundred spin-orbitals. We compared the coefficients in Eq.~\eqref{coul2} with Eqs.~\eqref{lauri2} and \eqref{llauri2} for various system sizes and fillings within the LLL and they were in exact agreement up to numerical precision errors. 
\subsection{Mapping the second quantized fermionic Hamiltonian to  the Pauli basis\label{jw}}
If one wants to simulate a fermionic system on a quantum computer, one needs to map the fermionic creation and annihilation operators onto qubit operators. Various such encodings have been studied, each with its own benefits and drawbacks \cite{jordan1993paulische,bravyi2002fermionic,bravyi2017tapering,verstraete2005mapping,seeley2012bravyi,tranter2015b,moll2016optimizing,whitfield2016local,tranter2018comparison,steudtner2017lowering}. Let $\sigma_j^{x,y,z}$ denote the Pauli-$X,Y,Z$ matrix. We choose the Jordan-Wigner transformation \cite{jordan1993paulische} where a single fermionic raising or lowering operator is mapped to a simple qubit raising or lowering operator $\sigma_j^{\pm}=(\sigma_j^x\mp i\sigma_j^y)/2$, at the cost of up to $N_{\text{so}}-1$ additional Pauli-$Z$ operators,
\begin{align}
c_j^\dag =& \sigma_j^+\prod_{k=j+1}^{N_{\text{so}}}\sigma_k^z\label{jw1}\\
c_j =& \sigma_j^-\prod_{k=j+1}^{N_{\text{so}}}\sigma_k^z\label{jw2}.
\end{align} 
The Pauli-$Z$ operator's role is to produce the sign factor that appears when acting with a fermionic operator on a Fock state  \cite{helgaker2014molecular}, leading to the canonical fermionic anti-commutation relations. Inserting Eqs.~\eqref{jw1} and \eqref{jw2} into Eq.~\eqref{sec1} results in the qubit Hamiltonian which is equivalent to the original system Hamiltonian and that can be written as a sum of positive real-valued coefficients $\omega_j$ (not to be confused with the cyclotron frequency $\omega_c$) times a phase factor $e^{i\theta_j}$ (whose sole purpose is to absorb the minus sign of negative Hamiltonian coefficients $f_{pq}$ and $h_{pqrs}$) times a tensor product of Pauli operators $P_j\in\{ \mathds 1_{2}, \sigma^x, \sigma^y, \sigma^z \}^{\otimes N_{\text{so}}}$,
\begin{align}
H=\omega_0e^{i\theta_0}\mathds 1_{2^{N_f}}+\sum_{j=1}^{\Sigma}\omega_j e^{i\theta_j} P_j\label{measurement1}.
\end{align}

\begin{figure}[h!]
	\centering
	\includegraphics[width=.49\textwidth]{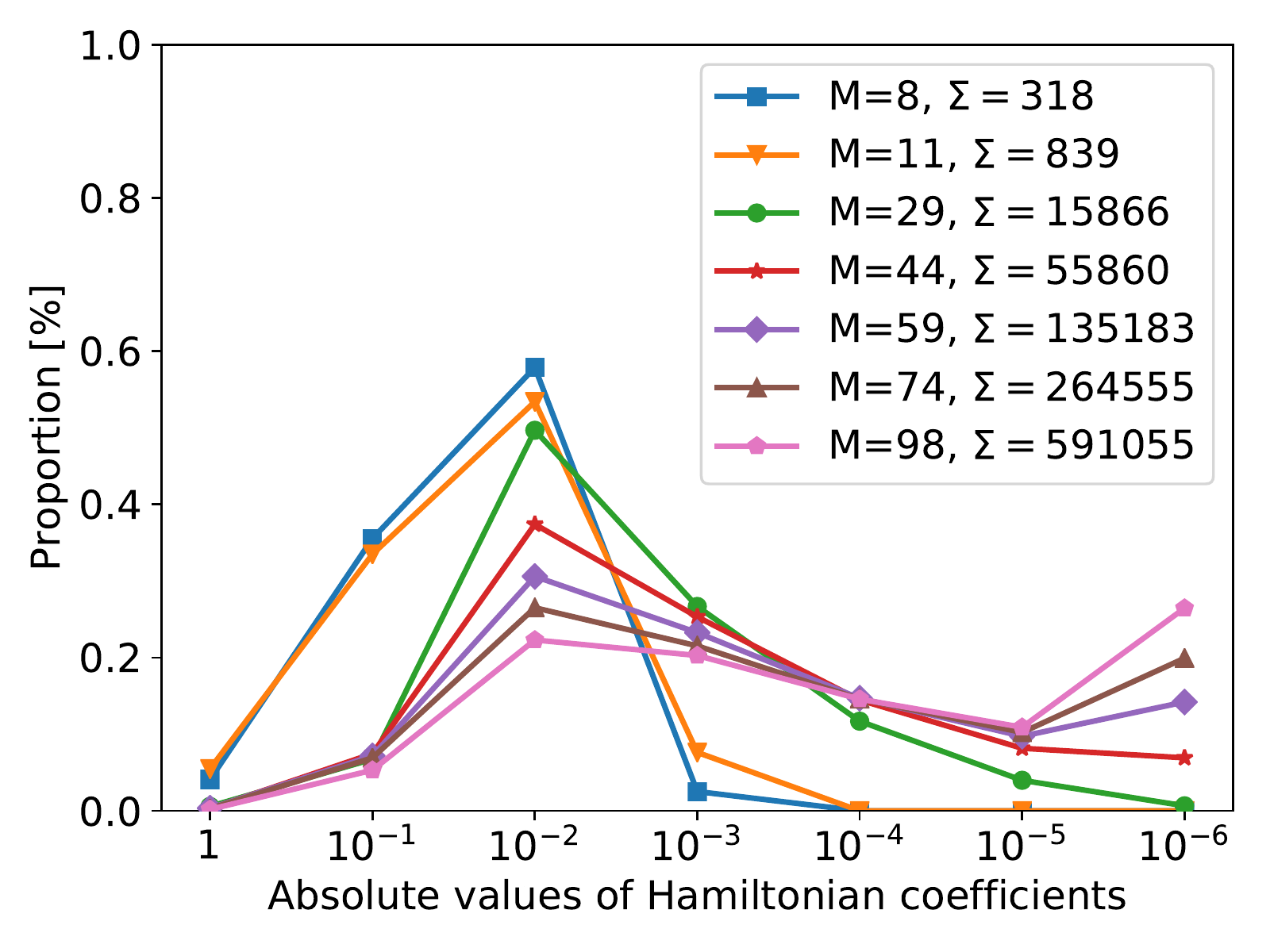}
	\caption{Scatter plot of the distribution of coefficient magnitudes of the real-valued coefficients of Eq.~\eqref{measurement1}  for various numbers of spin orbitals at a filling $\nu=1/3$ in the LLL $N=0$. Each discrete point on the $x$-axis describes the value range of $|\omega_j|^2$, for instance $10^{-2}$ contains all values $x$ within the range $x\in [10^{-3},10^{-2})$. Note that the largest coefficients are on the left-hand side, while the smallest coefficients are on the right-hand side of the graph. The $y$-axis displays the ratio of the number of terms within the range of a given $x$ w.r.t. the total number $\Sigma$ of non-zero Hamiltonian coefficients in Eq.~\eqref{measurement1}.\label{fig:ham_c}} 
\end{figure}

In Fig.~\ref{fig:ham_c}, we study the distribution of the range of values for the sum of squared coefficients for various system sizes. A general shift of the coefficients towards much smaller coefficient magnitudes with growing system size becomes apparent. Scaling analysis like these are important for determining upper bounds on the number of required measurements to estimate the ground state energy within a given precision and for various variational Ans\"atze $U(\boldsymbol{\theta})$ of the VQE, such as the Hamiltonian variational Ansatz~\cite{wecker2015progress}, as it depends on both the number and the relative weight of the non-zero terms appearing in the Hamiltonian of Eq.~\eqref{measurement1}.

Now that we have derived the system Hamiltonian of the FQH system in second quantization, the following section will give an estimate for the gate complexity to estimate its ground state energy using a state-of-the-art Hamiltonian simulation algorithm designed for an error-corrected universal quantum computer.
\section{Hamiltonian simulation through linear combination of unitaries\label{lcu}}
While Trotter based methods are likely the most efficient technique for implementing quantum simulations of the fractional quantum Hall effect on near-term quantum computers, other methods might be more competitive within cost models appropriate for error-corrected quantum computing. Within fault-tolerance the key cost model of interest is often the number of non-Clifford gates (usually T gates) required for the simulation because within error-correcting codes, T gates require orders of magnitude more resource to realize than Clifford gates and thus limit the calculation size \cite{Fowler2012}.

When studying quantum simulations of electronic structure within the context of error-correction we usually focus on state preparation using phase estimation. The quantum phase estimation algorithm \cite{Kitaev1995} allows one to measure the phase accumulated on a quantum register under the action of a unitary operator. To estimate this phase to within error $\epsilon$ one must apply the unitary a number of times scaling as ${O}(1/\epsilon)$. Furthermore, some varieties of phase estimation allow one to perform this measurement projectively, which enables sampling in the eigenbasis of the unitary. In the context of quantum simulation, this unitary usually corresponds to time evolution under the system Hamiltonian $H$ for time $t$ with eigenvalues $e^{-i H t}$ \cite{Abrams1999}. However, some recent papers \cite{berry2018improved,babbush2018encoding} have advocated instead that one perform phase estimation on a quantum walk with eigenvalues $e^{\pm i\arccos(H)}$ which is often possible to realize with lower overhead. Performing phase estimation on either operator will give the same information \cite{babbush2018encoding}. For either strategy, performing projective phase estimation on this operator will collapse the system register $\ket{\psi}$ to an eigenstate of the Hamiltonian with a probability that depends on the initial overlap between $\ket{\psi}$ and the eigenstate of interest. Thus, if $H \ket{n} = E_n \ket{n}$ then performing phase estimation will project the system register to the eigenstate $\ket{n}$, and readout the associated eigenvalue $E_n$ with probability $p_n =  \left\langle \psi\! \mid\! n \right \rangle \!\!\left\langle n \! \mid \! \psi \right \rangle$. Therefore, the number of times that one must repeat phase estimation to prepare eigenstate $\ket{n}$ with high probability scales as ${O}(1/p_n)$.
Here, we focus on the implementation of circuits that realize a quantum walk with eigenvalues $e^{\pm i\arccos(H)}$. The same strategies can be used to synthesize time evolution with additional logarithmic overheads, by using quantum signal processing \cite{low2017optimal}.

The FQHE Hamiltonian described in Section~\ref{system_hamiltonian} is a special case of the electronic structure Hamiltonian studied in quantum chemistry. Currently, the lowest T complexity quantum algorithms for simulating chemistry are all based on LCU methods \cite{childs2012hamiltonian}. LCU methods include Taylor series methods \cite{Berry2015}, qubitization \cite{low2016hamiltonian}, and Hamiltonian simulation in the interaction picture \cite{Low2018}. These methods were applied to realize quantum algorithms for electronic structure in Refs.~\cite{kivlichan2017bounding,BabbushSparse1,babbush2017exponentially,babbush2018quantum,babbush2018encoding,Berry2019} and elsewhere. All LCU methods involve simulating the Hamiltonian as a linear combination of unitaries,
\begin{align}
H = \sum_{\ell=1}^{L} \omega_\ell \, U_\ell, \qquad \lambda = \sum_{\ell=1}^L \left| \omega_\ell \right |,\label{ham_1}
\end{align}
where $U_\ell$ are unitary operators, $\omega_\ell$ are scalars, and $\lambda$ is a parameter that determines the complexity of these methods. The Hamiltonians in this paper satisfy this requirement once mapped to qubits (see Section~\ref{jw}) since strings of Pauli operators are unitary.

LCU methods perform quantum simulation in terms of queries to two oracle circuits defined as
\begin{align}
\textsc{select} \ket{\ell} \ket{\psi} & \mapsto \ket{\ell} U_\ell \ket{\psi},\label{select}\\
\textsc{prepare} \ket{0}^{\otimes \log(L)} & \mapsto \sum_{\ell=1}^L \sqrt{\frac{\omega_\ell}{\lambda}} \ket{\ell},\label{prepare}
\end{align}
where $\ket{\psi}$ is the system register and $\ket{\ell}$ is an ancilla register which usually indexes the terms in the linear combinations of unitaries in binary and thus contains $\log(L)$ ancillae. LCU methods can perform time-evolution with gate complexity scaling as
\begin{equation}
{\tilde O}\left( \left(C_S + C_P\right)\lambda\, t\right),\label{comp1}
\end{equation}
where $\tilde O$ indicates that polylogarithmic factors in the scaling are suppressed, $C_S$ and $C_P$ are the gate complexities of  $\textsc{select}$ and  $\textsc{prepare}$ respectively, and $t$ is time. Specifically, if the goal is to implement quantum phase estimation to estimate energies or project into an eigenstate of the Hamiltonian then the T cost (with constant factors) scales as
\begin{equation}
\frac{\sqrt{2}\pi \lambda \left(C_S + C_P\right)}{\Delta E},\label{scaling_1}
\end{equation}
where $\Delta E$ is the target precision in phase estimation (in the same units as $\lambda$) \cite{babbush2018encoding}.

In order to simplify scaling arguments, we will only consider scaling in terms of the cutoff in angular momentum $M$ and neglect the contribution due to the $N$ LLs in the following. In numerical studies of the FQHE, one typically only considers a handful of LLs (most of the times only a single one), while trying to push the state space describing each LL (described by $M$) as high as possible, thus $N\ll M$, which leads to $O(N_{\text{so}})\approx O(NM)\approx O(M)$. We also neglect the cost of performing the inverse quantum Fourier transformation, which is a negligible additive cost to the complexity of phase estimation ~\cite{nam2018approximate}.

To implement the LCU oracles one must be able to coherently (i.e., using a quantum circuit) translate the index $\ell$ into the associated $U_\ell$ and $\omega_\ell$. $U_\ell$ are related to the second quantized fermion operators (e.g., the $c_{{\bf P}}^\dag c_{{\bf Q}}^\dag c_{{\bf R}} c_{{\bf S}}$) and the $\omega_\ell$ are related to the coefficients (e.g., the $h_{{\bf P}{\bf Q}{\bf R}{\bf S}}$) described in Section~\ref{coulomb}. The $U_\ell$ have a structure that is straightforward to unpack in a quantum circuit using techniques described in Refs.~\cite{babbush2018encoding,Berry2019}. In particular, those papers show that one can implement the $\textsc{select}$ oracle with a complexity of ${O}(M)$ T gates and low constant factors in the scaling. In the context of quantum chemistry the $\omega_\ell$ are typically challenging to compute directly from this index. However, as described in the prior section, for the Hamiltonians of interest in this paper we are able to compute the $\omega_\ell$ efficiently from $\ell$ (which is essentially equivalent to computing the $h_{{\bf P}{\bf Q}{\bf R}{\bf S}}$ from the indices ${\bf P}, {\bf Q}, {\bf R}$ and ${\bf S}$). Still, the primary bottleneck for this implementation will be the realization of $\textsc{prepare}$ rather than $\textsc{select}$.

\begin{figure}[h!]
	\centering
	\includegraphics[width=0.49\textwidth]{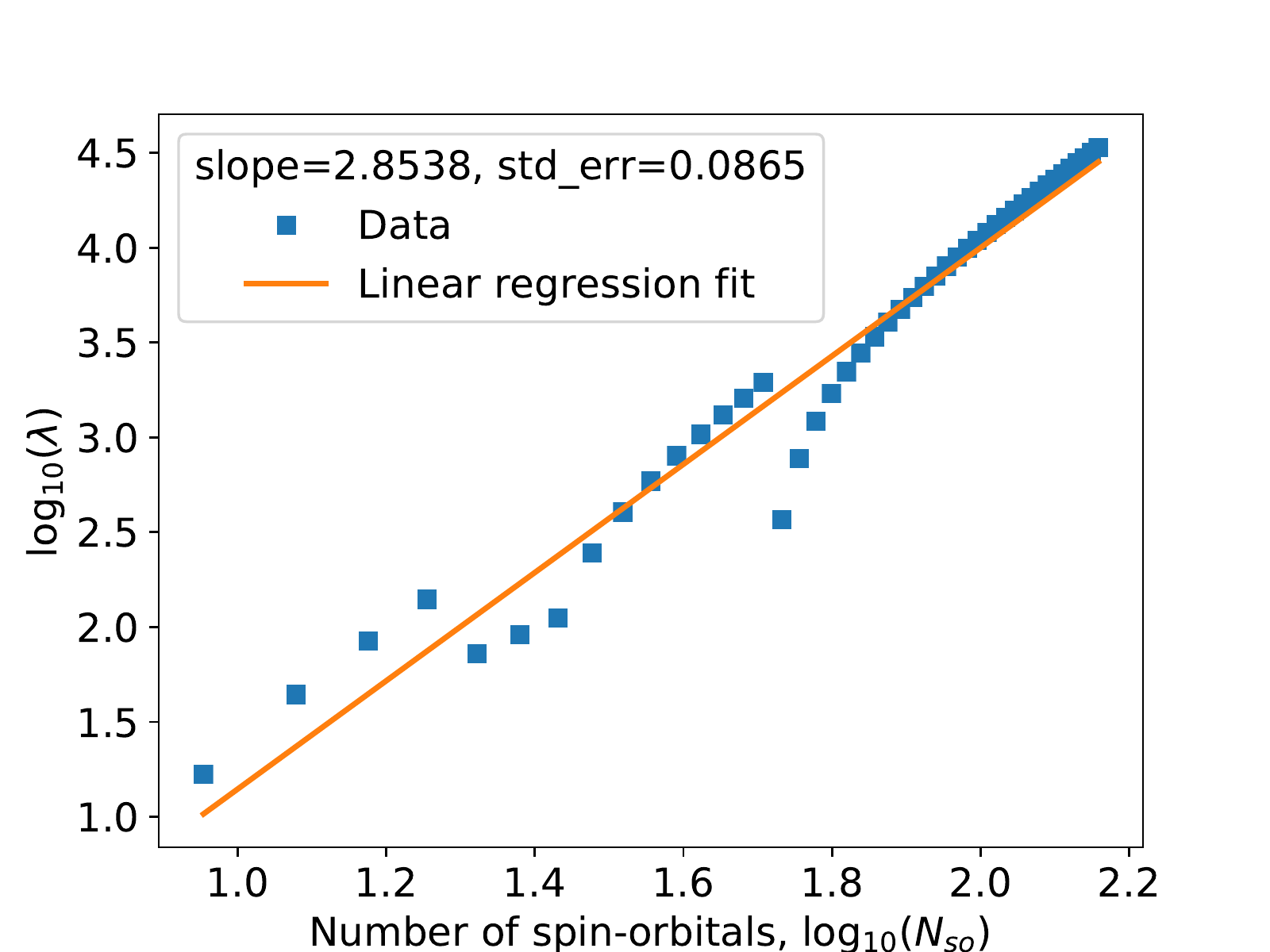}
	\caption{Linear regression fit of the scaling of the parameter $\lambda = \sum_{\ell=1}^L |\omega_\ell|$ for various system sizes in the LLL ranging from $M=8,11,14,\dots,144$, each blue square representing a system instance. Both, the $x$- and $y$-axes are on a (base-10) logarithmic scale.  \label{fig:scaling}} 
\end{figure}

The spectrum of the fractional quantum Hall effect Hamiltonian derived in Section~\ref{system_hamiltonian} can be simulated on a quantum computer using the low rank factorization strategy described in Ref.~\cite{Berry2019}. There, it is shown that one can perform phase estimation on an arbitrary basis electronic structure system with T complexity scaling as $O(N_{\text{so}}^{3/2} \lambda / \Delta E)$ where this $\lambda$ is the true 1-norm of the Hamiltonian as defined in Eq.~\eqref{ham_1}. In Fig~\ref{fig:scaling}, we plot the scaling of this quantity for various system sizes in the LLL, where $O(N_{\text{so}})=O(M)$ and $M$ again denoting the cutoff in angular-momentum. Empirically we find that in this context $\lambda = {O}(M^{2.85})$ which leads to an overall T complexity of $O(M^{4.35} / \Delta E)$. Since the approach described in ~\cite{Berry2019} is currently the lowest scaling approach to electronic structure simulations, the low rank factorization method with T complexity $O(M^{4.35} / \Delta E)$ is at present the most effective strategy in the current literature for simulating FQHE Hamiltonians restricted to the LLL. It should be noted that having a closed form for the one- and two-body Hamiltonian coefficients did not lead to a better scaling when we used an alternative simulation strategy, see Appendix~\ref{ryan} for more details. 
\section{Finding an initial state\label{init_state}}

In this section we focus on the preparation of initial states on a gate-model based quantum computer. Our aim is to find an initial state $\ket{\Psi_{\text{init}}}$ which approximates the true ground state $\ket{\Psi_0}$ of the system Hamiltonian and possesses a non vanishing overlap
\begin{align}
|\braket{\Psi_{\text{init}}|\Psi_{0}}|^2> 0,   \label{overlap}
\end{align}
where the left-hand side of the above equation defines the state fidelity. Moreover, we require these initial states to be both \textit{efficiently computable on classical computers} and \textit{efficiently preparable on a gate-based quantum computer}. Note that the initial state can serve as the starting point of quantum algorithms such as in Ref.~\cite{ge2019faster}, which is of course in general no longer efficiently simulatable on classical computers. The efficient construction of $\ket{\Psi_{\text{init}}}$ and the realization of $e^{-iHt}$, or in our case $e^{\pm i \arccos{H}}$ (neglecting the inverse quantum Fourier transform) are the main black box operations needed for Hamiltonian simulation. Even though one is in general not able to construct the accurate eigenstate, one can show that the success probability of measuring the desired energy using quantum phase estimation improves quadratically with the overlap of an initial state that is not the eigenstate of $H$ \cite{nielsen2002quantum}. 

Quantum algorithms designed to perform a digital quantum simulation of large system sizes often ignore the problem of finding an initial state fulfilling the above prerequisites with reasonable support on the ground state \cite{tubman2018postponing}, even though it is well-known that overlaps of approximate states will decrease exponentially with system size due to the Van Vleck catastrophe \cite{kohn1999nobel}. While it is unclear whether this orthogonality catastrophe can ever be overcome, it is possible to delay the vanishing of the overlap by using more elaborate initial states. 

We consider two algorithms to find a suitable initial state for our FQH system. The first algorithm, described in Section~\ref{hf}, makes use of generalized Hartree-Fock theory to find an initial state within the family of FGS following an imaginary time evolution \cite{kraus2010generalized}. The second algorithm, introduced in Section~\ref{asci}, uses a deterministic algorithm which samples from a large set of Slater determinants (which are contained in the family of FGS), to find a subset of determinants that are likely to have a large support on the exact ground state \cite{tubman2018postponing}. This state can be efficiently constructed using the \textsc{prepare} oracle defined in Eq.~\eqref{prepare}. While only the former algorithm is well-suited for NISQ era quantum computers, both algorithms may be used for state initialization of quantum phase estimation algorithms on error-corrected quantum computers.
\subsection{Single-reference state\label{hf}}
The goal of this section is to find and initial state within the family of pure FGS, since they can be prepared efficiently on a linearly connected qubit architecture \cite{ortiz2001quantum,wecker2015solving,jiang2018quantum,kivlichan2018quantum}. A FGS is defined as \cite{bravyi2004lagrangian,shi2018variational}
\begin{align}
\ket{\Psi_{\text{GS}}} = U_{\text{GS}}\ket{0},\label{single1}
\end{align}
where $\ket{0}$ is the fermionic vacuum and $U_{\text{GS}}$ is a unitary operator that can be written as an exponential of a quadratic Hamiltonian times an imaginary prefactor. FGS are the ground states of non-interacting fermionic systems and are uniquely described by the one-particle reduced density matrix, which in case of particle number conservation is identical to the reduced covariance matrix (CM)
\begin{align}
\Gamma_{ij} = \braket{\Psi_{\text{GS}}|c_j^\dag c_i|\Psi_{\text{GS}}}\label{single3},
\end{align}
where we want to highlight the (in the following derivation) convenient but unusual index ordering in the above definition. Since the CM is of dimension $(N_{\text{so}}\times N_{\text{so}})$, it can be efficiently computed on a classical computer, even though the state vector in Eq.~\eqref{single1} grows exponentially with system size. Since we consider number-conserving Hamiltonians, studying number-conserving FGS, for which the terms $\braket{\Psi_{\text{GS}}|c_j^\dag c_i^\dag|\Psi_{\text{GS}}}$ and $\braket{\Psi_{\text{GS}}|c_j c_i|\Psi_{\text{GS}}}$ vanish \cite{eisert2018entanglement} is sufficient. It is for this reason that we choose the CM definition as in Eq.~\eqref{single3}, which omits such correlators. Following Ref.~\cite{kraus2010generalized}, we describe in the remainder of this section how to find $\ket{\Psi_{\text{GS}}}$ as the lowest energy state which results from an imaginary time evolution of the CM.

Since our simulations are restricted to the LLL, we will neglect the quantum numbers indicating the LLs. The number-preserving system Hamiltonian can then be written as
\begin{align}
H=&\sum_{p,q=0}^{N_{\text{so}}}f_{pq}c_p^\dag c_q+\frac{1}{2}\sum_{p,q,r,s=0}^{N_{\text{so}}}h_{pqrs}c_p^\dag c_q^\dag c_r c_s.\label{single2}
\end{align}
We will use a short-hand notation for the above Hamiltonian that summarizes the quadratic and quartic terms to $H=T+V$. Due to the anti-commuting properties of fermionic raising and lowering operators, the above Hamiltonian can always be recast in a form where the two-body matrix elements $h_{pqrs}$ possess the following symmetries
\begin{align}
h_{pqrs}=-h_{qprs}=-h_{pqsr}=h_{qpsr}.\label{sym2}
\end{align}

Following ~\cite{kraus2010generalized}, the imaginary time evolution of the density matrix $\rho(\tau)$ of a Hamiltonian is given by 
\begin{align}
\rho(\tau)=\frac{e^{-H\tau}\rho(0)e^{-H\tau}}{\text{tr}[e^{-2H\tau}\rho(0)]},\label{single4}
\end{align}
and guides us to the ground state in the limit of $\tau$ going to infinity ($\tau$ denotes the imaginary time), provided the overlap of $\rho(0)$ with the ground state is non-zero \cite{lehtovaara2007solution}. Since the exponential contains quartic terms due to the interaction terms in Eq.~\eqref{single2}, the imaginary time evolution will in general take us out of the family of FGS. By imposing that Wick's theorem holds, we restrict the evolution of Eq.~\eqref{single4} to a state-dependent quadratic Hamiltonian. Therefore, the solution of the imaginary time evolution will be the lowest energy state of the state-dependent quadratic Hamiltonian. 

To derive an equation of motion for the CM, we first note that the time derivative of the density matrix is given by
\begin{align}
d_\tau{\rho}=&-\{H,\rho\}+2\rho\text{tr}[H\rho],\label{single5}
\end{align}
---where $\{A,B\}=AB+BA$ is the anti-commutator---by simply taking the time derivative $d_\tau=\tfrac{d}{d\tau}$ on both sides of  Eq.~\eqref{single4}. Since the time evolution of the expectation value of an (not explicitly time-dependent) operator $A$ is given by $d_\tau\braket{A}=\text{tr}\left[A\dot\rho(\tau)\right]$, where $\braket{A}=\text{tr}[A\rho]$, we arrive at the following expression for the time evolution of the CM,
\begin{align}
d_\tau\Gamma_{ji}=&-\text{tr}[\{H,c_i^\dag c_j\}\rho]+2\Gamma_{ji}\text{tr}[H\rho]\label{single6}.
\end{align}
By inserting the Hamiltonian of Eq.~\eqref{single2} into Eq.~\eqref{single6} and restricting the density matrix to be drawn from the family of number conserving FGS, we can express the time evolution of the CM in terms of a state-dependent mean-field term,
\begin{align}
d_\tau\Gamma_{ji}
=&-\{\Gamma, h_m(\Gamma)\}_{ji}+2[\Gamma h_m(\Gamma) \Gamma]_{ji},\label{single10}
\end{align} 
and where
\begin{align}
h_m(\Gamma)= f+2\text{tr}_{1,4}[h\Gamma]    \label{single9}
\end{align}
is the mean-field term describing the quadratic, but state-dependent Hamiltonian, where $f$ is a two dimensional matrix with entries $f_{pq}$, $h$ is a four-dimensional tensor with elements $h_{pqrs}$ and 
\begin{align}
\text{tr}_{1,4}[h\Gamma]=& \sum_{p,s=0}^{N_{\text{so}}}h_{pqrs}\Gamma_{sp}\label{single12}
\end{align}
is a partial trace operation. We present an explicit derivation of Eq.~\eqref{single10} in Appendix~\ref{cov} and note that our result is identical to the results in Refs.~\cite{kraus2010generalized,shi2018variational}. We solve Eq.\eqref{single10} numerically through a formal integration method as outlined in Appendix~\ref{formal_integration}. The energy of the mean field state is given by 
\begin{align}
E_{m}=\text{tr}[f\Gamma]+\text{tr}[\text{tr}_{1,4}[h\Gamma]\Gamma]\label{single11}.
\end{align}
Since the matrix $[h_m,\Gamma]$ is anti-symmetric, $[h_m,\Gamma]^2$ is negative definite and leads to a monotonic decrease of the energy in time,
\begin{align}
d_\tau E_m =& 2\text{tr}\left[ ([h_m,\Gamma])^2 \right]\leq 0,\label{ssingle12}
\end{align}
which is also observed in the numerical simulations, see Fig.~\ref{fig:num_cons} in Appendix~\ref{formal_integration}. The imaginary time evolution will thus lead us to a (local) minimum in the energy landscape of a quadratic, but state-dependent Hamiltonian described by $h_m$ in Eq.~\eqref{single9}. If we denote with $O_\Gamma$ the $(N_{\text{so}}\times N_{\text{so}})$ orthogonal matrix which diagonalizes the CM through
\begin{align}
\Gamma = O_\Gamma \begin{pmatrix}
0&&&&&\\
&\ddots&&&&\\
&&0&&&\\
&&&1&&\\
&&&&\ddots&\\
&&&&&1
\end{pmatrix}O_\Gamma^T,\label{single13}
\end{align}
where the number of 1s on the diagonal corresponds to the number of electrons $N_{\text{el}}$ in the system, we can write the result of the imaginary time evolution in the basis where the FGS is a single Slater determinant of the form
\begin{align}
\ket{\Psi_{\text{init}}}=\tilde c_1^\dag \cdots \tilde c_{N_{\text{el}}}^\dag \ket{0},\label{single14}
\end{align}
where we defined a new set of fermionic creation and annihilation operators in the rotated spin-orbital basis
\begin{align}
\tilde c_j = \sum_{i} (O_{\Gamma})_{ij} c_i.\label{single15}
\end{align}
Using the generalized Hartree-Fock method of Ref.~\cite{kraus2010generalized} as summarized in this section, one can readily apply the constructions scheme of e.g. Ref.~\cite{kivlichan2018quantum} to implement a single Slater determinant as in Eq.~\eqref{single14} on a quantum computer in $N_{\text{so}}/2$ circuit depth using $\binom{N_{\text{so}}}{2}$ Givens rotations.
\subsection{Multi-reference state\label{asci}}
A single Slater determinant (as introduced in Section~\ref{hf}) is a state of independent particles and from the particle's perspective, it is unentangled \cite{somma2002simulating}. Since the ground state of the FQH system is expected to be a highly entangled state, eventually, a single Slater determinant will have a poor overlap with the exact ground state. In order to simulate larger system sizes, one faces the challenge of improving the state overlap using a method complementary to the generalized Hartree-Fock approach, which is both, efficiently computable on a classical computer and efficiently implementable on a quantum computer. One way of improving the initial state overlap is by generating a multi-reference state, i.e. a linear combination of Slater determinants similar to Eq.~\eqref{prepare},
\begin{align}
\ket{\Psi_{\text{init}}} = \sum_{i=1}^LC_i\ket{D_i}\label{asci1},
\end{align}
where the sum runs over $L\ll 2^{N_{\text{so}}}$ values, $C_i$ are real-valued coefficients with $\sum_i|C_i|^2=1$ and $\ket{D_i}$ are the "most important" Slater determinants according to a physically motivated ranking criterion (the symbol $L$ used here should not be confused with the identical symbol we used to denote the number of terms of the LCU Hamiltonian in Eq.~\eqref{ham_1}). We will study the performance of the Adaptive Sampling Configuration Interaction (ASCI) algorithm \cite{tubman2016deterministic,schriber2016communication,tubman2018modern,schriber2017adaptive} in the FQH setting, which is a state-of-the-art algorithm used in quantum chemistry calculations to obtain highly accurate energy estimates for strongly correlated molecules, competitive with full configuration quantum Monte Carlo and density matrix renormalization group methods \cite{tubman2016deterministic}. At the core of the algorithm lies a ranking criterion for the expansion coefficients $C_i$ that determines which determinants $\ket{D_i}$ should be included in Eq.~\eqref{asci1}. We will give a brief overview of ASCI following Ref.~\cite{tubman2016deterministic} in Section~\ref{asci_explain}, explain how we derive the fidelity of the resulting state in Section~\ref{overlap_estimation}, and conclude with how a linear combination of Slater determinants could efficiently be implemented on a quantum computer in Section~\ref{lin_prep}.
\subsubsection{The ASCI algorithm\label{asci_explain}}
The ASCI algorithm is an iterative method to find the most important Slater determinants by sampling determinants based on a ranking criterion derived from conditions on a steady-state solution following an imaginary time evolution. Two determinant subspaces define the ASCI algorithm, namely, the core space and the target space, each containing \textit{cdets}- and \textit{tdets}-many determinants ($tdets\leq cdets$), respectively. 

In the first iteration step the core space consists only of a single Slater determinant $\ket{\Psi_{\text{GS}}}$ obtained from the method outlined in Section~\ref{hf}, with corresponding energy $E_m$ as given by Eq.~\eqref{single11}. The first step in each iteration consists of computing the space of all determinants which are connected with the core space through single- and double excitations, e.g. determinants generated by applying $c_p^\dag c_q$ and $c_p^\dag c_q^\dag c_rc_s$. For all determinants generated in that manner one has to compute the coefficients
\begin{align}
A_i=\sum_{\substack{j\neq i\\ j\in \textit{cdets}}}\frac{H_{ij}C_j}{H_{ii}-E}.\label{asci2}
\end{align}
Here, $E$ describes the lowest energy eigenvalue from the previous diagonalization and $H_{ij}=\braket{D_i|H|D_j}$ are off-diagonal Hamiltonian matrix elements. In the first iteration we set $E=E_m$. 

The computation of the amplitudes in Eq.~\eqref{asci2} is motivated by the stationary state solution of an imaginary time propagation of a state Ansatz of the form defined by Eq.~\eqref{asci1}. One then chooses the largest \textit{tdets} determinants from the sets $\{|C_i|\}$ and $\{|A_i|\}$ of core space and single- and double-excited core space determinants and diagonalizes the $(\textit{tdets}\times\textit{tdets})$-dimensional reduced system Hamiltonian, keeping only the eigenvector belonging to the lowest eigenvalue $E$ \footnote{Clearly, if you take a core determinant $\ket{C_k}$ and search all single- and double excitations of that determinant, chances are high that you will obtain determinants which are also elements of the core set. In that case, we keep the coefficient with the largest value (by magnitude) and discard the rest.}. This eigenvector will have entries $(C_1,C_2,\dots,C_{\textit{tdets}})^T$, with each entry belonging to a unique Slater determinant of the target space. The \textit{cdets} largest coefficients are kept and re-normalized and their respective determinants form the new core space in the next iteration step. One repeats these steps until the energy converges, which we generally observe after around four to five iterations for all system sizes studied (see Fig.~\ref{fig:energy} in Appendix~\ref{additional_plots}).

One of the computationally more costly steps is the evaluation of the overlaps $H_{ij}$, which we discuss in more detail in Appendix~\ref{hii} and \ref{hij}. For all our ASCI simulations, we choose the core space to be identical to the target space of the previous iteration step, $L=\textit{tdets}=\textit{cdets}$. As outlined in Appendix~\eqref{eigenbasis_cov}, we transformed the Hamiltonian in Eq.~\eqref{asci2} for the ASCI simulation into the eigenbasis of the CM using the transformation given by Eq.~\eqref{single15}, where the Hartree-Fock state is a simple tensor product of $N_{\text{el}}$ distinct fermionic creation operators acting on the fermionic vacuum state.
\begin{figure*}
	\centering
	\begin{subfigure}
		\centering
		\includegraphics[width=.475\textwidth]{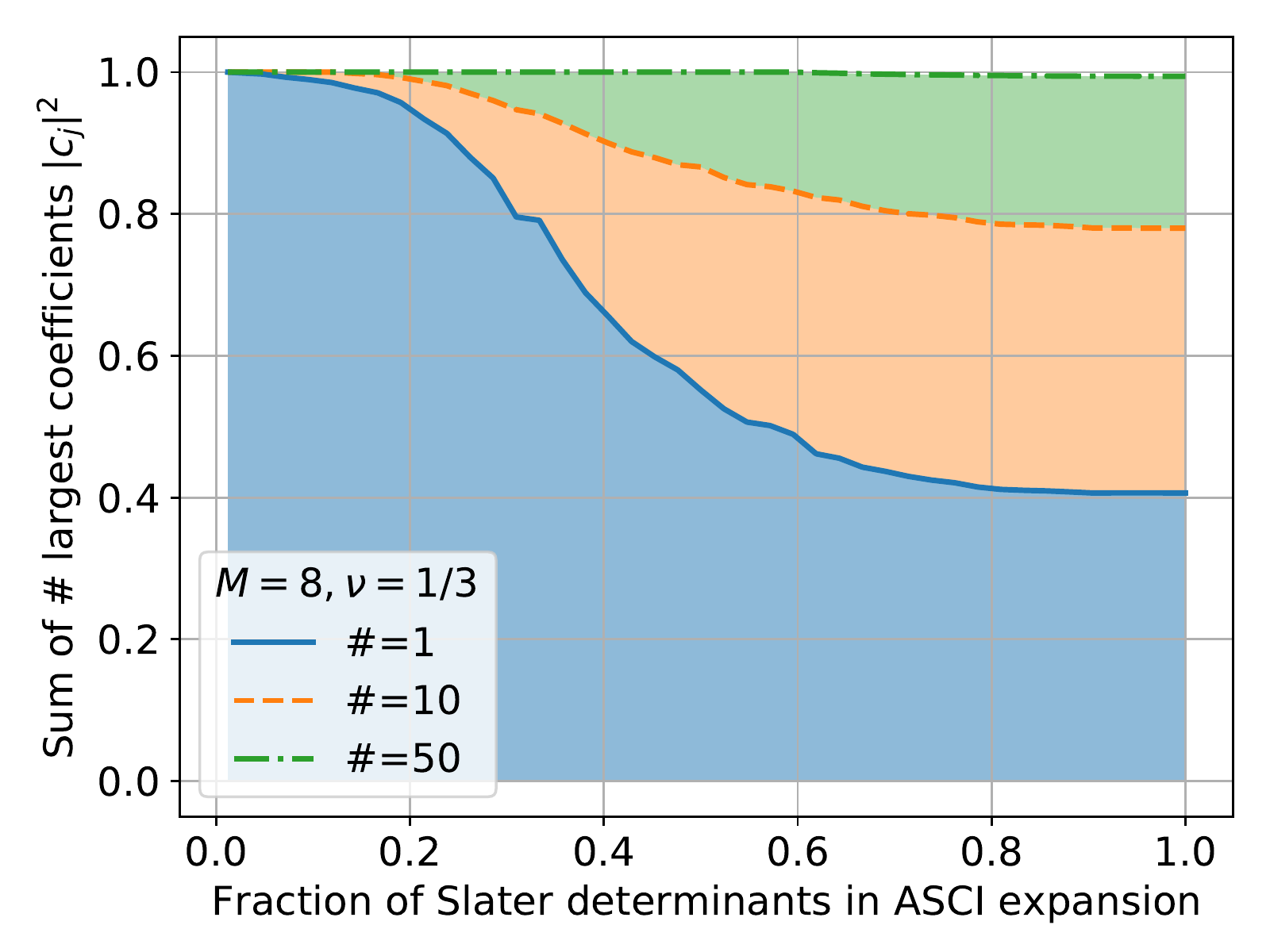}
	\end{subfigure}
	\begin{subfigure}
		\centering 
		\includegraphics[width=.475\textwidth]{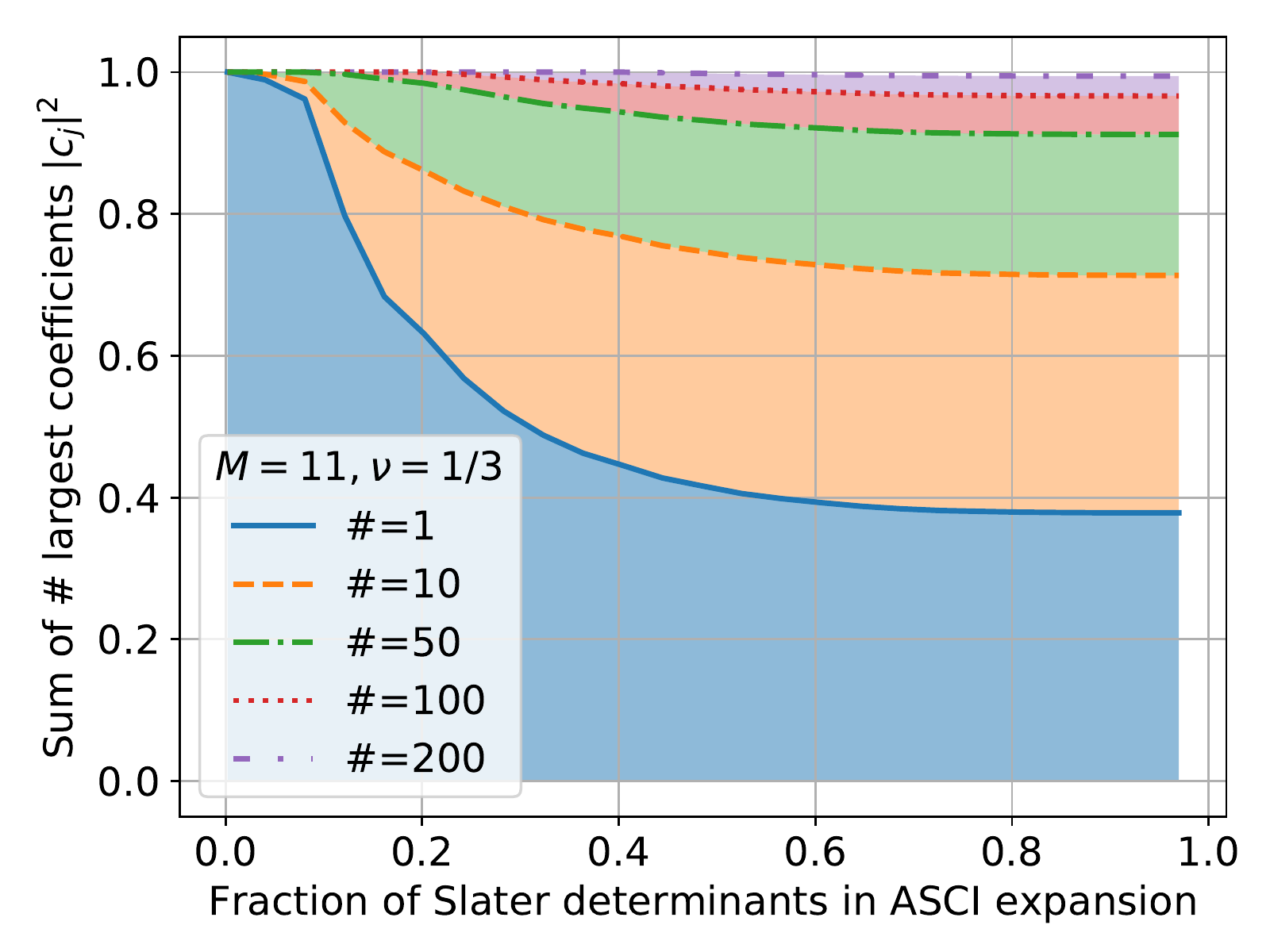}
	\end{subfigure}

	\vskip\baselineskip
	\begin{subfigure}
		\centering 
		\includegraphics[width=.475\textwidth]{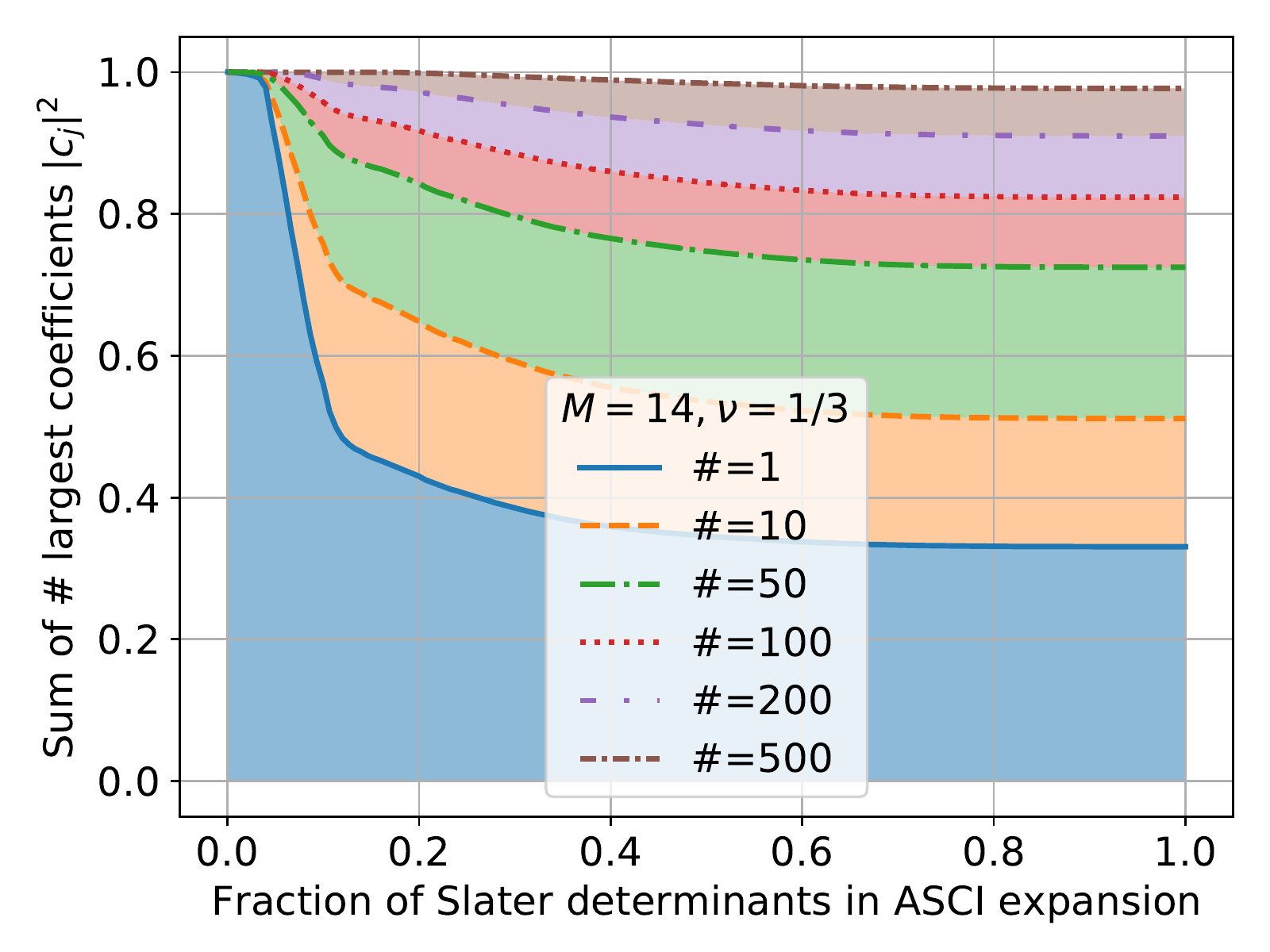}
	\end{subfigure}
	\begin{subfigure}
		\centering 
		\includegraphics[width=.475\textwidth]{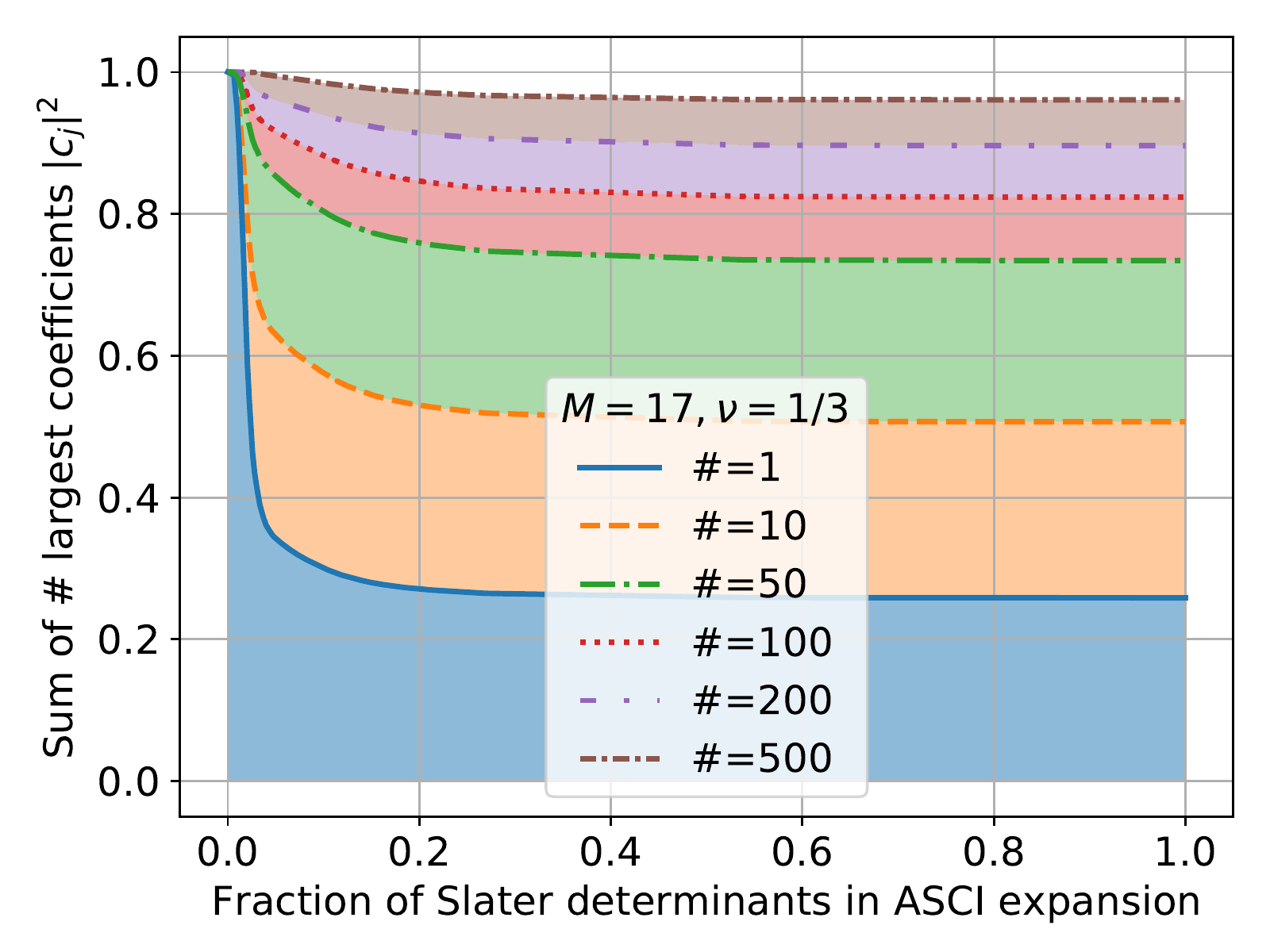}
	\end{subfigure}
	\caption{Scatter plots showing the sum of the squared coefficients in the ASCI expansion for various numbers of $tdets(=cdets)$ for system sizes of $N_{\text{so}}=9,12,15,18$ spin-orbitals at filling $\nu=1/3$ in ascending order from the upper-left to the lower-right figure.. The blue curve (solid line) shows the behaviour of the mean-field solution $\ket{\Psi_{\text{GS}}}$. The rightmost points (where the fraction of Slater determinants in ASCI expansion is identical to 1) within each figure corresponds to the FCI expansion, i.e. all $\binom{N_{\text{so}}}{N_{\text{el}}}$ relevant Slater determinants are taken into account for those points and the sum of the $\#$ (where $\#$ is to be replaced with the number indicated in the grey box) is identical to the fidelity defined on the left-hand side of Eq.~\eqref{overlap}. \label{asci_convergence_plots}}
\end{figure*}
\subsubsection{Overlap estimation\label{overlap_estimation}}
If the ASCI expansion in Eq.~\eqref{asci1} includes all $\binom{N_{\text{so}}}{N_{\text{el}}}$ Slater determinants containing $N_{\text{el}}$ electrons, the ASCI solution is identical to the Full Configuration Interaction (FCI) solution and will give the exact ground state of the system Hamiltonian \footnote{FCI in our case refers to including all number-conserving determinants in the ASCI expansion---which grows exponentially with system size---and provides an exact solution, see e.g. Ref.~\cite{helgaker2014molecular}.}. We expand the exact solution as 
\begin{align}
\ket{\Psi_{0}}=\sum_{k=1}^{\text{FCI}} \tilde C_k \ket{D_k}
\end{align}
and compute the squared overlap w.r.t. the ASCI state in Eq.~\eqref{asci1} containing $L\leq \text{FCI}$ determinants, which is identical to the support of the ASCI expansion on the exact solution, i.e. the state fidelity defined on the left-hand side of Eq.~\eqref{overlap}. Since the number of determinants in a FCI expansion grows exponential with system size, once we go beyond exactly solvable system sizes, we will no longer be able to talk about the support of a subset of determinants on the exact ground state of the system Hamiltonian, but rather on the ground state of the reduced system Hamiltonian which is spanned by the $tdets$ determinants of the ASCI expansion.

\begin{figure*}
	\centering
	\begin{subfigure}
		\centering
		\includegraphics[width=.475\textwidth]{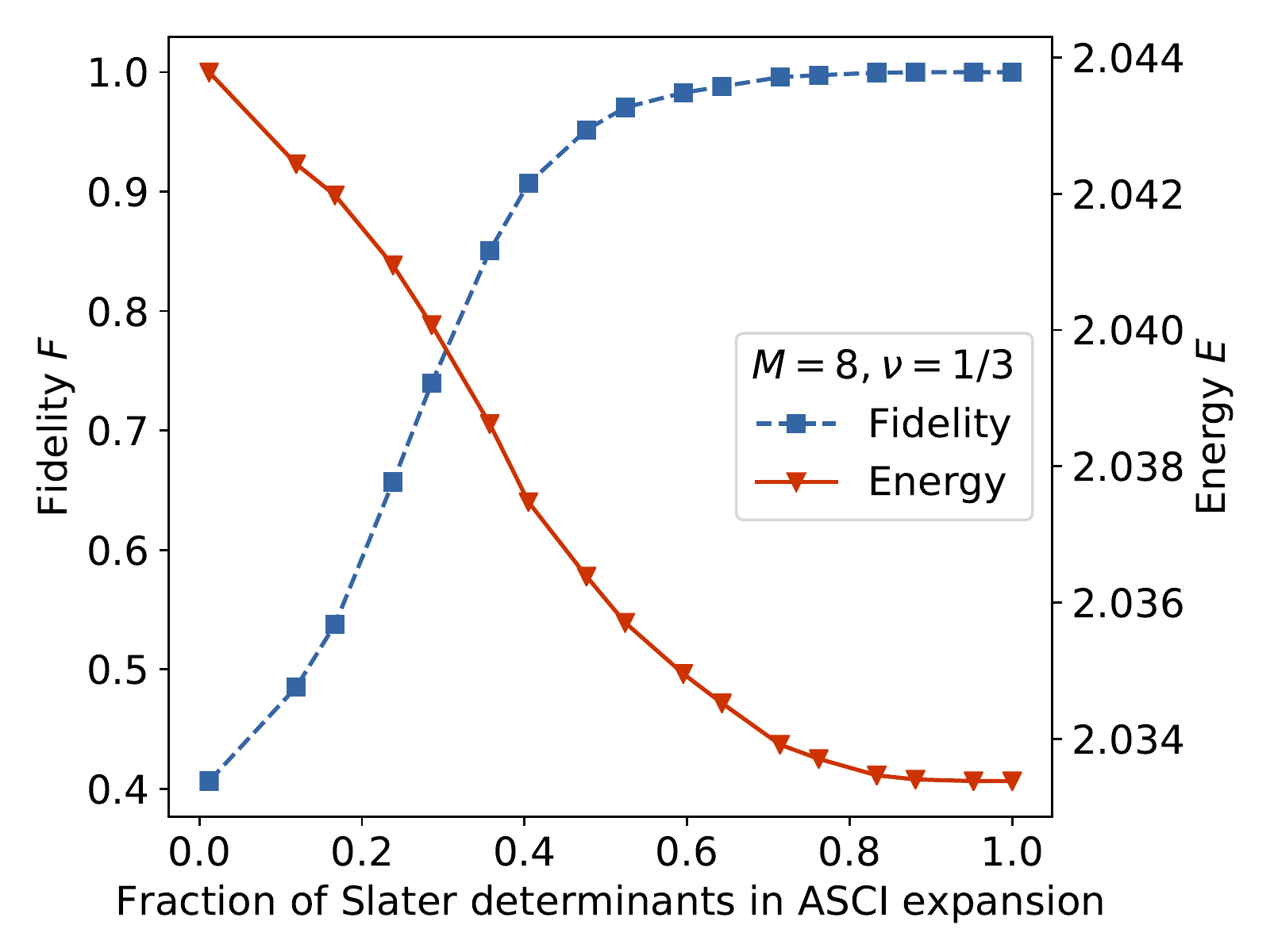}
	\end{subfigure}
	\begin{subfigure}
		\centering 
		\includegraphics[width=.475\textwidth]{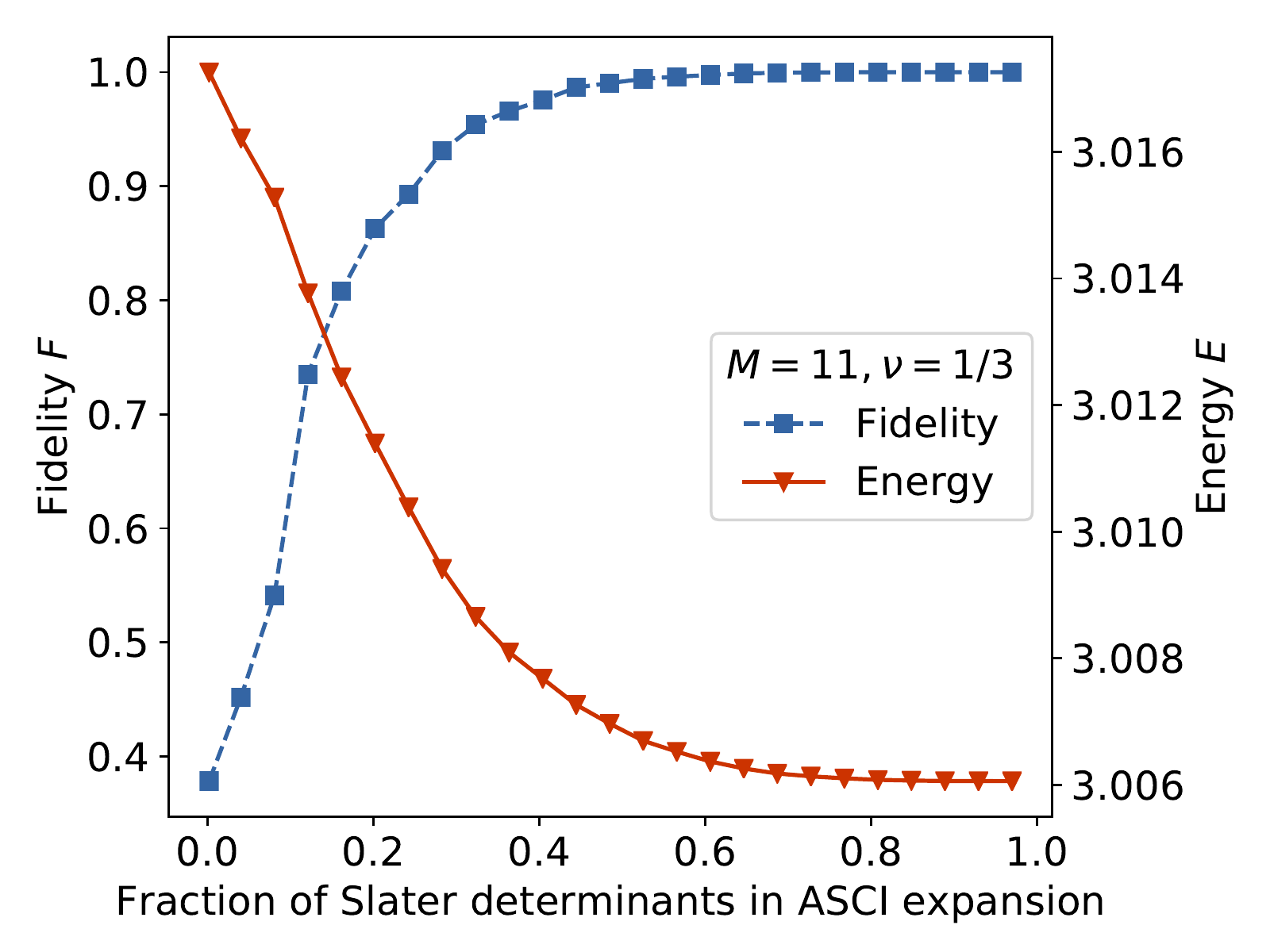}
	\end{subfigure}

	\vskip\baselineskip
	\begin{subfigure}
		\centering 
		\includegraphics[width=.475\textwidth]{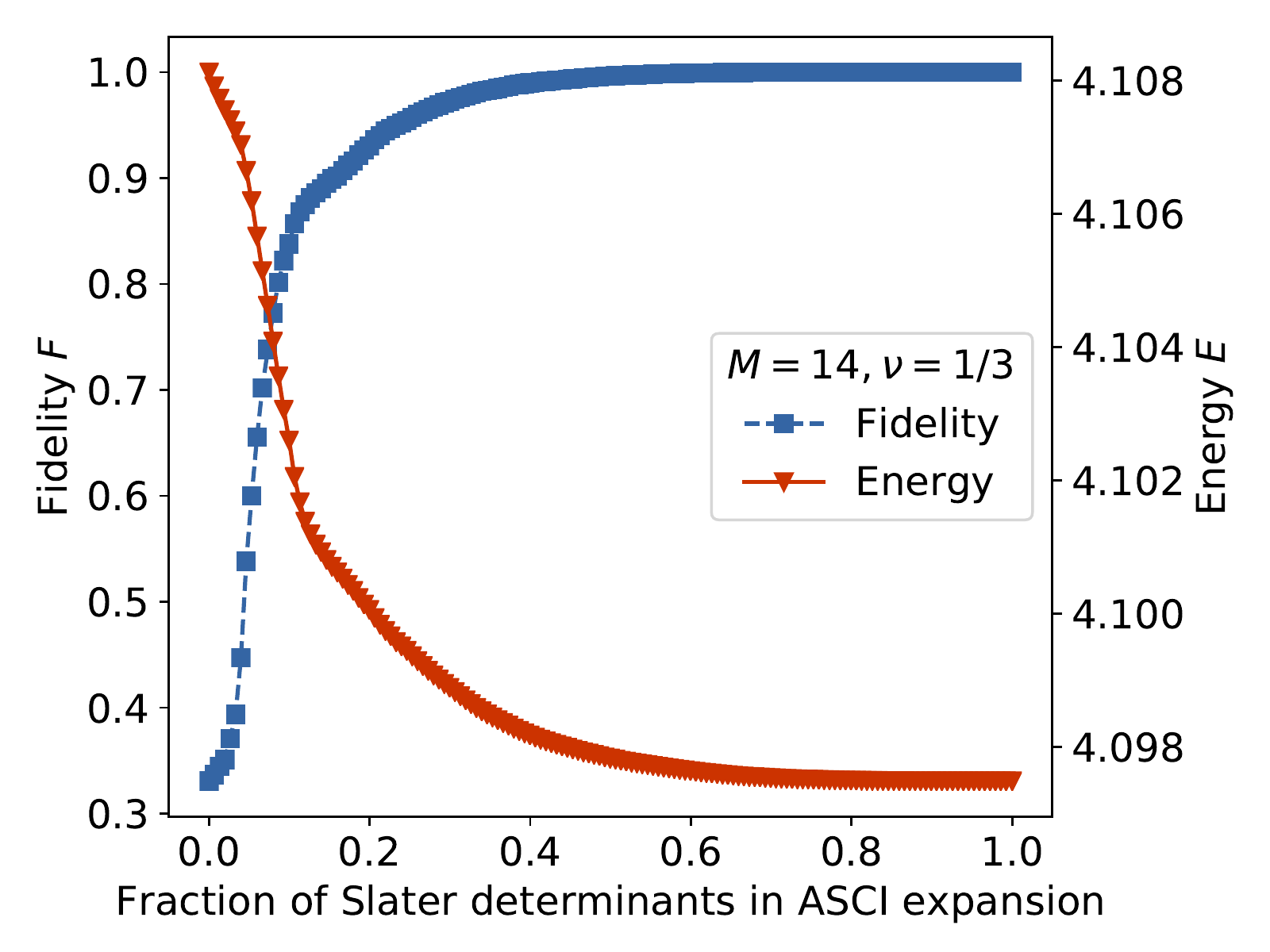}
	\end{subfigure}
	\begin{subfigure}
		\centering 
		\includegraphics[width=.475\textwidth]{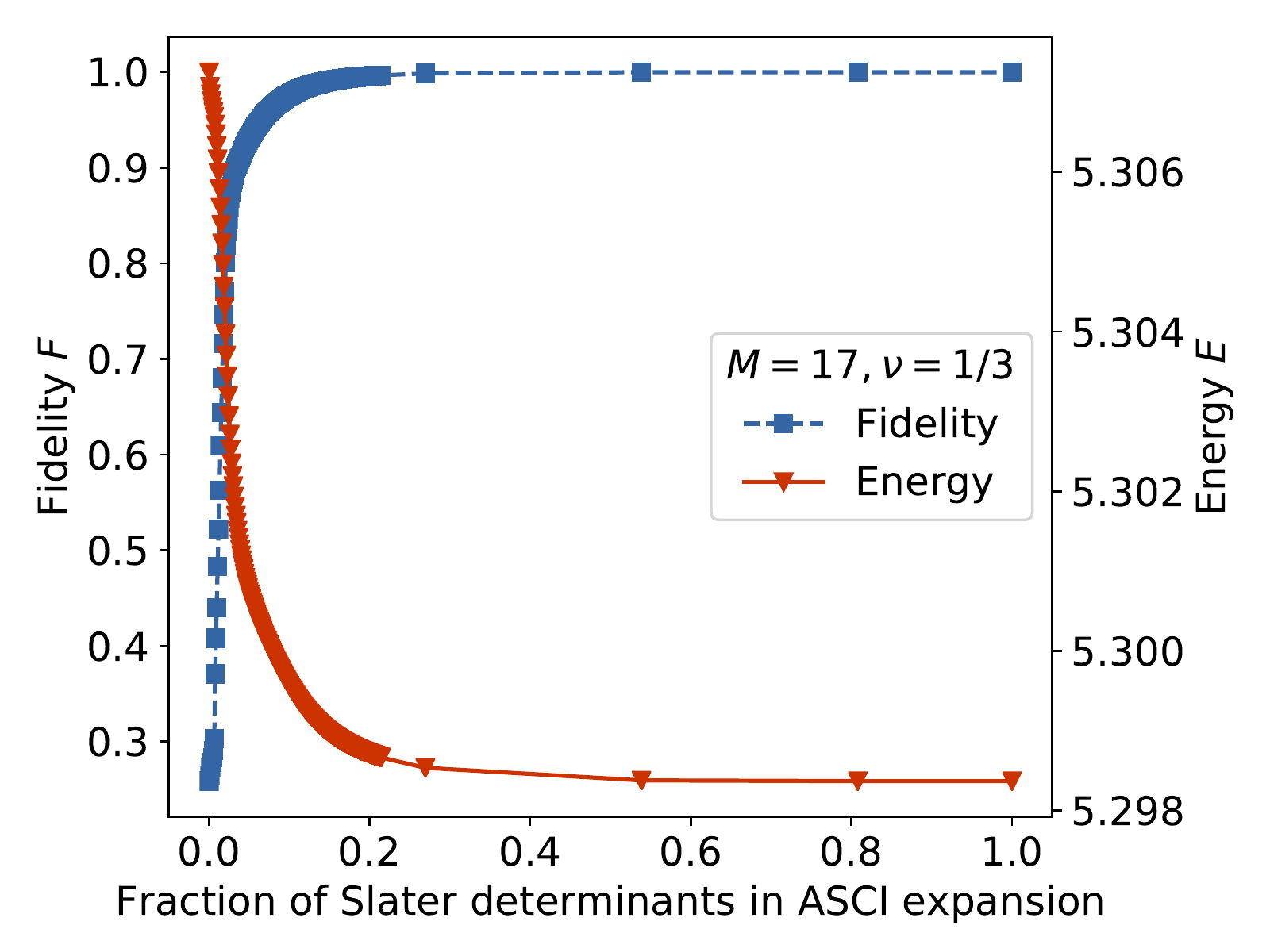}
	\end{subfigure}
	\caption{Scatter plots of the fidelity $F=|\braket{\Psi_{\text{init}}|\Psi_0}|^2$ (blue colored squares) and the convergence of the energy $E$ (red colored triangles) for various numbers of determinants in the ASCI expansion of Eq.~\eqref{asci1}. Instead of the total number of Slater determinants in the expansion, we plot the ratio w.r.t. the FCI expansion on the x-axis, with the first data point corresponding to the single reference state $\ket{\Psi_{\text{GS}}}$ and the last to the FCI expansion. The four plots show system sizes with $N_{\text{so}}=9,12,15,18$ spin orbitals at filling $\nu=1/3$ in ascending order from the upper-left to the lower-right figure.  \label{fidelity}}
\end{figure*}

\subsubsection{Preparing a linear combination of Slater determinants on a quantum computer\label{lin_prep}}
Recent work showed that a linear combination of Slater determinants, required e.g. for realizing the mapping described by the \textsc{prepare} oracle in Section~\ref{lcu}, could be implemented efficiently on a quantum computer through the use of a quantum read-only memory, whose purpose is to read classical data indexed by a quantum register \cite{babbush2018encoding}. The construction scheme was improved upon by reducing the number of ancillary qubits needed to 1, resulting in a state preparation protocol, where $\ket{\Psi_{\text{init}}}$ can be constructed using only $O(N_{\text{so}}L)$ gates \cite{tubman2018postponing}, where $L$ is here identical to the number of core and target space determinants in the ASCI expansion. As previously stated, while the single reference state method introduced in Section~\ref{hf} is suitable for NISQ devices, the preparation of linear combination of Slater determinants outlined in Section~\ref{asci} will require error-corrected quantum computers, as it demands the implementation of many layers of multi-qubit Toffoli-type gates, which are costly to implement \cite{motzoi2017linear}. 
\section{Numerical results\label{numerics}}
In this section we present our numerical results for implementing a FGS state and a multi-reference state as proposed in Sections~\ref{hf} and \ref{asci} for small instances. 

We study the quality of the initial state Ansatz of a system containing $N_{\text{el}}$ electrons in $N_{\text{so}}=3N_{\text{el}}$ spin-orbitals, which corresponds to a filling of $\nu=1/3$ in the LLL. This corresponds to a fixed electron density, which can be seen from Eqs.~\eqref{nu1}-\eqref{nu2}.

By performing a formal integration of the equations of motion of the CM given by Eq.~\eqref{single10}, we obtain the mean-field solution $\ket{\Psi_{\text{GS}}}$ of the system Hamiltonian. The numerical method is detailed in Appendix~\ref{formal_integration} and was performed using $10^5$ time steps at step size $\Delta\tau=0.01$ for all simulation results in Fig.~\ref{asci_convergence_plots} and Fig.~\ref{fidelity}, as well as in the simulations shown in Appendix~\ref{additional_plots}. The mean-field energy converges for all cases well before the end of the imaginary time evolution and the number of particles is conserved throughout the simulation, as exemplified in Fig.~\ref{fig:num_cons} in the Appendix.

In Fig.~\ref{asci_convergence_plots}, we study the support of the most important Slater determinants (i.o.w. those carrying the largest coefficients $|C_i|$) in the ASCI expansion of Eq.~\eqref{asci1} for system sizes $N_{\text{so}}=9,12,15,18$. For each set of data points, we study how the support changes when enlarging the space of core determinants, keeping in mind that we set $cdets=tdets$. The horizontal axis displays the fraction of core determinants in the current ASCI expansion w.r.t. the FCI expansion. The very last data point in each of the plots compares the sum of the squared coefficients to the FCI expansion and the corresponding value is thus equivalent to the state fidelity $F$ defined in Eq.\eqref{overlap} of the ASCI expansion. The single determinant expansion is equivalent to $\ket{\Psi_{\text{GS}}}$ and thus describes the mean-field behaviour. It drops from around $F\approx 0.4$ for the smallest system size in the upper-left corner to $F\approx 0.25$ for the largest simulated system size in the lower-right corner of Fig.~\ref{asci_convergence_plots}. For all simulations, constructing an ASCI expansion of ten Slater determinants guarantees an initial state fidelity well above $F=0.5$, where we assumed an error-free construction of the linear combination of Slater determinants.

In Fig.~\ref{fidelity}, we investigate the convergence of both, the fidelity $F$, as well as the energy $E$---which corresponds to the lowest energy eigenvalue obtained from diagonalizing the reduced system Hamiltonian in the ASCI algorithm---for system sizes $N_{\text{so}}=9,12,15,18$. The first (last) data point in each individual plot corresponds to the mean-field solution (FCI expansion / exact ground state). Each marker in Fig.~\ref{fidelity} corresponds to an individual ASCI simulation. The convergence of the energy of the reduced Hamiltonian  for each individual ASCI simulation is displayed in Appendix~\ref{additional_plots} in Fig.~\ref{fig:energy} for a variety of core determinants, which shows that ASCI typically converges after about five iterations for the respective system sizes. 

\begin{figure}[h!]
	\centering
	\includegraphics[width=.49\textwidth]{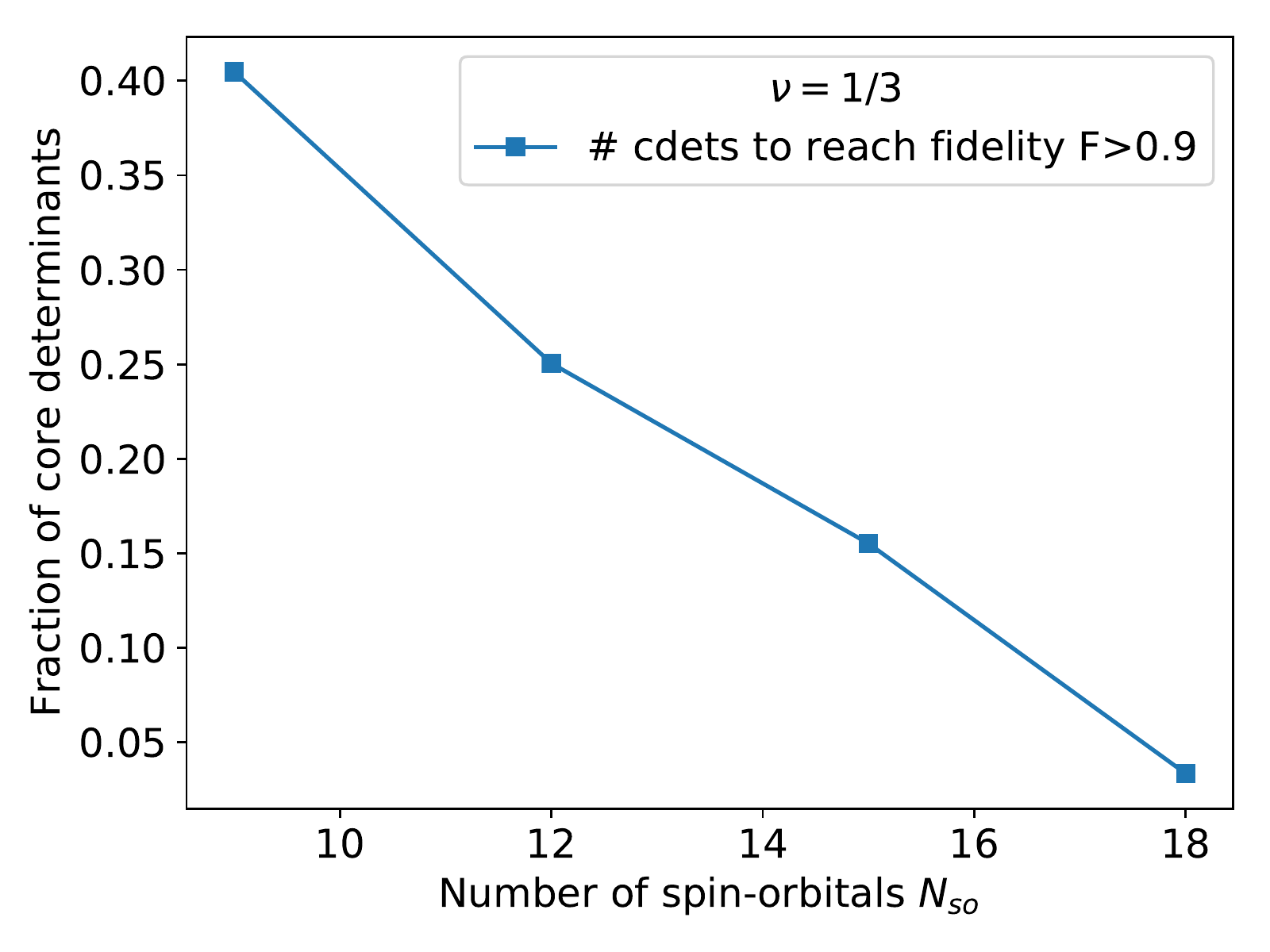}
	\caption{Scatter plot of the minimal number of ASCI core determinants (in terms of its ratio to the FCI expansion) needed to obtain state fidelities $F>0.9$ for system sizes $N_{\text{so}}=9,12,15,18$. The values were obtained by linear extrapolation of the two data sets belonging to the largest (lowest) fidelity below (above) the threshold value $F=0.9$. We note that the size of the determinant space corresponding to a FCI expansion displayed on the x-axis grows exponentially.    \label{fig:fidelity_threshold}} 
\end{figure}

One can observe from Fig.~\ref{fidelity}, that the fidelity does not converge much faster than the energy, which makes ASCI an unsuitable candidate for estimating state overlap for intractable system sizes (given that this trend continuous) unlike the findings observed for  the various physical systems studied in Ref.~\cite{tubman2018postponing}. There, the argument is that if the fidelity were to converge much faster than the energy \textit{and} the latter would start to converge already at reasonable system sizes, one would have a heuristic argument that supports the legitimacy of approximating the overlap of the initial state with the true ground state by using the largest possible ASCI expansion instead of $\ket{\Psi_0}$, since the latter is unknown. However, for the system sizes studied here, this behaviour was not observed. 

In Fig.~\ref{fig:fidelity_threshold} we show how the minimal number of determinants needed to reach a fidelity of at least $F=0.9$ scales with system size. The horizontal axis shows the number of spin-orbitals studied, where the number of FCI determinants grows exponentially, while the vertical axis displays the number of core determinant size w.r.t. the FCI to reach the desired fidelity, where the latter was obtained by linear extrapolation of the two simulated core sets displaying the largest (lowest) fidelity below (above) the threshold value $F=0.9$. The close-to-linear behavior in Fig.~\ref{fig:fidelity_threshold} shows that for the system sizes studied here, only a sub-exponential increase in terms of the number $L=cdets=tdets$ of Slater determinants in the ASCI expansion is required to obtain an initial state $\ket{\Psi_{\text{init}}}$ with an overlap of at least $F=0.9$ with the true ground state $\ket{\Psi_0}$. Larger-scale numerical simulation are needed to vindicate or disprove the observed trend for increasing values of $N_{\text{so}}$.

\section{Discussion\label{discussion}}
In the following section, we discuss various avenues that could be explored in future studies, such as improving the model FQHE Hamiltonian, and choosing different geometries and basis sets for the system Hamiltonian, to using the Laughlin state as a proving ground to test heuristic Ans\"atze for NISQ algorithms. For completeness, we show how correlation functions  (which contain all information about the respective physical system) can be computed for both the FGS and the multi-reference expansion. 
\subsection{Finite size studies\label{finite_size}}
A natural question to ask, is whether it makes sense to perform a digital quantum simulation of a FQH system on a non error-corrected architecture, where me might be restricted to anywhere between tens up to a few hundreds of qubits. Current exact simulations of FQH systems are restricted to a handful of particles, but it turns out that the largest computer simulations today of around 50 spin-orbitals already exceeds the typical length scale (which is given by the magnetic length $l_B\approx 25\text{nm}/ \sqrt{B[\text{T}]}$ of the problem considerably and it is therefore sensible to assume that one can make simulations that reflect properties which may extend to the thermodynamic limit even for relatively small numbers of particles. The goal of a digital quantum simulation of a quantum system can not be to try to simulate the actual system size, as the quantum resource requirements would be astronomical. As a small example taken from Ref.~\cite{jain2007composite}, a typical $1(\text{mm})^2$ sample contains roughly $10^9$ electrons. A toy system of 100 electrons distributed among 250 spin-orbitals in the LLL (corresponding to $\nu=0.4$) would lead to $10^{72}$ distinct ground state configurations, a number comparable to the number of particles in our universe. An error-corrected quantum computer would however only require 250 logical qubits (neglecting additional qubits required for the employed quantum algorithm) to represent this state.
\subsection{Augmenting the model \label{augmenting}}
In our discussion, we focused on the Coulomb interaction as it provides the key to the understanding of the FQHE. In order to make the system more realistic by taking into account effects that play a subdominant role in comparison to the electron-electron interaction described by $H_2$, one can add additional terms to the system Hamiltonian of Eq.~\eqref{sh}. 

A two-dimensional electron gas is typically realized in experiments in dirty samples where random one-particle potentials of e.g. positive donor ions are scrambled across the probe (this is known as disorder). To account for their effect on the electrons, one therefore  has to include one body potentials $\sum_jU(\mathbf r_j)$ as well, whose specific form depend on material properties. By computing the one-body coefficients due to the disorder terms, its effect could as well be included at a free cost in terms of qubit resources. 

The role of the electron spin has been neglected in our derivations entirely, since we assumed that the magnetic field is large enough that all spin degrees of freedom are frozen. In order to account for the effect of the spin, one would have to add the Zeeman term $g\mu \mathbf B\cdot \sum_{k=1}^{N_{\text{el}}}\left(\mathbf S_z\right)_j$, where $\left(\mathbf S_z\right)_j$ is the $z$-component of the spin of electron $j$, $\mu$ is the Bohr magneton and $g$ the Land\'{e} g-factor. This would double the number of required qubits, since an additional register for each state would be required as a placeholder for the orbital spin component.

We have chosen a "soft" boundary (it is not a physical boundary) by introducing a cutoff in angular momentum. By using an harmonic trapping potential instead, one can simulate a physical boundary that allows one to exert pressure on the system by tuning the strength of the trapping potential. 

We restrict ourselves to the disk geometry in symmetric gauge, but one could have also chosen a different gauge, such as the Landau gauge  $\mathbf A = B(-y,0,0)^T$. Similarly, one can choose other geometries, for instance geometries which do not possess a boundary and are useful when studying bulk properties. Two prominent examples of such geometries are a two-dimensional sheet of electrons wrapped around the surface of a sphere, known as the Haldane sphere, or a two-dimensional sheet of electrons wrapped around a cylinder with periodic boundary conditions, which constitutes a torus geometry. See e.g. Ref.~\cite{fremling2013coherent} for more details on the torus geometry and Ref.~\cite{wooten2014configuration} for Hamiltonians that incorporate LL mixing within the Haldane sphere geometry.
\subsection{Using the Laughlin wave function as a sanity check for the variational Ansatz\label{laughlin}}
It is well known that for small system sizes the  Laughlin wave function has a large overlap with the ground state of the FQH Hamiltonian in the LLL \cite{laughlin1983anomalous}. However, it is not the exact ground state of the FQH Hamiltonian, but rather the ground state of different, so-called parent Hamiltonian \cite{trugman1985exact,kapit2010exact,lee2015geometric,glasser2016lattice}. To our knowledge, there has yet to appear a quantum circuit that efficiently constructs the Laughlin wave function for various filling factors $\nu$, with the exception of integer filling factors  \cite{latorre2010quantum}. Even though a Fock-space representation of the Laughlin state exists \cite{di2017unified}, it is not clear to us how this could be efficiently mapped onto a quantum circuit. An efficient quantum algorithm for generating the Laughlin state (or related states describing higher filling factors such as the Moore-Read state \cite{moore1991nonabelions}) would most likely be of vital importance for digital quantum simulations of the FQHE Hamiltonian. Furthermore, a recent paper introduced a classically efficient variational method going beyond FGS to enable the study of FQHE systems in the spirit of composite fermions \cite{shi2018variational}, but so far this method has not yet been applied to FQH systems and is not clear how well it will improve over a generalized Hartree-Fock Ansatz.

Even without an efficient algorithm for the implementation of the Laughlin state at hand, it could still play an important role for choosing appropriate variational Ans\"{a}tze of the VQE algorithms. If a variational Ansatz would approximate the Laughlin wave function (by performing a VQE simulation with its corresponding parent Hamiltonian), it would be a strong indicator that the variational Ansatz can construct states that lie in the same universality class as the Laughlin wave function. Since the Laughlin wave function is an analytic expression, one can compare the results measured by a quantum computer with the theoretically predicted behavior even for large system sizes. 
\subsection{Computing correlation functions\label{corr_fct}}
In order to be able to extract ground state properties, such as the one-particle reduced density matrix, the pair correlation function and static structure factor, one has to compute the expectation values of products of the fermionic field operators, which can be performed efficiently on a quantum computer \cite{wecker2015solving,kivlichan2018quantum}. We define the fermionic field operators
\begin{align}
\hat\Psi^\dag(\mathbf r)=&\sum_p\eta_p^*(\mathbf r)c_p^\dag\label{meas1}\\
\hat\Psi(\mathbf r)=&\sum_p\eta_p(\mathbf r)c_p,\label{meas2}
\end{align}
and the one-particle reduced density matrix and pair correlation function 
\begin{align}
G_1(\mathbf r, \mathbf r')=&\langle \hat\Psi^\dag(\mathbf r) \hat\Psi(\mathbf r') \rangle\label{meas3}\\
G_2(\mathbf r, \mathbf r')=&\langle \hat\Psi^\dag(\mathbf r)\hat\Psi^\dag(\mathbf r') \hat\Psi(\mathbf r')\hat\Psi(\mathbf r) \rangle.\label{meas4}
\end{align}
The one-particle reduced density matrix $G_1(\mathbf r, \mathbf r')$ measures the values of the fermionic field operators at points $\mathbf r$ and $\mathbf r'$ and is identical to the electron density for $\mathbf r=\mathbf r'$. Thus, the number of electrons is given by $N_{\text{el}}=\text{tr}[G_1(\mathbf r, \mathbf r)]$. The pair correlation function $G_2(\mathbf r, \mathbf r')$ is a measure of the density correlations and is proportional to the pair distribution function. By combining Eqs.~\eqref{meas1}-\eqref{meas4}, the measurement of correlation functions can be broken down into measurements of sums of quartic and quadratic fermionic operator expectation values.

For FQH states and more specifically for states describing a uniform density (at least inside the disk) isotropic liquid, one expects from extrapolation of finite system results that the one-particle reduced density matrix has an absence of off-diagonal long-range order \cite{girvin1987off}, 
\begin{align}
\lim_{|\mathbf r-\mathbf r'|\rightarrow \infty}G_1(\mathbf r, \mathbf r')=0\label{meas5}
\end{align}
and that the FQH state is a quantum liquid, which is characterized by \cite{kamilla1997fermi}
\begin{align}
\lim_{|\mathbf r-\mathbf r'|\rightarrow \infty}G_2(\mathbf r, \mathbf r')=\text{constant}\label{meas6},    
\end{align}
(see e.g. chapters 8 and 12 in ~\cite{jain2007composite}) as opposed to the mean-field solution that produces a crystal and whose pair correlation function oscillates all the way to infinity. Any approximate ground state generated either through VQE approaches on NISQ devices or more elaborate methods such as ASCI (or the method introduced in ~\cite{ge2019faster}) should be able to reproduce the characteristic behavior as predicted by Eqs.~\eqref{meas5}-\eqref{meas6}. In Appendix~\ref{corr}, we give analytic expressions on how $G_2(\mathbf r,\mathbf r')$ may be efficiently computed on a classical computer for the FGS and show how multi-reference state approaches can be computed, given that the latter is kept to tractable system sizes. We also show the crystal-like patterns observed in the pair correlation function for a FGS Ansatz in Fig.~\ref{fig:g_2} of Appendix~\ref{corr}.

Another physical quantity of interest regarding FQH states is the R\'enyi entropy, which contains information about whether the underlying entanglement obeys an area or volume law and whether the system is in a insulating or conducting phase. An explicit quantum circuit for measuring the R\'enyi entropy w.r.t. the Laughlin state on a quantum computer is given in Ref.~\cite{johri2017entanglement}.

\section{Conclusion and outlook\label{conclusion}}
We have presented an \textit{ab-initio} roadmap to simulate the FQH Hamiltonian. We derived efficiently computable analytical expressions for the  respective one- and two-body Hamiltonian coefficients which allow for LL mixing. Using the the low-rank factorization method of ~\cite{Berry2019} to extract the Hamiltonian eigenspectrum, we found a T gate complexity of $O(M^{4.35}/\Delta E)$ to estimate the energy to precision $\Delta E$. This presents the current most efficient method to simulate the spectrum of the FQH Hamiltonian on an error-corrected quantum computer. We performed small-scale numerical simulations within the LLL to investigate the initial state fidelities of two efficiently computable and preparable Ans\"{a}tze based on the generalized Hartree-Fock method and the ASCI algorithm, suitable for NISQ and error-corrected quantum processors, respectively. While the latter method shows a sub-exponential scaling in the required number of determinants to reach high fidelity initial states, larger scale numerical simulations are needed to better determine the large system-size behavior. In addition, scaling analysis for the parameter $\lambda$ for systems including higher LLs are needed to discover the respective gate complexity for simulations beyond the LLL. To further improve the initial state Ansatz, an efficient implementation of Laughlin-type states should be a major focus of future work. 
\section{Acknowledgement}
The authors thank Ryan Babbush for initiating the discussion on Hamiltonian simulation, suggesting to use a strategy that exploits the analytical form of the Hamiltonian coefficients and for critically reviewing parts of the draft. MK thanks Giovanna Morigi for supporting the project and Daniela Pfannkuche for helpful discussions and hospitality. This work has been supported by the EU through the FET flagship project OpenSuperQ, the German Research Foundation (priority program No. 1929 GiRyd) and by the German Ministry of Education and Research (BMBF) via the QuantERA project NAQUAS. Project NAQUAS has received funding from the QuantERA ERA-NET Cofund in Quantum Technologies implemented within the European Union’s Horizon 2020 program. 

\bibliography{mybib}{}
\bibliographystyle{unsrt}
\appendix

\section{Derivation of analytical result for the Coulomb matrix elements including arbitrary LL mixing in the symmetric gauge disk geometry using the angular momentum eigenbasis\label{app:analytical_matrix_elements}}
We will evaluate Eq.~\eqref{hpqrs}, which displays the trivial symmetry $h_{\mathbf{PQRS}}=h_{\mathbf{QPSR}}$ due to the indistinguishability of electrons. For the evaluation of Eq.~\eqref{hpqrs}, we use the Fourier representation of the Coulomb operator \cite{tsiper2002analytic}, more specifically, 
\begin{align}
\frac{1}{|\mathbf r_{1}-\mathbf r_2|}=\frac{1}{2\pi}\int d\mathbf q \frac{1}{q}e^{i\bf q(\bf r_1-\bf r_2)}\label{derivation1}.
\end{align}
For simplicity, we introduce a shorthand notation for a product of Laguerre polynomials of two quantum number tuples belonging to the same particle \footnote{ In this work, $\mathbf P$ ($\mathbf Q$) and $\mathbf S$ ($\mathbf R$) belong to particle 'one' ('two').}, 
\begin{align}
L_{[S_j,R_k,\dots]}(x)= L_{S_j}^{(P_j-S_j)}(x)L_{R_k}^{(Q_k-R_k)}(x)\cdots.\label{derivation2}
\end{align}
In the following, one has to distinguish between two cases: case (i), where $P_2-S_2\geq 0$ and case (ii), where $P_2-S_2<0$. We will however only have to consider case (i), since case (ii) follows from the integral symmetry $h_{\mathbf{PQRS}}^{(ii)}={h^{(i)}_{\mathbf{SRQP}}}^*$. Moving to complex plane by substituting $\mathbf r_j$ with $r_je^{-i\theta_j}$ and $\mathbf q$ with $qe^{-i\alpha}$, we insert the Fourier transformation defined in Eq.~\eqref{derivation1} into  Eq.~\eqref{hpqrs} and write $\mathbf q\cdot \mathbf r_i=qr_i\cos(\alpha-\theta_i)$, which results in
\begin{align}
h_{\mathbf{PQRS}}=&\frac{e^2}{2\pi\epsilon}\iint_0^\infty dr_1 dr_2 dq\iint_0^{2\pi} d\theta_1 d\theta_2 d\alpha \nonumber\\
&\times r_1r_2\psi_{\mathbf P}^*(\mathbf{r_1})\psi_{\mathbf Q}^*(\mathbf{r_2})\psi_{\mathbf S}(\mathbf{r_1})\psi_{\mathbf R}(\mathbf{r_2})\nonumber \\ 
&\times e^{iq(r_1\cos(\alpha-\theta_1)-r_2\cos(\alpha-\theta_2))}\label{derivation3}.
\end{align}
First, the integration w.r.t. the polar variable $\alpha$ is performed. The result of this integration is a manifestation of the conservation of angular momentum due to the appearance of the delta function $\delta_{P_2-S_2,R_2-Q_2}$. Note that due to the conservation of angular momentum, the choice of $P_2-S_2\geq 0$ also implies that $R_2-Q_2\geq 0$. The expression after integrating out the polar degree of freedom reads
\begin{align}
h_{\mathbf{PQRS}}
=&\frac{e^2\mathcal C}{\epsilon}\int_0^\infty dqK_{\mathbf P, \mathbf S}(q)K_{\mathbf R, \mathbf Q}(q)^*\delta_{P_2-S_2,R_2-Q_2} \label{derivation7},
\end{align}
where $K_{\mathbf P, \mathbf S}(q)$ is defined as
\begin{align}
K_{\mathbf P, \mathbf S}(q)=&\int_0^\infty dr_1\int_{0}^{2\pi}d\tilde \theta_1r_1^{P_2+S_2+1}e^{i\tilde \theta_1(P_2-S_2)}e^{-\tfrac{1}{2}r_1^2}\nonumber\\ &\times L_{P_1}^{(P_2)}\left(\tfrac{r_1^2}{2}\right)L_{S_1}^{(S_2)}\left(\tfrac{r_1^2}{2}\right)e^{iqr_1\cos(\tilde\theta_1)}\label{derivation5}.
\end{align}
We use the integral representation of the Bessel function of Eq.~\eqref{derivation8} to rewrite Eq.~\eqref{derivation5}. For $y>0$ and complex parameters $\alpha$ and $\nu$, satisfying $\text{Re}\{\alpha\}>0$ and $\text{Re}\{\nu\}>-1$ \cite{kolbig1996hankel,shukla2012integral} (the results for this type of integral given in standard  literature \cite{gradshteyn2014table,erdelyi1954tables} are incorrect as they contain sign errors), we have
\begin{align}
&\int_0^\infty dx x^{\nu+1}e^{-\alpha x^2}L_m^{(\nu-\sigma)}(\alpha x^2)L_n^{(\sigma)}(\alpha x^2)J_\nu(xy)\nonumber\\
=&(-1)^{m+n}(2\alpha)^{-\nu-1}y^\nu e^{-\tfrac{y^2}{4\alpha}}L_m^{(\sigma-m+n)}\left(\tfrac{y^2}{4\alpha}\right)\nonumber\\
&\times L_n^{(\nu-\sigma+m-n)}\left(\tfrac{y^2}{4\alpha}\right)\label{derivation10},
\end{align}
which has the same functional form as the integral in Eq.~\eqref{derivation9}. We will further need the following identity for Laguerre polynomials, for $a,b\in\mathds Z$,
\begin{align}
\frac{(-x)^a}{a!}L_b^{(a-b)}(x)=\frac{(-x)^b}{b!}L_a^{(b-a)}(x),\label{derivation11}
\end{align}
which can be proven by simply inserting the definition of Laguerre polynomials into Eq.~\eqref{bs3}. Using the integral identity of Eq.~\eqref{derivation10}, we can bring Eq.~\eqref{derivation5} into the form displayed in Eq.~\eqref{derivation12} and the explicit form of the integral for case (i) in Eq.~\eqref{derivation7} reduces to
\begin{align}
h_{\mathbf{PQRS}}^{\text{(i)}}=\frac{e^2\mathcal C^{(i)}}{\epsilon}\int_0^\infty dq f^{(1)}(q)f^{(2)}(q)\delta_{P_2-S_2,R_2-Q_2}\label{derivation14},
\end{align}
where the constant $\mathcal C^{(i)}$ is given in Table~\ref{tab:integ1}.
\begingroup
\squeezetable
\begin{table*}
	\begin{centering} 
		\caption{\raggedleft In order to use the integral formula of Eq.~\eqref{derivation19}, one has to study the various parameter regimes and apply the transformation given in Eq.~\eqref{derivation11} to ensure that all requirements for using the integral formula are met. Note that the argument of the Laguerre polynomials are omitted.\raggedright}
		\renewcommand{\arraystretch}{0.1}
		\begin{tabular}{|c|| l| l|}
			\hline
			case&parameter regime& integrand substitutions \\ 
			\hline 
			&&\\
			(i.i)& $\begin{aligned}
			&(S_1-P_1\geq 0)\land(P_\Sigma-S_\Sigma<0)\\ &\land(Q_1-R_1\geq 0)\land(R_\Sigma-Q_\Sigma<0)
			\end{aligned}$ &
			$\begin{aligned}
			&L_{S_\Sigma}^{(P_\Sigma-S_\Sigma)}=\tfrac{P_\Sigma!}{S_\Sigma!}(-x)^{S_\Sigma-P_\Sigma}L_{P_\Sigma}^{(S_\Sigma-P_\Sigma)},\ L_{Q_\Sigma}^{(R_\Sigma-Q_\Sigma)}=\tfrac{R_\Sigma!}{Q_\Sigma!}(-x)^{Q_\Sigma-R_\Sigma}L_{R_\Sigma}^{(Q_\Sigma-R_\Sigma)}
			\end{aligned}$ \\ 
			\hline 		&&\\
			(i.ii)& $\begin{aligned}
			&(S_1-P_1\geq 0)\land(P_\Sigma-S_\Sigma<0)\\ &\land(Q_1-R_1\geq 0)\land(R_\Sigma-Q_\Sigma\geq0)
			\end{aligned}$ &
			$\begin{aligned}
			&L_{S_\Sigma}^{(P_\Sigma-S_\Sigma)}=\tfrac{P_\Sigma!}{S_\Sigma!}(-x)^{S_\Sigma-P_\Sigma}L_{P_\Sigma}^{(S_\Sigma-P_\Sigma)}
			\end{aligned}$ \\ 
			\hline 		&&\\
			(i.iii)& $\begin{aligned}
			&(S_1-P_1\geq 0)\land(P_\Sigma-S_\Sigma<0)\\ &\land(Q_1-R_1< 0)
			\end{aligned}$ &
			$\begin{aligned}
			&L_{S_\Sigma}^{(P_\Sigma-S_\Sigma)}=\tfrac{P_\Sigma!}{S_\Sigma!}(-x)^{S_\Sigma-P_\Sigma}L_{P_\Sigma}^{(S_\Sigma-P_\Sigma)},\ L_{R_1}^{(Q_1-R_1)}(x)=\tfrac{Q_1!}{R_1!}(-x)^{R_1-Q_1}L_{Q_1}^{(R_1-Q_1)}
			\end{aligned}$ \\ 
			\hline 		&&\\
			(i.iv)& $\begin{aligned}
			&(S_1-P_1\geq 0)\land(P_\Sigma-S_\Sigma\geq0)\\ &\land(Q_1-R_1\geq 0)\land(R_\Sigma-Q_\Sigma<0)
			\end{aligned}$ &
			$\begin{aligned}
			&L_{Q_\Sigma}^{(R_\Sigma-Q_\Sigma)}=\tfrac{R_\Sigma!}{Q_\Sigma!}(-x)^{Q_\Sigma-R_\Sigma}L_{R_\Sigma}^{(Q_\Sigma-R_\Sigma)}
			\end{aligned}$ \\ 
			\hline 		&&\\
			(i.v)& $\begin{aligned}
			&(S_1-P_1\geq 0)\land(P_\Sigma-S_\Sigma\geq0)\\ &\land(Q_1-R_1\geq 0)\land(R_\Sigma-Q_\Sigma\geq0)
			\end{aligned}$ &
			$\begin{aligned}
			&\text{no substitution necessary}
			\end{aligned}$ \\ 
			\hline 		&&\\
			(i.vi)& $\begin{aligned}
			&(S_1-P_1\geq 0)\land(P_\Sigma-S_\Sigma\geq0)\\ &\land(Q_1-R_1< 0)
			\end{aligned}$ &
			$\begin{aligned}
			&L_{R_1}^{(Q_1-R_1)}=\tfrac{Q_1!}{R_1!}(-x)^{R_1-Q_1}L_{Q_1}^{(R_1-Q_1)}
			\end{aligned}$ \\ 
			\hline 		&&\\
			(i.vii)& $\begin{aligned}
			&(S_1-P_1< 0)\land(Q_1-R_1\geq 0)\\
			&\land (R_\Sigma-Q_\Sigma<0)
			\end{aligned}$ &
			$\begin{aligned}
			&L_{P_1}^{(S_1-P_1)}=\tfrac{S_1!}{P_1!}(-x)^{P_1-S_1}L_{S_1}^{(P_1-S_1)},\ L_{Q_\Sigma}^{(R_\Sigma-Q_\Sigma)}=\tfrac{R_\Sigma!}{Q_\Sigma!}(-x)^{Q_\Sigma-R_\Sigma}L_{R_\Sigma}^{(Q_\Sigma-R_\Sigma)}
			\end{aligned}$ \\ 
			\hline 		&&\\
			(i.viii)& $\begin{aligned}
			&(S_1-P_1< 0)\land(Q_1-R_1\geq 0)\\
			&\land (R_\Sigma-Q_\Sigma\geq0)
			\end{aligned}$ &
			$\begin{aligned}
			&L_{P_1}^{(S_1-P_1)}=\tfrac{S_1!}{P_1!}(-x)^{P_1-S_1}L_{R_1}^{(P_1-S_1)}
			\end{aligned}$ \\ 
			\hline 		&&\\
			(i.ix)& $\begin{aligned}
			&(S_1-P_1< 0)\land(Q_1-R_1< 0)
			\end{aligned}$ &
			$\begin{aligned}
			&L_{P_1}^{(S_1-P_1)}=\tfrac{S_1!}{P_1!}(-x)^{P_1-S_1}L_{S_1}^{(P_1-S_1)},\ L_{R_1}^{(Q_1-R_1)}=\tfrac{Q_1!}{R_1!}(-x)^{R_1-Q_1}L_{Q_1}^{(R_1-Q_1)}
			\end{aligned}$ \\ 
			\hline
		\end{tabular}
		\label{tab:regime}
	\end{centering}
\end{table*}
\endgroup   
In order to compute the integrals in Eq.~\eqref{derivation14}, we will have to make a small detour into the properties of hypergeometric functions. We define the Pochhammer symbol (also known as the rising factorial) $(\lambda)_n= \Gamma(\lambda+n)/\Gamma(\lambda)$ and the generalized hypergeometric series \cite{niukkanen1983generalised}
\begin{align}
{_p}F_q\left[\begin{matrix}
&a_1,\dots,a_p;&\\
&b_1,\dots,b_q;&
\end{matrix}z\right]=\sum_{j=0}^\infty \frac{(a_1)_j\cdots (a_p)_j}{(b_1)_j\cdots (b_q)_j}\frac{z^j}{j!}.\label{derivation15}
\end{align}
Two properties of the generalized hypergeometric series are noteworthy: First, as soon as at least one of the numerator parameters $a_k$ is a non-positive integer, the series terminates and becomes a finite polynomial in $z$. Second, if one of the denominator parameters $b_l$ is non-positive, there will appear a zero in the denominator due to the properties of the Pochhammer symbol and Eq.~\eqref{derivation15} is no longer a well defined expression. As we will see, the requirement $b_l>0$ is the reason our final expression for $h_{\mathbf{PQRS}}$ will be case-sensitive to the values of the quantum numbers (it is the reason we have to consider all the various parameter regimes in Table~\ref{tab:regime}). There are also higher-order hypergeometric functions which possess more than one variable, such as the first Lauricella function \cite{lauricella1893sulle,poh2001some}
\begin{align}
&F_A^{(r)}\left[\begin{matrix}
a,b_1,\dots,b_r;\\c_1,\dots,c_r;
\end{matrix}z_1,\dots,z_r\right]\nonumber\\
=&\sum_{k_1,\dots,k_r=0}^\infty\frac{(a)_{k_1+\dots+k_r}(b_1)_{k_1}\cdots(b_r)_{k_r}}{(c_1)_{k_1}\cdots(c_r)_{k_r}}\frac{z_1^{k_1}}{k_1!}\cdots\frac{z_r^{k_r}}{k_r!},\label{derivation16}
\end{align}
with the constraint $(|z_1|+\dots+|z_r|)<1$, which is only of significance for the convergence of non-terminating hypergeometric series. As we shall see, all hypergeometric sums we encounter terminate, meaning that we do not need to worry about any convergence issues.  

We simplify the following expressions by omitting the arguments of the Laguerre polynomials, writing $L_n^{(\alpha)}(x)= L_n^{(\alpha)}$, where $x=q^2/2$ is a substitution used for $q$ in Eq.~\eqref{derivation14}. 
This allows us to simplify Eq.~\eqref{derivation14}, after substitution of the integration variable, to 
\begin{align}
h_{\mathbf{PQRS}}=&\frac{e^2}{\epsilon}\sqrt{\frac{P_1!Q_\Sigma!S_\Sigma!R_1!}{2P_\Sigma!Q_1!S_1!R_\Sigma!}}\int_0^\infty dx x^{(P_2-S_2+1/2)-1}\nonumber\\
&\times e^{-2x}L_{[P_1,S_\Sigma,R_1,Q_\Sigma]}\delta_{P_2-S_2,R_2-Q_2}.\label{derivation18}
\end{align}
The integral in Eq.~\eqref{derivation18} is a Laplace transform of a product of Laguerre polynomials which has been thoroughly studied, e.g. in Refs.~\cite{poh2001some,erdelyi1936einige,mayr1935integraleigenschaften,srivastava1984treatise}. For $\text{Re}\{p\}>0$, $\text{Re}\{s\}>0$, and $n_j\in\mathds N_0$ for $j=1,\dots,r$, the following integral identity holds \cite{poh2001some},
	\begin{align}
	&\int_0^\infty dx  x^{p-1}e^{-sx} L_{n_1}^{(\alpha_1)}(\lambda_1x)\cdots L_{n_r}^{(\alpha_r)}(\lambda_rx)\nonumber\\=&\frac{\Gamma(p)}{s^p}\left(\prod_{k=1}^r{{n_k+\alpha_k}\choose{n_k}}\right)\nonumber\\&\times F_A^{(r)}\left[\begin{matrix}
	p,-n_1,\dots,-n_r;\\\alpha_1+1,\dots,\alpha_r+1;
	\end{matrix}\frac{\lambda_1}{s},\dots,\frac{\lambda_r}{s}\right].\label{derivation19}
	\end{align}
By setting $r=4$, $s=2$ and $\lambda_j=1$ for $j=1,\dots,4$, we can use Eq.~\eqref{derivation19} to solve the integral in Eq.~\eqref{derivation18}. One hast to exert caution, since just as hypergeometric series must not have negative integers in its lower set of parameters, the same holds for Lauricella functions---they are a generalization of the former. A further prerequisite for applying the integral formula is that the real part of $p$ must be larger than zero and $n_j\in \mathds N_0$. In addition,  the Lauricella function is well defined only for $\alpha_j\geq0$. Using again the transformation between Laguerre polynomials given in Eq.~\eqref{derivation11}, we can bring the integral~\eqref{derivation18} to a form which meets all prerequisites for using the integral formula. 

In Table~\ref{tab:regime} we consider all possible parameter regimes for $\mathbf{P},\mathbf{Q},\mathbf{R},\mathbf{S}$ that would allow the order of any Laguerre polynomial appearing in Eq.~\eqref{derivation18} to become negative. Then, one can flip the sign of the negative exponent via Eq.~\eqref{derivation11} and the additional polynomial in $x$ ensures that the power $p$ of the final polynomial is positive. The results for all possible parameter regime choices are summarized in Tables~\ref{tab:integ1} and \ref{tab:regime}. Eq.~\eqref{derivation18} can thus be recast into the following form,
\begin{align}
h_{\mathbf{PQRS}}^{(i)}=&\frac{e^2\mathcal C^{(i)}}{\epsilon}\int_0^\infty dx x^{p-1}e^{-2x}L_{[n_1,n_2,n_3,n_4]}(x)\nonumber\\&\times\delta_{P_2-S_2,R_2-Q_2}\label{derivation20},
\end{align}
where $\mathcal C^{(i)}$, $p$ and $[n_1,n_2,n_3,n_4]$ are given in Table~\eqref{tab:integ1} and the solution of the integral is given by Eq.~\eqref{derivation19}.

The numerical challenge is thus to either find a fast and reliable implementation of the Lauricella function, or to break the Lauricella function down into lower-order hypergeometric functions. We will give an explicit solution which computes a Coulomb matrix element as a simple scalar product between two vectors. One of these two vectors contains entries which are the results of sums of fractions of rising factorials.
There are numerous ways to break down the Lauricella function into lower-order hypergeometric expressions, and which one to pic should depend on which expressions one can compute fast and reliable. In the last part of this section, we give an example on how this can be achieved. We consider the integral representation of the Lauricella function $F_A^{(r)}$\cite{padmanabham2000summation}
\begin{align}
&F_A^{(r)}\left[\begin{matrix}
a,b_1,\dots,b_r;\\
c_1,\dots,c_r;
\end{matrix}x_1,\dots,x_r\right]\nonumber\\
=&\int_0^{\infty}dte^{-t}\frac{t^{a-1}}{\Gamma(a)}\prod_{j=1}^r{_1}F_{1}\left[\begin{matrix}
b_j;\\
c_j;
\end{matrix}x_jt\right]\label{derivation21},
\end{align}
where $\text{Re}\{a\}>0$ and $\text{Re}\{x_1+\dots+x_r\}<1$ and we are only interested in the case where $r=4$. Clearly, in order to avoid zeros in the denominator, we require all $c_j$ to be positive integers. Using the definition of the hypergeometric function and the Cauchy-product formula, the product of two such hypergeometric functions results in
\begin{align}
&{_1}F_{1}\left[\begin{matrix}
b_1;\\
c_1;
\end{matrix}x_1t\right]{_1}F_{1}\left[\begin{matrix}
b_2;\\
c_2;
\end{matrix}x_2t\right]\nonumber\\
=&\sum_{k=0}^{\infty}\left(\sum_{l=0}^k\frac{(b_1)_{k-l}(b_2)_{l}}{(c_1)_{k-l}(c_2)_l}\frac{x_1^{k-l}x_2^l}{(k-l)!l!}\right)t^k.\label{derivation22}
\end{align}
We perform the product of all four hypergeometric functions of Eq.~\eqref{derivation21} in the above manner and the resulting coefficient of the resulting polynomial is given by
\begin{align}
\xi_k=\sum_{p=0}^k&\sum_{q=0}^p\sum_{r=0}^q \frac{(-n_1)_r(-n_2)_{q-r}}{(\alpha_1+1)_{r}(\alpha_2+1)_{q-r}(\alpha_3+1)_{p-q}}\nonumber\\
&\times \frac{(-n_3)_{p-q}(-n_4)_{k-p}}{(\alpha_4+1)_{k-p}r!(q-r)!(p-q)!(k-p)!2^k}\label{derivation23},
\end{align}
which we call the convolution coefficient. This coefficient allows us to compute the expression in Eq.~\eqref{derivation21},
\begin{align}
&\int_0^\infty dt e^{-t}\frac{t^{p-1}}{\Gamma(p)}\prod_{i=1}^4{_1}F_{1}\left[\begin{matrix}
-n_i;\\
\alpha_i+1;
\end{matrix}\tfrac{t}{2}\right]\nonumber\\
=&\sum_{k=0}^{n_1+n_2+n_3+n_4}\frac{\xi_k}{\Gamma(p)}\int_0^\infty dt e^{-t} t^{p+k-1}\nonumber\\
=&\sum_{k=0}^{n_1+n_2+n_3+n_4}\xi_k(p)_k.\label{derivation24}
\end{align}
We recognize the definition of the Gamma function in the last integral. By using the definition of the Pochhammer symbol and defining the two column vectors $\boldsymbol \xi=(\xi_0,\xi_1,\dots,\xi_{n_1+n_2+n_3+n_4})^T$ and $\boldsymbol{(p)}=((p)_0,(p)_1,\dots,(p)_{n_1+n_2+n_3+n_4})^T$, we have
\begin{align}
&F_A^{(r)}\left[\begin{matrix}
a,b_1,\dots,b_r;\\
c_1,\dots,c_r;
\end{matrix}x_1,\dots,x_r\right]
= \boldsymbol \xi \cdot \boldsymbol{(p)}
\label{derivation25}.
\end{align}
As a sanity check, we compared the values of Eq.~\eqref{derivation25} combined with the additional prefactors appearing in Eqs.~\eqref{derivation19} and \eqref{derivation20},  with $M^l_{mn}$ in Eq.~\eqref{coul2} for the LLL for various system sizes and they are---up to numerical precision error---in exact agreement with one another. 
\section{Hamiltonian simulation through linear combination of unitaries using the self-inverse matrix decomposition strategy\label{ryan}}
In this section, we use an alternative algorithm to the low rank factorization algorithm of \cite{Berry2019} used in Section~\ref{lcu} to sample the eigenspectrum of the Hamiltonian $H$. This algorithm is based on the self-inverse matrix decomposition strategy first described in Ref.~\cite{Berry2013} and makes use of the fact that we have an analytical form for the matrix elements of the Hamiltonian as derived in Section~\ref{system_hamiltonian} and that within the LLL, the largest matrix element $\max_\ell(\omega_\ell)$ of the Hamiltonian elements is constant $O(1)$ \footnote{The largest matrix element at fixed filling factor $\nu=1/3$ turns out to be identical in the LLL for all studied angular momenta cutoffs.}. As we will see, this approach scales considerably worse than the  low rank factorization method of Ref.~\cite{Berry2019} presented in Section~\ref{lcu}, unless one is able to considerably lower the computational cost of evaluating sums of products of factorials as given by Eq.~\eqref{lauri2} and Eqs.~\eqref{coul2}-\eqref{coul5}. In the approach used in this section one will dramatically increase the number of terms in the Hamiltonian but with the advantage that the coefficients of the state $\textsc{prepare} \ket{0}^{\otimes \log(L)}$ will only be $1$ or $i$, which makes the state much simpler to prepare. 

As described in Section~\ref{system_hamiltonian}, for the Hamiltonians of interest in this paper we are able to compute the $\omega_\ell$ efficiently from $\ell$ (which is essentially equivalent to computing the $h_{{\bf P}{\bf Q}{\bf R}{\bf S}}$ from the indices ${\bf P}, {\bf Q}, {\bf R}$ and ${\bf S}$). Several steps are required in order to go from computing these coefficients to implementing the $\textsc{prepare}$ operator. The ability to compute the coefficients essentially allows us to prepare the state
\begin{equation}
   \sqrt{\frac{1}{L}} \sum_{\ell=1}^L  \ket{\ell} \ket{\omega_\ell}.\label{pr1}
\end{equation}
But to translate this into the desired state,
\begin{equation}
   \sqrt{\frac{1}{\lambda}} \sum_{\ell=1}^L  \sqrt{\omega_\ell} \ket{\ell},
\end{equation}
we will use the self-inverse matrix decomposition strategy first described in Ref.~\cite{Berry2013}. This is described in the context of simulating electronic structure in Section 4.4 of Ref.~\cite{babbush2017exponentially}. The result is that $C_P$ (again, the cost to implement $\textsc{prepare}$) ends up scaling like the cost to compute the coefficient mentioned above, but the $\lambda$ value is increased to scaling like $O(L \max_\ell(\omega_\ell))$.

Essentially, the strategy which leads to this scaling is as follows. First, one re-imagines the Hamiltonian as being a sum of a very large number of terms where each term has a coefficient that is the same magnitude; specifically, the coefficient of each term is either $+ \zeta$ or $-\zeta$ where $\zeta$ is thus chosen to limit the precision of the Hamiltonian representation, i.e., $\zeta = O(\epsilon)$. The largest term in the original Hamiltonian will have the property that each of the subterms into which it is decomposed in the new Hamiltonian has the same coefficient; thus, each term consists of $O(\max_\ell(\omega_\ell) / \epsilon)$ subterms of magnitude $\zeta$. The advantage of this is that the coefficients of the state that we must realize with $\textsc{prepare}$ are now all either $1$ or $i$. For details of how this is realized, see Ref.~\cite{Berry2013}. In order to determine whether we should phase by $1$ or $i$ we need to compute the coefficient and compare it to a coin register. Essentially, we are dramatically increasing the number of terms in the Hamiltonian by decomposing each term into a sum of small terms that all have the same coefficient up to a sign. The difficult part of realizing prepare is thus simply to decide which sign is associated with each computational basis.

In the following, we will give an estimate of the time complexity for computing the two-body Hamiltonian matrix elements $h_{\mathbf{PQRS}}$ for the LLL approximation, where $h_{\mathbf P\mathbf Q\mathbf R\mathbf S}=M_{mn}^l$ is given by Eq.~\eqref{coul2}. These coefficients correspond to the $\omega_\ell$ mentioned before. We ignore the complexity of the one-body matrix elements of Eq.~\eqref{kin1}, as they are trivial to compute and there are many fewer of them so they are easier to simulate as well. We expect a slightly worse, but similar complexity scaling when computing $h_{\mathbf{PQRS}}$ via Eq.~\eqref{lauri2}, which takes into account LL mixing. We consider Eq.~\eqref{coul2}, and observe that it is sufficient to only consider the complexity of computing $C_{lmn},A_{mn}^l$ and $B_{nm}^l$ as given by Eqs.~\eqref{coul3}-\eqref{coul5}, respectively and then singling out the term possessing the largest complexity. While $C_{lmn}$ is dominated by the cost of computing the Gamma function, $A_{mn}^l$ and  $B_{nm}^l$ require the evaluation of a finite sum of division and products of Gamma functions, which will thus have a larger complexity than $C_{lmn}$.

The problem is that to compute the value of $A_{mn}^l$ and  $B_{nm}^l$ to within precision $\epsilon$ it is required to use a number of bits that scales as $O(M \log M)$ where $M$ is the cutoff in angular momentum). This is because $A_{mn}^l$ and  $B_{nm}^l$ involve computing factorials of $M$ and we know from Stirling's approximation that $\log (M!) = {O}(M \log M)$. We then need to multiply these numbers together, which gives us complexity $O(M^2 \text{polylog}(M)) = \tilde{O}(M^2)$. If one then explicitly evaluates the sum the complexity becomes $\tilde{O}(M^3)$, which is very bad. One could choose to expand the sum using LCU methods (by which we mean, one can consider each term in the sum as a distinct term in the Hamiltonian) but this will dramatically increase $\lambda$. We note that the complexity bound on computing the coefficients $\omega_\ell$ can in principle be further reduced using algorithms designed for computing linearly convergent series as in Refs.~\cite{haible1998fast,karatsuba1991fast}.

Unfortunately, the number of times we must repeat this primitive is $\lambda = O(L \max_\ell (\omega_\ell))$. In our context, when restricting ourselves to the LLL, $\omega_\ell = O(1)$ for the two-body operator and we have $L = O(M^3)$ for that operator \footnote{The scaling in terms of $N$ (which we neglect in that section) and $M$ is however not $O(N^3M^3)$, as one might think, but $O(N^4M^3)$, since the conservation of angular momentum only reduces the $M$ scaling by one order.}. For the one-body operator $\max_\ell \omega_\ell = O(N)$, where $N$ is the number of LLs, but there are only $L = O(N_{\text{so}})$ terms due to the delta function in Eq.~\eqref{kin1}. Furthermore, when considering only the LLL, we have $\max_\ell \omega_\ell = O(1)$ and $L = O(M)$ for the two-body operators. Thus, overall we have that $\lambda = O(N_{\rm so} N + M^3) = O(M^3)$ when restricting ourselves to the LLL.

Putting this all together then, we see that using Eq.~\eqref{scaling_1}, the total complexity of deploying phase estimation to estimate the ground state energy within precision $\Delta E$ is $\tilde{O}(M^6 / \Delta E)$ up to log factor, which is considerably worse than the $O(N^{4.35} / \Delta E)$ T gate complexity obtained when using the low rank factorization of Ref.~\cite{Berry2019} presented in Section~\ref{lcu}. Overall, it is surprising that we cannot exceed this more generic strategy despite having a closed form for the coefficients. The reason is ultimately because of the extremely high precision required to compute the Gamma functions in the coefficients.
\section{Derivations for the equations of motion of the CM \label{cov}}
For FGS (as defined in Eq.~\eqref{single1}) with fixed particle number, Wick's theorem gives 
\begin{align}
\braket{c_p^\dag c_i^\dag c_q c_j}=&-\Gamma_{qp}\Gamma_{ji}+\Gamma_{jp}\Gamma_{qi}\label{single7}\\
\braket{c_p^\dag c_q^\dag c_i^\dag c_r c_s c_j}=&\Gamma_{jp}\Gamma_{sq}\Gamma_{ri}-\Gamma_{jp}\Gamma_{rq}\Gamma_{si}+\Gamma_{sp}\Gamma_{rq}\Gamma_{ji}\nonumber\\
&-\Gamma_{sp}\Gamma_{jq}\Gamma_{ri}+\Gamma_{rp}\Gamma_{jq}\Gamma_{si}-\Gamma_{rp}\Gamma_{sq}\Gamma_{ji}\label{single8},
\end{align}
since the pairing terms $\braket{c_ic_j}$ and $\braket{c_i^\dag c_j^\dag}$ vanish \cite{eisert2018entanglement}. We compute the quadratic contribution to the imaginary-time evolution of Eq.~\eqref{single6} using Eqs.~\eqref{single7} and \eqref{single8},
\begin{align}
\text{tr}[\{T,c_i^\dag c_j\}\rho]=&\sum_{p,q}f_{pq}\braket{c_p^\dag c_q c_i^\dag c_j+ c_i^\dag c_j c_p^\dag c_q}\nonumber\\
=&-2\sum_{p,q}f_{pq}\braket{c_p^\dag c_i^\dag c_q c_j}+\sum_{p}f_{pi}\braket{c_p^\dag c_j}\nonumber
\\
&+\sum_{q}f_{jq}\braket{c_i^\dag c_q}\nonumber\\
=&2\sum_{p,q}f_{pq}(\Gamma_{qp}\Gamma_{ji}-\Gamma_{jp}\Gamma_{qi})+\sum_{p}f_{pi}\Gamma_{jp}\nonumber\\
&+\sum_{q}f_{jq}\Gamma_{qi}\nonumber\\
=&2\text{tr}[f\Gamma]\Gamma_{ji}-2[\Gamma f \Gamma]_{ji}+[\{\Gamma, f\}]_{ji}.\label{cov1}
\end{align} 
The contributions from the quartic interaction term are given by 
	\begin{align}
	&\text{tr}[\{V,c_i^\dag c_j\}\rho]\nonumber\\
	=&\sum_{p,q,r,s}h_{pqrs}\braket{c_p^\dag c_q^\dag c_i^\dag c_r  c_s c_j}-\tfrac{1}{2}\sum_{p,q,s}h_{pqis}\braket{c_p^\dag c_q^\dag c_s c_j}\nonumber\\&+\tfrac{1}{2}\sum_{p,q,r}h_{pqri}\braket{c_p^\dag c_q^\dag c_r c_j}-\tfrac{1}{2}\sum_{p,r,s}h_{pjrs}\braket{c_i^\dag c_p^\dag c_r c_s}\nonumber\\
	&+\tfrac{1}{2}\sum_{q,r,s}h_{jqrs}\braket{c_i^\dag c_q^\dag c_r c_s}.\label{cov2}
	\end{align}
The first term on the right-hand side of Eq.~\eqref{cov2} gives
\begin{align}
&\sum_{p,q,r,s}h_{pqrs}\braket{c_p^\dag c_q^\dag c_i^\dag c_r  c_s c_j}\nonumber\\
=&\sum_{p,q,r,s}h_{pqrs}(\Gamma_{jp}\Gamma_{sq}\Gamma_{ri}-\Gamma_{jp}\Gamma_{rq}\Gamma_{si}+\Gamma_{sp}\Gamma_{rq}\Gamma_{ji}\nonumber\\[-10pt]
&\qquad\qquad\quad -\Gamma_{sp}\Gamma_{jq}\Gamma_{ri}+\Gamma_{rp}\Gamma_{jq}\Gamma_{si}-\Gamma_{rp}\Gamma_{sq}\Gamma_{ji})\nonumber\\
=&-4[\Gamma \text{tr}_{1,4}[h\Gamma]\Gamma]_{ji}+2\text{tr}[\text{tr}_{1,4}[h\Gamma]\Gamma]\Gamma_{ji},\label{cov3}
\end{align}
the remaining terms contribute
\begin{align}
&-\tfrac{1}{2}\sum_{p,q,s}h_{pqis}\braket{c_p^\dag c_q^\dag c_s c_j}+\tfrac{1}{2}\sum_{p,q,r}h_{pqri}\braket{c_p^\dag c_q^\dag c_r c_j}\nonumber\\
&-\tfrac{1}{2}\sum_{p,r,s}h_{pjrs}\braket{c_i^\dag c_p^\dag c_r c_s}+\tfrac{1}{2}\sum_{q,r,s}h_{jqrs}\braket{c_i^\dag c_q^\dag c_r c_s}\nonumber\\
=&2[\{\Gamma,\text{tr}_{1,4}[h\Gamma]\}]_{ji}.\label{cov4}
\end{align}
The second term in Eq.~\eqref{single6} simplifies to 
\begin{align}
2\Gamma_{ji}\text{tr}[H\rho(t)]=&2\text{tr}[f\Gamma]\Gamma_{ji}+2\text{tr}[\text{tr}_{1,4}[h\Gamma]\Gamma]\Gamma_{ji}.\label{cov5}
\end{align}
Defining the mean field term
\begin{align}
h_m(\Gamma)= f+2\text{tr}_{1,4}[h\Gamma],\label{cov6}
\end{align}
the imaginary-time evolution of $\Gamma$ is given by 
\begin{align}
d_\tau\Gamma_{ji}
=&-[\{\Gamma, h_m(\Gamma)\}]_{ji}+2[\Gamma h_m(\Gamma) \Gamma]_{ji}.\label{cov7}
\end{align}
\section{Rotating the system Hamiltonian into the eigenbasis of the CM \label{eigenbasis_cov}}
For the ASCI algorithm described in Section~\ref{asci_explain}, we rotate the Hamiltonian of Eq.~\eqref{sec1} into the eigenbasis of the CM $\Gamma$ at the end of the imaginary time evolution. Let $O$ be the matrix which diagonalizes the CM as in Eq.~\eqref{single13} and corresponding orbital rotations as given in Eq.~\eqref{single15}. The system Hamiltonian in the mean-field eigenbasis is then given by 
\begin{align}
H =& \sum_{i,j}\tilde f_{ij}\tilde c_i^\dag\tilde c_j + \frac{1}{2}\sum_{i,j,k,l}\tilde h_{ijkl}\tilde c_i^\dag \tilde  c_j^\dag \tilde c_k\tilde c_l\label{eigen1}
\end{align}
where
\begin{align}
\tilde f_{ij}=& \sum_{p,q}O_{pi} f_{pq} O_{qj}\label{eigen2}\\
\tilde h_{ijkl} =& \sum_{p,q,r,s}h_{pqrs}O_{pi}O_{qj}O_{rk}O_{sl}.\label{eigen3}
\end{align}
Note that under the orthogonal transformations, the integrals are still anti-symmetric w.r.t. index permutation (however, the conservation of angular momentum is now no longer visible in the indices). 
\section{Imaginary time evolution - formal integration \label{formal_integration}}
\begin{figure}[h!]
	\includegraphics[width=.49\textwidth]{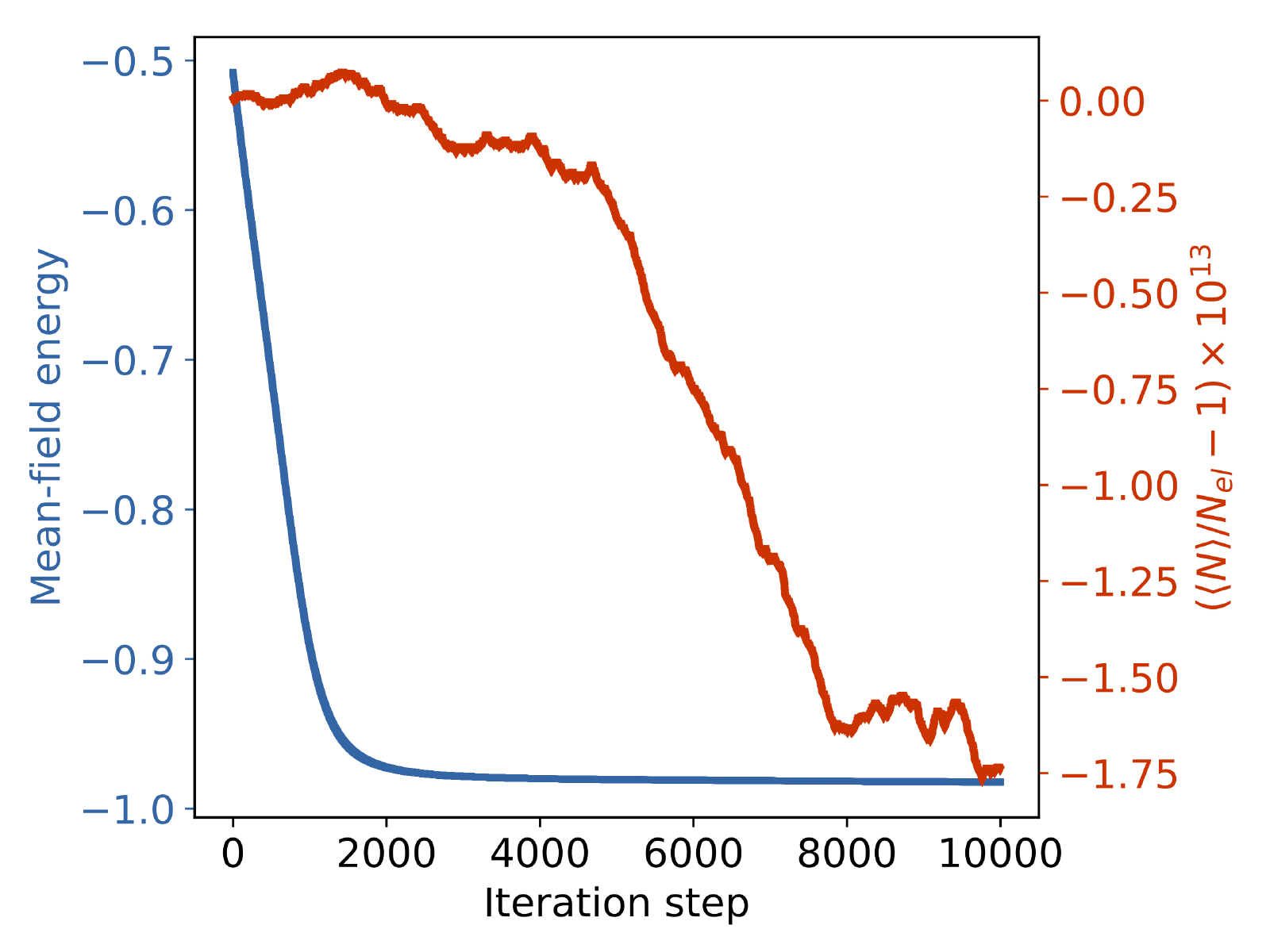}
	\caption{Scatter plot of the monotonic decrease of the mean-field energy (blue, left $y$-axis) as defined in Eq.~\eqref{single11} at each iteration step for the imaginary time evolution of the CM for a system of $N_{\text{el}}=4$ electrons distributed among $N_{\text{so}}=12$ spin-orbitals at a chemical potential $\mu=1$ and step size $\Delta\tau=0.01$.  We also calculated the deviation of the number of particles present at each iteration step by plotting $(\braket{N}/N_{\text{el}}-1)\times 10^{13}$ (red line, right $y$-axis), where $\braket{N} = \text{tr}(\Gamma)$ is the expectation value of the particle number operator. Note that the particle number changes from $N_{\text{el}}=4$ only in the 13. decimal place.     \label{fig:num_cons}} 
\end{figure}

We aim to solve the differential equation
\begin{align*}
d_\tau\Gamma = -h_m(\Gamma)-\Gamma h_m(\Gamma)\Gamma.
\end{align*}
Following Ref.~\cite{kraus2010generalized}, we formally integrate the equation of motion, which result in
\begin{align}
\Gamma(\tau) = O(\tau) \Gamma(0) O(\tau)^T,
\end{align}
where $O(\tau)$ is an orthogonal matrix (if the transformation were not orthogonal, it could take us out of the family of FGS, where Wick's theorem no longer applies) given by 
\begin{align}
O(\tau)=\mathcal T\exp\left(\int_0^\tau d \tau'A(\Gamma(\tau'))\right),
\end{align}
with $\mathcal T$ denoting the time-ordering operator. For a small time step $\Delta \tau$, we can expand $O(\tau)$ using the orthogonality property $A(\Gamma(\tau))^T=-A(\Gamma(\tau))$ to get in first order $\Delta \tau$
\begin{align}
\Gamma(\tau+\Delta \tau) =& \Gamma(\tau) - \Gamma(\tau)A(\Gamma(\tau))\Delta \tau\nonumber\\ &+ A(\Gamma(\tau))\Gamma(\tau)\Delta \tau+O(\Delta \tau^2).
\end{align}
This leads to 
\begin{align}
\frac{\Gamma(\tau+\Delta\tau)-\Gamma(\tau)}{\Delta\tau}= [A(\Gamma(\tau)),\Gamma(\tau)]
\end{align}
which in the limit of small $\Delta\tau$ should be equal to the right-hand side of Eq.\eqref{single6}. Together with the fact that $\Gamma^2=-\mathds 1_{2N_f}$, this allows us to find an explicit expression for $A(\Gamma(\tau))$, namely
\begin{align}
A(\Gamma(\tau))=\frac{1}{2}[h_m(\Gamma(\tau)),\Gamma(\tau)], 
\end{align}
since 
\begin{align}
[A,\Gamma]
=&\frac{1}{2}\left(-h_m -\Gamma h_m\Gamma-\Gamma h_m\Gamma-h_m\right) \nonumber\\
=& -h_m -\Gamma_mh_m\Gamma_m.
\end{align}
Thus, for small time steps $\Delta\tau$, we can compute the CM by an orthogonal transformation of the prior CM via
\begin{align}
\Gamma(\tau+\Delta\tau) \approx& \exp\left(A(\Gamma(\tau))\Delta \tau\right)\Gamma(\tau)\nonumber\\
&\times \exp\left(-A(\Gamma(\tau))\Delta\tau\right).\label{ite99}
\end{align}
While in the Dirac representation of fermionic creation and annihilation operators the particle number is conserved, small numerical fluctuations will lower the number of particles when working in a Majorana representation, where only parity is a conserved quantity. One would have to introduce a chemical potential in order to enforce particle number conservation during the iterative process in the latter case. As shown in Fig.~\ref{fig:num_cons}, particle number is conserved when solving the equations of motions in the fermionic basis following the imaginary time evolution as defined in Eq.~\eqref{ite99}.
\section{Computing correlation function for single-reference and multi-reference states \label{corr}}
If the initial state is a single Slater determinant $\ket{\Psi_{\text{GS}}}$, we have 
\begin{align}
G_1(\mathbf r, \mathbf r')=&
\sum_{p,q}\eta_p^*(\mathbf r)\eta_q(\mathbf r')\bra{\Psi_{\text{GS}}}c_p^\dag c_q\ket{\Psi_{\text{GS}}}\nonumber\\
=&\boldsymbol{\eta}(\mathbf r')^T\Gamma \boldsymbol{\eta}^*(\mathbf r)\label{corr1}
\end{align}
where $\boldsymbol{\eta}(\mathbf r)=(\eta_0(\mathbf r), \eta_1(\mathbf r),\dots,\eta_{M}(\mathbf r))^T$ is a vector of the basis functions chosen and $M$ is the angular momentum cutoff. The mean-field density correlations are given by 
\begin{align}
G_2(\mathbf r, \mathbf r') 
=&\boldsymbol{\eta}(\mathbf r)^T\Gamma \boldsymbol{\eta}^*(\mathbf r)\boldsymbol{\eta}(\mathbf r')^T\Gamma \boldsymbol{\eta}^*(\mathbf r')\nonumber\\
&-\boldsymbol{\eta}(\mathbf r')^T\Gamma \boldsymbol{\eta}^*(\mathbf r)\boldsymbol{\eta}(\mathbf r)^T\Gamma \boldsymbol{\eta}^*(\mathbf r').\label{corr2}
\end{align}

Both correlation functions in Eqs.~\eqref{corr1} and \eqref{corr2} can be computed efficiently. One can compute the one-particle reduced density matrix and the pair correlation function also for the multi-reference state of Eq.~\eqref{asci1} through the CM  using Wick's theorem. 

\begin{figure}[h!]
	\centering                                              
	\includegraphics[width=.49\textwidth]{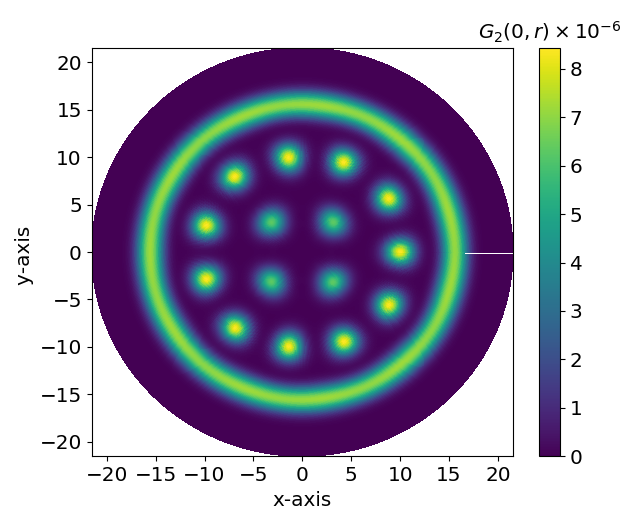}
	\caption{Plot of the pair correlation function $G_2(\mathbf 0, \mathbf r)$, with $N_{\text{so}}=138$ at filling $\nu=1/3$ ($N_{\text{el}}=46$), as defined in Eq.~\eqref{meas4} and expressed in terms of the CM as outlined in Section~\ref{corr} for a single-reference state $\ket{\Psi_{\text{GS}}}$ obtained from imaginary time evolution via the formal integration method of Section~\ref{formal_integration}, with $10^4$ steps and step size $\Delta t=0.1$. The first particle was located at the origin. A crystal structure emerges in the bulk, while the disk boundary is centered at a circle of radius $R_d=\sqrt{2(M+1)}\approx 16.6$, where the magnetic length is set to unity.\label{fig:g_2}}
\end{figure}

In Fig.~\ref{fig:g_2}, we show the pair correlation function for the FGS solution obtained from imaginary time evolution as introduced in Section~\ref{hf} for $N_{\text{so}}=138$ spin-orbitals at filling $\nu=1/3$ in the LLL, with an apparent crystal-like structure emerging as expected for mean-field solutions (see e.g. chapter 4 in Ref.~\cite{yoshioka2013quantum}). 

\section{Computing the overlaps}
The following subsections detail some of the computational steps that were used for implementing the ASCI algorithm.
\subsection{Computing the diagonal terms\label{hii}}
One of the computationally more costly steps in the ASCI algorithm is getting the diagonal Hamiltonian term 
\begin{align}
H_{ii}=\braket{A_i|H|A_i},\label{over1}
\end{align}
where $\ket{A_i}$ is a determinant from the set of determinants $H\ket{\{C\}}$ (which are all unique determinants that have non-zero coefficients when acting with $H$ on the core determinants $\ket{\{C\}}=\{\ket{C_1},..
.,\ket{C_{\text{cdets}}}\}$). By inserting the Hamiltonian of  Eq.~\eqref{eigen1}, we get
\begin{align}
H_{ii} 
=&\sum_{p,q}\tilde f_{pq}\bra{A_i}\tilde c_p^\dag\tilde c_q\ket{A_i} \nonumber\\
&+ \frac{1}{2}\sum_{p,q,r,s}\tilde h_{pqrs}\bra{A_i}\tilde c_p^\dag \tilde  c_q^\dag \tilde c_r\tilde c_s\ket{A_i}.\label{over2}
\end{align}
Let us first look at the second term in Eq.~\eqref{over2}. The only non-vanishing terms are given by either (i) $p=r$ and $q=s$, or (ii) $p=s$ and $q=r$. Furthermore, both $p$ and $q$ must be occupied in $\ket{A_i}$, a condition we will denote as $\{p,q\}\in A_{i,\text{occ}}$. Since $p\neq q$ (a fermionic mode can only contain zero or one particle) and due to the anti-symmetry in $\tilde h_{pqrs}$, we can unify cases (i) and (ii), which results in a factor of two, 
\begin{align}
\sum_{p,q,r,s}\tfrac{\tilde h_{pqrs}}{2}\bra{A_i}\tilde c_p^\dag \tilde  c_q^\dag \tilde c_r\tilde c_s\ket{A_i}
=& \sum_{p,q}\tilde h_{pqpq}\bra{A_i}\tilde c_p^\dag \tilde  c_q^\dag \tilde c_p\tilde c_q\ket{A_i}\nonumber\\
=& -2\sum_{\substack{\{p,q\}\in A_{i,\text{occ}}\\p<q}}\tilde h_{pqpq},\label{over3}
\end{align}
where the minus sign is due to the Jordan-Wigner transformation (which we will explain a couple of lines below in more detail). Similarly, we get for the first term in Eq.\eqref{over2}
\begin{align}
\sum_{p,q}\tilde f_{pq}\bra{A_i}\tilde c_p^\dag\tilde c_q\ket{A_i}=\sum_{p\in A_{i,\text{occ}}}\tilde f_{pp}, \label{over4}
\end{align}
thus resulting in a simple formula for the diagonal terms
\begin{align}
H_{ii} = \sum_{p\in A_{i,\text{occ}}}\tilde f_{pp}-2\sum_{\substack{\{p,q\}\in A_{i,\text{occ}}\\p<q}}\tilde h_{pqpq},\label{over5}
\end{align}
where the dependence on index $i$ results in the task of finding all occupied orbitals $A_{i,\text{occ}}$ in the determinant $\ket{A_i}$, which is a simple problem.
\subsection{Computing the off-diagonal terms\label{hij}}
We consider the action of the Hamiltonian operator on a core determinant
\begin{align}
H \ket{C_j} = \left( \sum_{p,q}\tilde f_{pq}\tilde c_p^\dag\tilde c_q + \frac{1}{2}\sum_{p,q,r,s}\tilde h_{pqrs}\tilde c_p^\dag \tilde  c_q^\dag \tilde c_r\tilde c_s \right)\ket{C_j}.\label{over6}
\end{align}
We are going to treat the one- and two-body operators separately. 
\subsubsection{One-body terms}
Since we have real coefficients, $\tilde f_{pq} = \tilde f_{qp}$ and thus
\begin{align}
\sum_{p,q}\tilde f_{pq}\tilde c_p^\dag\tilde c_q\ket{C_j} =& 2\sum_{p<q}\tilde f_{pq}\tilde c_p^\dag\tilde c_q\ket{C_j}+\sum_p\tilde f_{pp}\tilde c_p^\dag \tilde c_p \ket{C_j}.\label{over7}
\end{align}
In our case, the one-body terms are diagonal, therefore 
\begin{align}
\sum_{p,q}\tilde f_{pq}\tilde c_p^\dag\tilde c_q\ket{C_j} =& \sum_p\tilde f_{pp}\tilde c_p^\dag \tilde c_p \ket{C_j}.\label{over8}
\end{align}
\subsubsection{Two-body terms}
Due to the real-valued, anti-symmetric nature of the two-body coefficients, we have 
\begin{align}
\sum_{p,q,r,s} \tfrac{\tilde h_{pqrs}}{2}\tilde c_p^\dag \tilde  c_q^\dag \tilde c_r\tilde c_s\ket{C_j} =&
2\sum_{p<q;r<s}\tilde h_{pqrs}\tilde c_p^\dag \tilde  c_q^\dag \tilde c_r\tilde c_s\ket{C_j}, \label{over9}
\end{align}
since $p\neq q$ and $r\neq s$ (otherwise, they would give zero). We can break the operator into parts of single and double excitations (we will omit the factor $2$ in front of the sum for now and add it later),
	\begin{align}
	&\sum_{p<q,r<s}\tilde h_{pqrs}\tilde c_p^\dag \tilde  c_q^\dag \tilde c_r\tilde c_s \nonumber\\
	=&\sum_{\substack{p<q;r<s\\ p\neq \{r,s\};q\neq \{r,s\} }}\tilde h_{pqrs}\tilde c_p^\dag \tilde  c_q^\dag \tilde c_r\tilde c_s  
	+ \sum_{\substack{q<p;r<p\\ q\neq r }}\tilde h_{qprp}\tilde c_q^\dag \tilde  c_p^\dag \tilde c_r\tilde c_p  \nonumber\\
	&
	+ \sum_{\substack{q<p<r}}\tilde h_{qppr}\tilde c_q^\dag \tilde  c_p^\dag \tilde c_p\tilde c_r+ \sum_{\substack{r<p<q}}\tilde h_{pqrp}\tilde c_p^\dag \tilde  c_q^\dag \tilde c_r\tilde c_p
	\nonumber\\
	&+ \sum_{\substack{p<q;p<r\\ q\neq r }}\tilde h_{pqpr}\tilde c_p^\dag \tilde  c_q^\dag \tilde c_p\tilde c_r
	+ \sum_{\substack{p<q }}\tilde h_{pqpq}\tilde c_p^\dag \tilde  c_q^\dag \tilde c_p\tilde c_q.\label{over10}
	\end{align}
In the following, we will discuss how we can numerically get the matrix elements. 
\subsection{Identity map}
Some terms in $H$ map the input determinant $\ket{D_i}$ back onto itself. Since in Eq.~\eqref{asci2}, the sum is taken over all input core determinants that do not map onto itself via the action of $H$, such amplitudes will be set to zero. 
\subsection{One-body excitations}
We have four terms in Eq.~\eqref{over10} that create single excitations. Such determinants which differ only by a single creation-annihilation pair are called single-connected. In order to know which terms to take into account when going from a determinant $\ket{D_1}$ to a single-connected determinant $\ket{D_2}$ via application of $H\ket{D_1}$, we first determine the pair $[i,j],\quad \text{with}\ i<j$, that indicates the spin-orbitals $i$ and $j$ where the two determinants differ. We let $k_j\in\{0,1\}$ denote the occupation of spin-orbital $j$ and consider two distinct cases.

We denote with $\alpha_{\text{JW}}$ the phase factor due to the Jordan-Wigner transformation as introduced in Section~\ref{jw} and first let $k_i=0$ in $\ket{D_1}$, which leads to $k_j=1$ in $\ket{D_2}$ and thus
\begin{align}
&\sum_{\substack{q<p;r<p\\ q\neq r }}\tilde h_{qprp}\tilde c_q^\dag \tilde  c_p^\dag \tilde c_r\tilde c_p = \sum_{\substack{p>j\\p\in N_{\text{el}}^{\ket{D_1}}}}\tilde h_{ipjp}\tilde c_i^\dag \tilde  c_p^\dag \tilde c_j\tilde c_p\\
&\qquad\text{with}\ \alpha_{\text{JW}} = -(-1)^{k_{i+1}+...+k_{j-1}}\nonumber\\
&\sum_{\substack{q<p<r}}\tilde h_{qppr}\tilde c_q^\dag \tilde  c_p^\dag \tilde c_p\tilde c_r =
\sum_{\substack{i<p<j\\p\in N_{\text{el}}^{\ket{D_1}}}}\tilde h_{ippj}\tilde c_i^\dag \tilde  c_p^\dag \tilde c_p\tilde c_j\\
&\qquad\text{with}\ \alpha_{\text{JW}}  = (-1)^{k_{i+1}+...+k_{p-1}+1_p+k_{p+1}+...+k_{j-1}}\nonumber\\
&\sum_{\substack{r<p<q}}\tilde h_{pqrp}\tilde c_p^\dag \tilde  c_q^\dag \tilde c_r\tilde c_p = \sum_{\substack{j<p<i\\p\in N_{\text{el}}^{\ket{D_1}}}}\tilde h_{pijp}\tilde c_p^\dag \tilde  c_i^\dag \tilde c_j\tilde c_p = 0 \\ &\qquad\text{since} \ i<j\nonumber\\
&\sum_{\substack{p<q;p<r\\ q\neq r }}\tilde h_{pqpr}\tilde c_p^\dag \tilde  c_q^\dag \tilde c_p\tilde c_r = \sum_{\substack{p<i\\p\in N_{\text{el}}^{\ket{D_1}}}}\tilde h_{pipj}\tilde c_p^\dag \tilde  c_i^\dag \tilde c_p\tilde c_j\\
&\qquad\text{with}\ \alpha_{\text{JW}} = -(-1)^{k_{i+1}+...+k_{j-1}},\nonumber
\end{align}
where $p\in N_{\text{el}}^{\ket{D_1}}$ are indices belonging to occupied spin orbitals. 

Second, if $k_i=1$ in $\ket{D_1}$, we have $k_j=1$ in $\ket{D_2}$ and thus
\begin{align}
&\sum_{\substack{q<p;r<p\\ q\neq r }}\tilde h_{qprp}\tilde c_q^\dag \tilde  c_p^\dag \tilde c_r\tilde c_p = \sum_{\substack{p>j\\p\in N_{\text{el}}^{\ket{D_1}}}}\tilde h_{jpip}\tilde c_j^\dag \tilde  c_p^\dag \tilde c_i\tilde c_p\label{apx1}\\
&\qquad\text{with}\ \alpha_{\text{JW}}= -(-1)^{k_{i+1} +...+k_{j-1}}\nonumber\\
&\sum_{\substack{q<p<r}}\tilde h_{qppr}\tilde c_q^\dag \tilde  c_p^\dag \tilde c_p\tilde c_r = \sum_{\substack{j<p<i\\p\in N_{\text{el}}^{\ket{D_1}}}}\tilde h_{jppi}\tilde c_j^\dag \tilde  c_p^\dag \tilde c_p\tilde c_i= 0 \label{apx2}\\ &\qquad\text{since} \ i<j\nonumber\\
&\sum_{\substack{r<p<q}}\tilde h_{pqrp}\tilde c_p^\dag \tilde  c_q^\dag \tilde c_r\tilde c_p = \sum_{\substack{i<p<j\\p\in N_{\text{el}}^{\ket{D_1}}}}\tilde h_{pjip}\tilde c_p^\dag \tilde  c_j^\dag \tilde c_i\tilde c_p\label{apx3}\\
&\qquad\text{with}\ \alpha_{\text{JW}} = -(-1)^{k_{i+1}\nonumber +...+k_{p-1}+k_{p+1}+...+k_{j-1}}\\
&\sum_{\substack{p<q;p<r\\ q\neq r }}\tilde h_{pqpr}\tilde c_p^\dag \tilde  c_q^\dag \tilde c_p\tilde c_r = \sum_{\substack{p<i\\p\in N_{\text{el}}^{\ket{D_1}}}} \tilde h_{pjpi}\tilde c_p^\dag \tilde  c_j^\dag \tilde c_p\tilde c_i\label{apx4}\\
&\qquad\text{with}\ \alpha_{\text{JW}} = -(-1)^{k_{i+1} +...+k_{j-1}}.\nonumber
\end{align}	
Eqs.~\eqref{apx1}-\eqref{apx4} give three non-vanishing terms from Eq.~\eqref{over10} which can create single-connected determinants.
\subsection{Two-body excitations}
Only the first term on the right hand side of Eq.~\eqref{over10} gives rise to two-body excitations. There are $\binom{4}{2}$ possible non-vanishing determinants, all having an identical form for the Jordan-Wigner phase factors.
\section{Convergence of the ASCI energy\label{additional_plots}}
\begin{figure*}
	\centering
	\begin{subfigure}
		\centering
		\includegraphics[width=.475\textwidth]{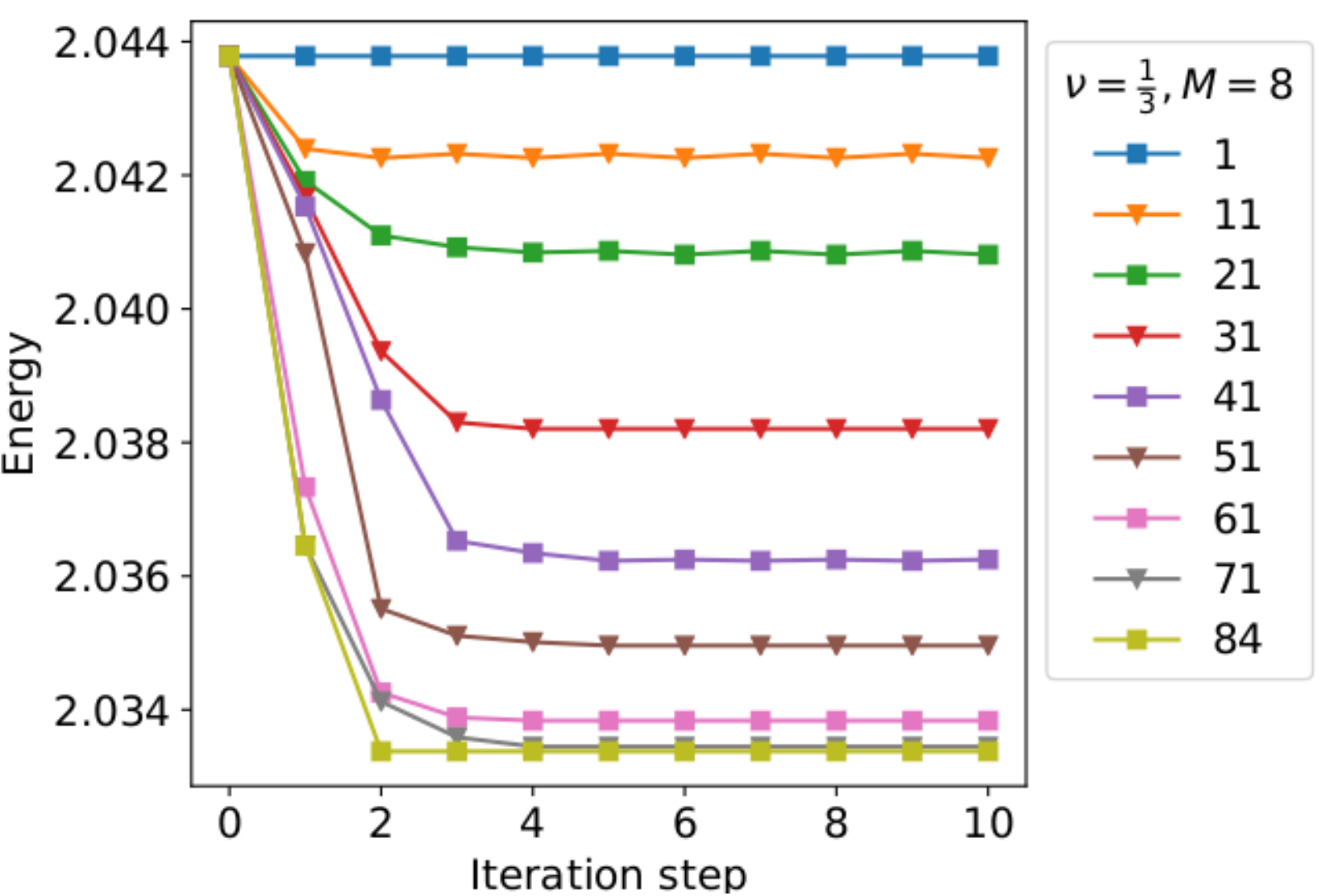}
	\end{subfigure}
	\begin{subfigure}
		\centering 
		\includegraphics[width=.475\textwidth]{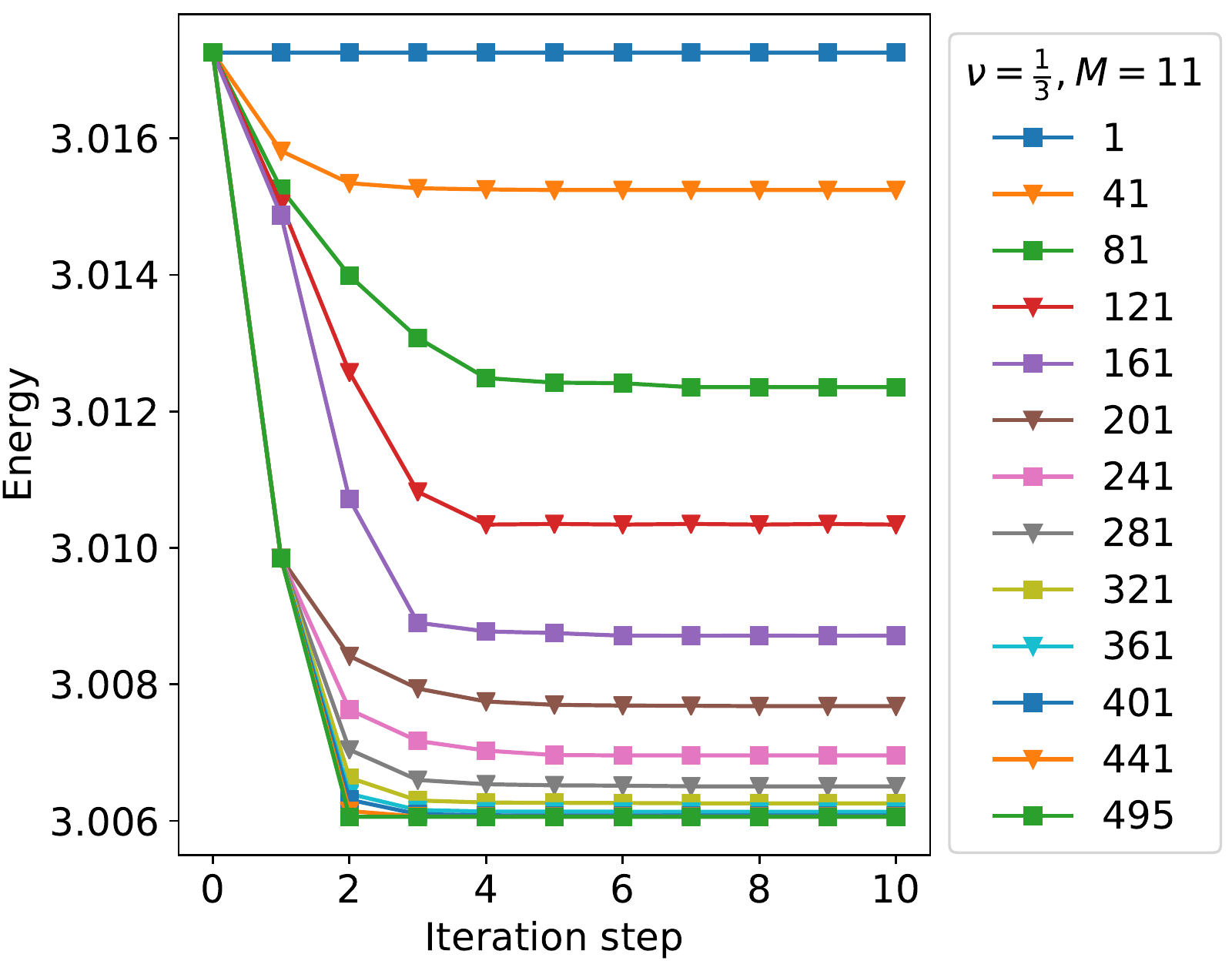}
	\end{subfigure}

	\vskip\baselineskip
	\begin{subfigure}
		\centering 
		\includegraphics[width=.475\textwidth]{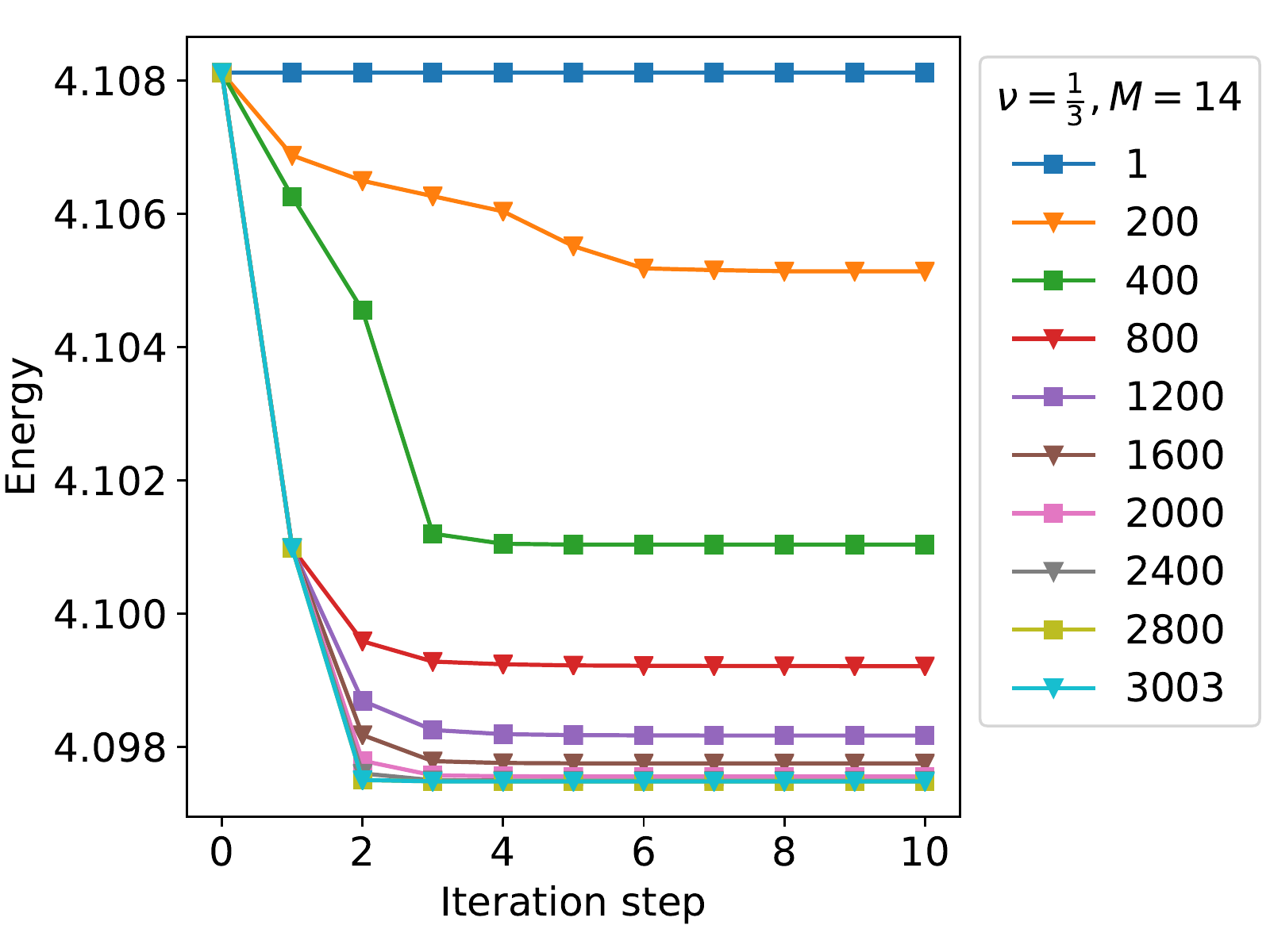}
	\end{subfigure}
	\begin{subfigure}
		\centering 
		\includegraphics[width=.475\textwidth]{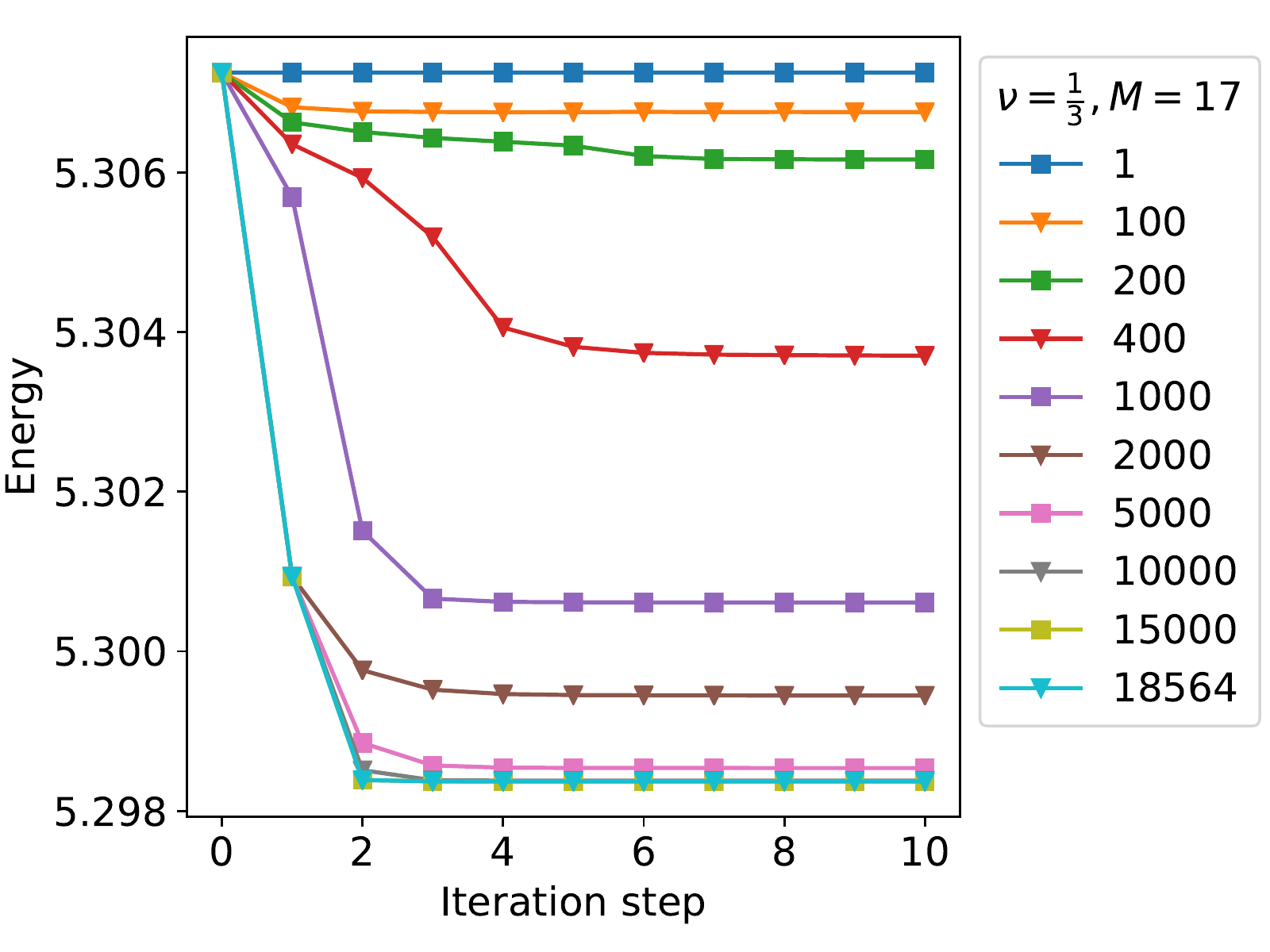}
	\end{subfigure}
	\caption{Scatter plots of the convergence of the energy obtained from diagonalizing the reduced system Hamiltonian at each of the ASCI iteration steps for a filling factor $\nu=1/3$ and systems containing $N_{\text{so}}=9,12,15,18$ spin-orbitals in the LLL. The top-most data points in each plot belong to the single reference state $\ket{\Psi_{\text{GS}}}$ obtained from the method presented in Section~\ref{hf}, while the bottom-most correspond to the FCI ASCI expansion, i.e. the exact solution, where $tdets=cdets=\binom{N_{\text{so}}}{N_{\text{el}}}$. \label{fig:energy}}
\end{figure*}

In Fig~\ref{fig:energy}, we show how the energy of the reduced system Hamiltonian converges to the exact ground state energy when increasing the number of determinants in the ASCI expansion.  A monotonic decreasing behavior as well as a convergence after about five ASCI iterations for all system sizes is clearly visible. 

\end{document}